\documentclass[11pt]{article}

\usepackage{jheppub} 

\usepackage{amsfonts, amsmath, amssymb, mathrsfs, amsthm, mathtools, array, enumerate, epsf, graphicx, grffile}

\usepackage[dvipsnames]{xcolor}

\usepackage[compatibility=false]{caption}

\usepackage{subcaption}

\PassOptionsToPackage{svgnames}{xcolor}

\usepackage{breqn}

\usepackage{xcolor}

\usepackage{breqn}

\definecolor{LightBlue}{rgb}{0.678, 0.847, 0.902}

\definecolor{cornellred}{rgb}{0.7, 0.11, 0.11}

\definecolor{darktangerine}{rgb}{1.0, 0.66, 0.07}

\definecolor{LightGreen}{RGB}{224,255,224}

\definecolor{LightYellow}{RGB}{255,255,217}

\definecolor{VeryLightBlue}{RGB}{222,239,255}

\definecolor{LightGray}{rgb}{0.85, 0.85,0.85}

\definecolor{DarkGreen}{rgb}{0,0.66, 0}

\definecolor{DarkYellow}{rgb}{0.8, 0.8, 0.}

\definecolor{LightPurple}{RGB}{241,224,240}

\usepackage{tikz}
\usetikzlibrary{decorations.pathmorphing}
\tikzset{node distance=2cm, auto}
\tikzset{snake it/.style={decorate, decoration=snake}}

\usetikzlibrary{decorations.pathmorphing}

\newcommand{\rmd}{{\mathrm{d}}}

\newenvironment{lcases}
  {\left\lbrace\begin{aligned}}
  {\end{aligned}\right.}

  \allowdisplaybreaks

\newcommand{\ket}[1]{|#1\rangle}
\newcommand{\bra}[1]{\langle #1|}
\newcommand{\braket}[2]{\langle #1|#2 \rangle}

\renewcommand\d{\text{d}}
\newcommand\e{\text{e}}
\newcommand\Tr{\mathrm{Tr}}

\hypersetup{
    pdfencoding=unicode,
	colorlinks=true,
	urlcolor=Maroon,
	linkcolor=RoyalBlue,
	citecolor=Maroon,
	pdftitle={Asymptotic bootstrap and instantons}, 				
	pdfauthor={David Berenstein, Joao Rodrigues, Victor A. Rodriguez},
	pdfdisplaydoctitle=true,
	pdfstartview=FitH,
	linktocpage=true
}

\definecolor{bleudefrance}{rgb}{0.19, 0.55, 0.91}
\definecolor{candyapplered}{rgb}{1.0, 0.03, 0.0}

\title{Asymptotic bootstrap for unitary matrix integrals at complex coupling}

\author[1,2,3]{David Berenstein}\emailAdd{dberens@physics.ucsb.edu}
\author[4,1]{\!\!, Jo\~ao Rodrigues}\emailAdd{joao.carlos.rodrigues@tecnico.ulisboa.pt}
\author[1]{\!\!, Victor A. Rodriguez}\emailAdd{varodriguez@ucsb.edu}
\affiliation[1]{Department of Physics, University of California, Santa Barbara, CA 93106, USA}
\affiliation[2]{Institute of Physics, University of Amsterdam, Science Park 904, PO Box 94485, 1090 GL Amsterdam, The Netherlands}
\affiliation[3]{Delta Institute for Theoretical Physics, Science Park 904, PO Box 94485, 1090 GL Amsterdam, The Netherlands}
\affiliation[4]{CAMGSD, Departamento de Matem\'atica, Instituto Superior T\'ecnico, Universidade de Lisboa, 1049-001 Lisboa, Portugal}

\abstract{
We apply an asymptotic bootstrap estimate method to the non-perturbative study of unitary matrix integrals. The method combines exact recursion relations with asymptotic control of large modes to achieve very high numerical precision without relying on positivity or semidefinite programming. We demonstrate its effectiveness in large-$N$ unitary matrix models by computing Wilson loop expectation values with sensitivity to exponentially small instanton effects and validating them against analytical instanton calculations. We further use the method to explore phase diagrams of unitary matrix models in complex 't Hooft coupling space, where positivity is absent, and observe that Stokes lines provide a useful proxy for additional phase boundaries. Our results show that asymptotic bootstrap estimates offer a practical and precise tool for probing the non-perturbative structure of unitary matrix integrals.
}

\begin{document}
\maketitle

\section{Introduction}
\label{sec:introduction}

Over the past several years, the bootstrap approach to strongly coupled systems has expanded beyond its original incarnation in the conformal bootstrap \cite{Belavin:1984vu,Rattazzi:2008pe,El-Showk:2012cjh} (see \cite{Poland:2018epd,Rychkov:2023wsd} for comprehensive reviews).
Applications now span a broad range of areas, including matrix models \cite{Anderson:2016rcw,Lin:2020mme,Kazakov:2021lel}, lattice gauge theory and statistical systems \cite{Anderson:2016rcw,Kazakov:2022xuh,Cho:2022lcj,Kazakov:2024ool}, as well as single-particle and matrix quantum mechanics \cite{Han:2020bkb,Berenstein:2021dyf,Lin:2023owt,Cho:2024kxn,Fawzi2024,Lawrence:2024mnj,Cho:2024owx,Lin:2025srf,Cho:2025vws}, among many others. 

At a conceptual level, a bootstrap problem is typically built from two key ingredients: (1) exact relations among the observables of interest, and (2) a positivity condition constraining these observables. 
In favorable situations, these ingredients can be formulated as a semidefinite programming (SDP) problem, which can be solved efficiently to obtain sharp, two-sided bounds.

In practice, however, realistic physical problems often require large-scale SDPs that rapidly become computationally expensive. Motivated by this limitation, a simple approximate bootstrap-like method was introduced in \cite{Berenstein:2025itw} (and reviewed below) for probability measures on the circle, which does not require solving an SDP. 
Instead, the method exploits the expected asymptotic decay of observables to produce very high numerical precision estimates. 
From this perspective, the approach may be viewed as a truncation scheme that leverages asymptotic behavior to accurately reconstruct observables.\footnote{In spirit, this method is similar to the Gliozzi truncation approach \cite{Gliozzi:2013ysa} developed in the context of the conformal bootstrap. See also \cite{MARCHESINI1985225,Anderson:2016rcw,Li:2017ukc,Bender:2023ttu,Hu:2025yrs}, for example, for related truncation-based methods applied to models similar to those considered here. A notable difference is that the present method leads to a
purely linear computation of the estimates.} 
Importantly, this truncation method does not rely on positivity, and is therefore applicable in a wider range of physical settings.

The purpose of \textbf{this paper} is to illustrate the effectiveness and high numerical precision of this method --- which we refer to as the \emph{asymptotic bootstrap estimate}\footnote{We use this terminology to distinguish it from SDP approaches that yield rigorous bootstrap \emph{bounds}.} --- in a nontrivial setting of physical interest: unitary matrix integrals. 
First, we give an argument for the exponentially fast convergence of the method and derive asymptotic estimates of the numerical error.
In particular, we show that the method is sufficiently precise to resolve exponentially small instanton effects. 
Moreover, it provides a practical tool for exploring phase diagrams of unitary matrix models in complex coupling space, where no obvious positivity condition holds and SDP-based approaches are not available.

In \textbf{section~\ref{sec:method}}, we introduce the asymptotic bootstrap estimate method in the context of simple probability measures on the circle and analyze its asymptotic error.
We show that the method achieves exponentially high numerical precision even in the absence of positivity. 

In \textbf{section~\ref{sec:Wilson loops and instantons}}, we demonstrate the effectiveness of the method by
performing a detailed comparison with non-perturbative instanton contributions
to Wilson loop expectation values in large-$N$ unitary matrix integrals, see figures \ref{fig:Instantons10} and \ref{fig:Instantons3}.
For these instanton effects, we also provide a complementary perspective based on a Lefschetz-thimble analysis, presented in appendix~\ref{app:LefschetzThimbleAnalysis}.

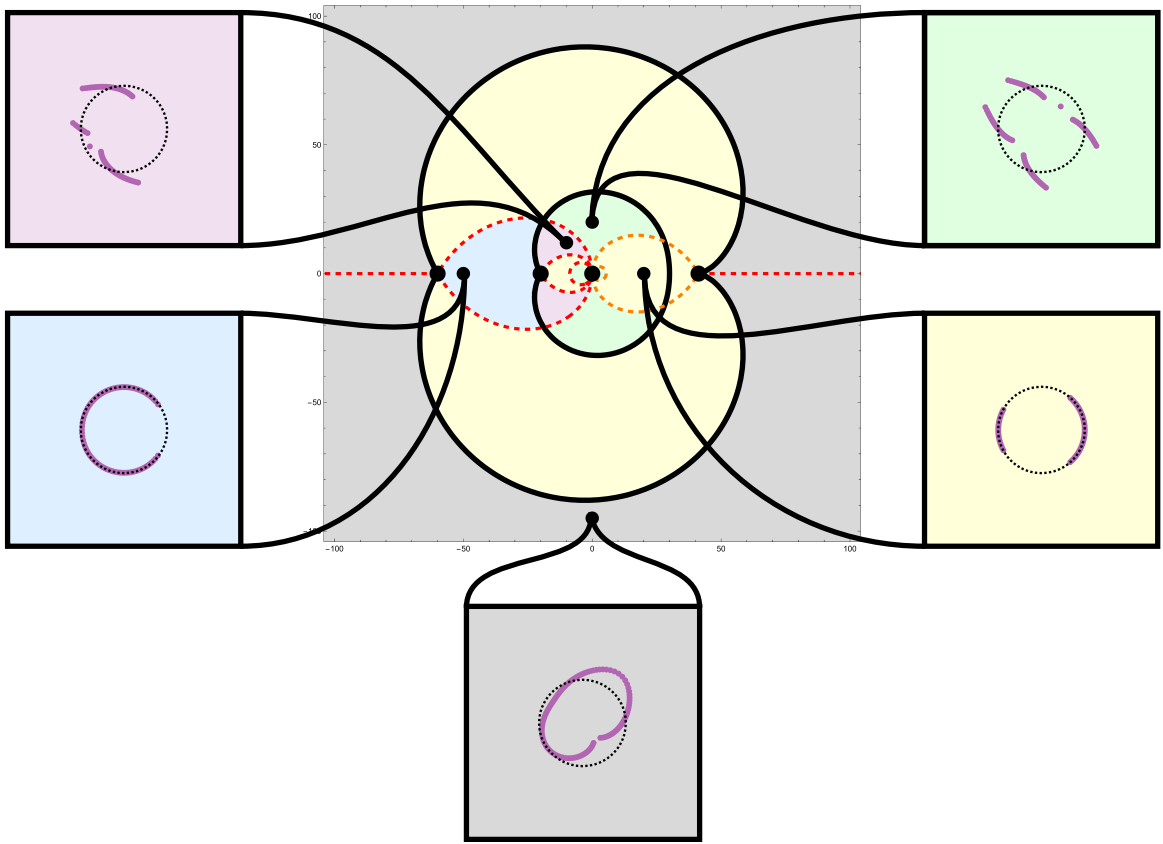
\begin{figure}
    \centering
    \begin{tikzpicture}
        \node at (0,0) {\includegraphics[scale = 0.30]{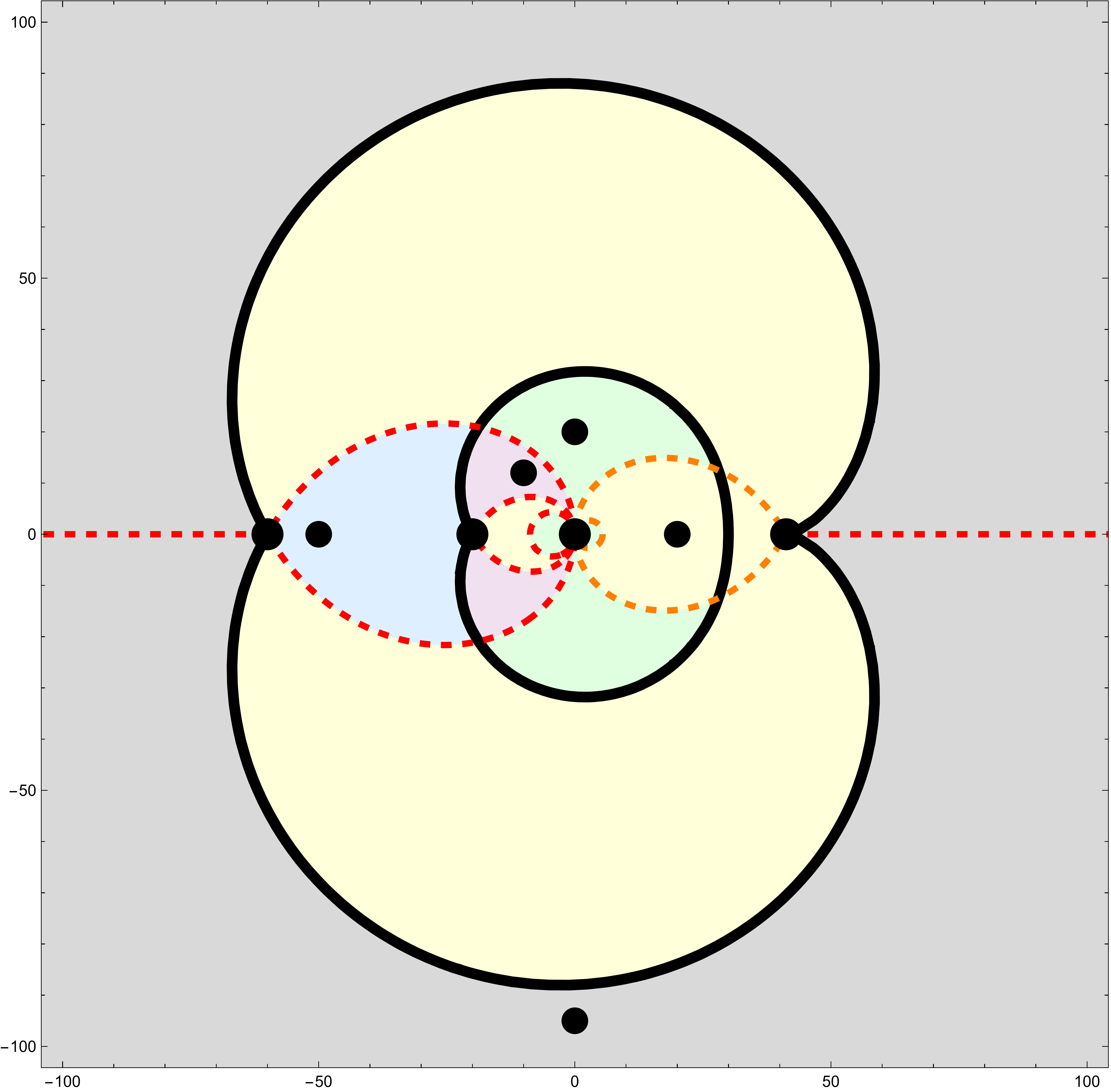}};

\node[anchor=center] at (6.1,2) {\includegraphics[scale = 0.22]{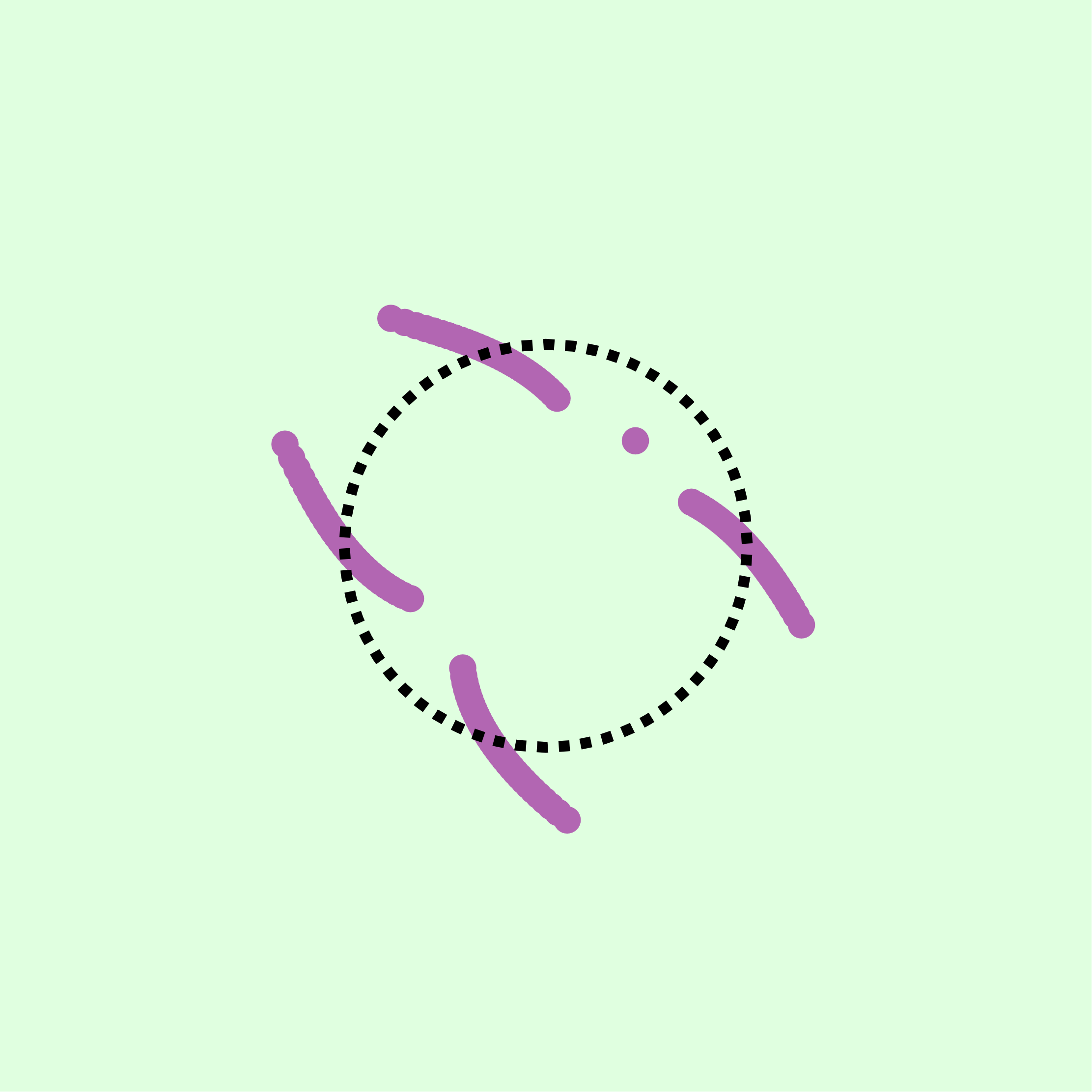}};

\node[anchor=center] at (6.1,-2) {\includegraphics[scale = 0.22]{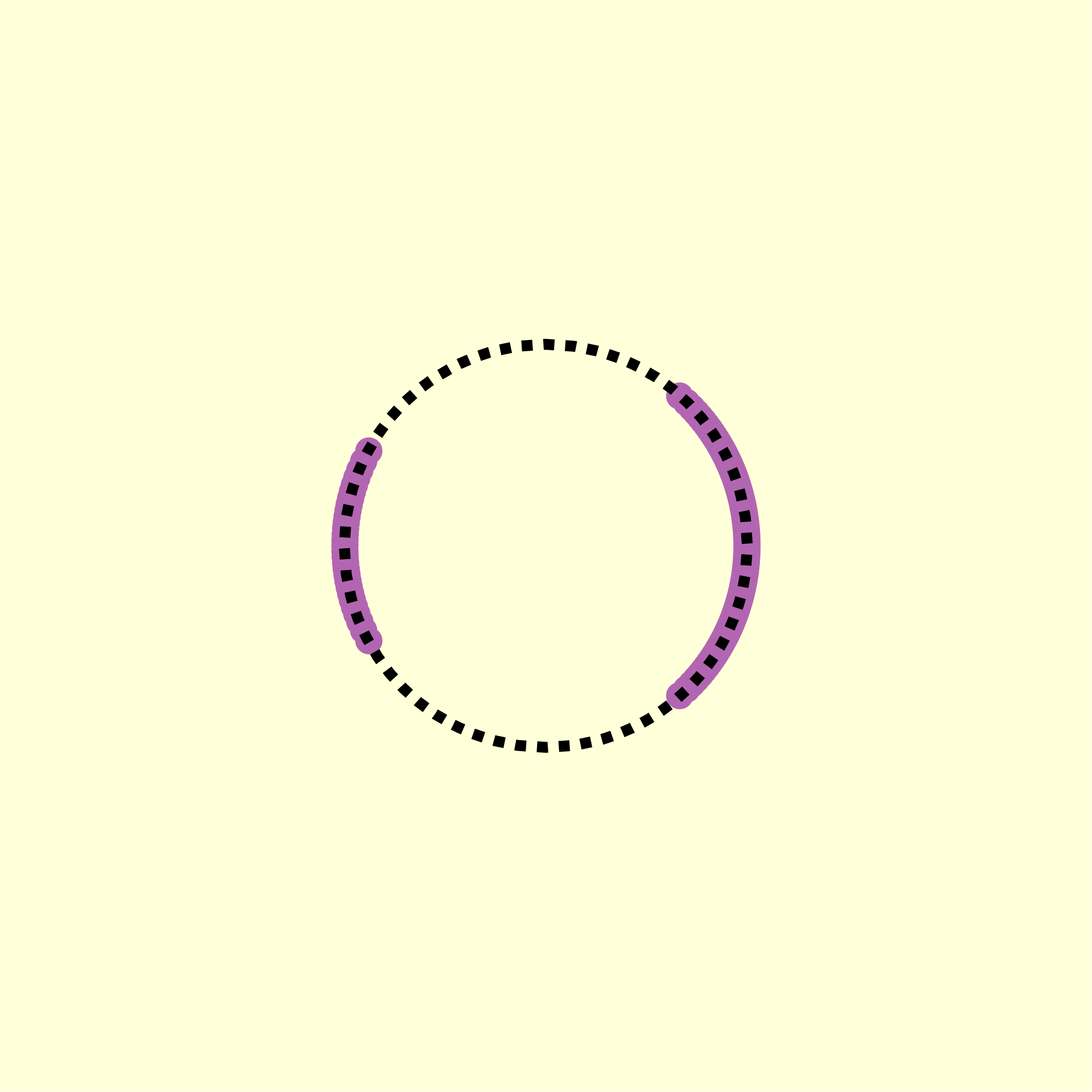}};

\node[anchor=center] at (-6.1,2) {\includegraphics[scale = 0.22]{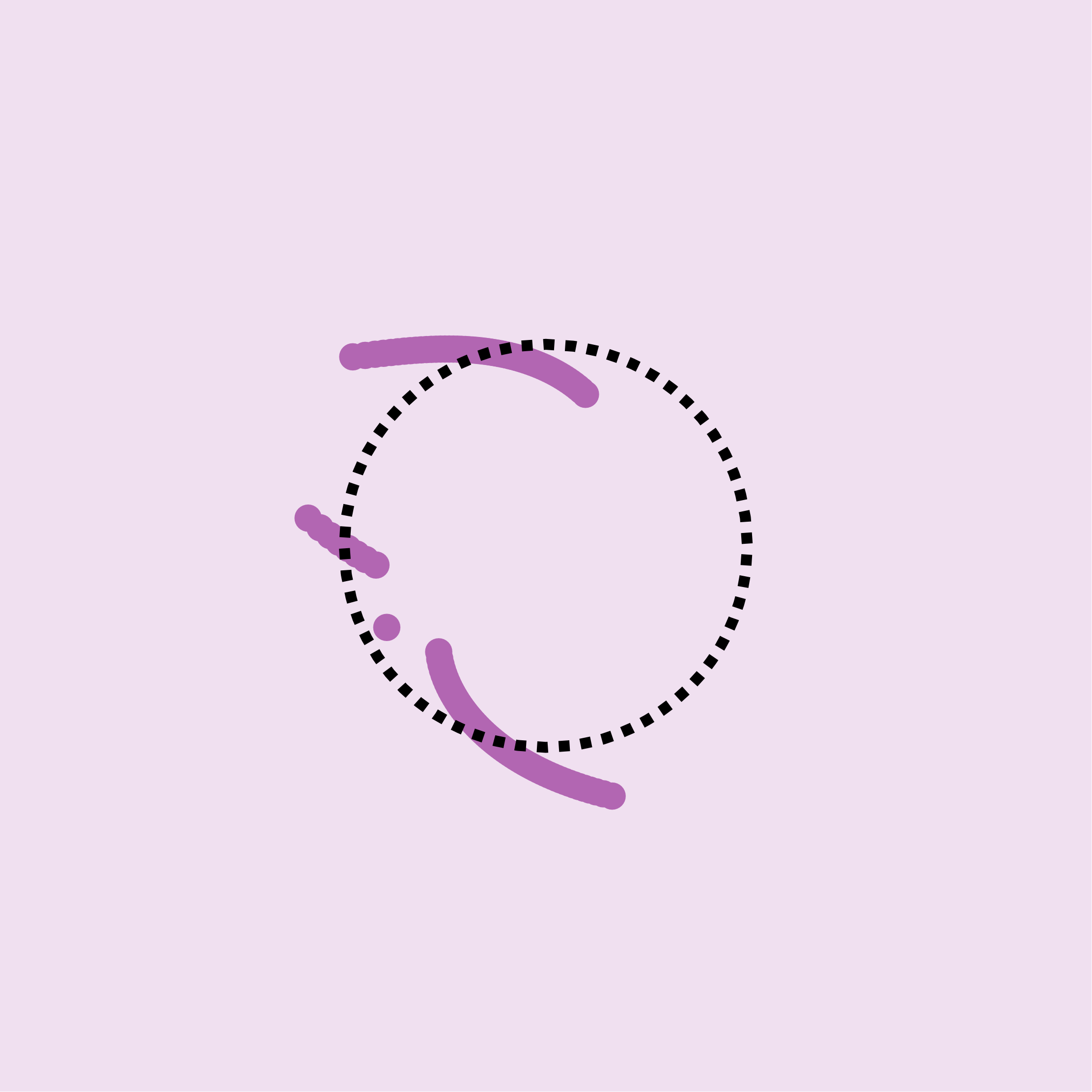}};

\node[anchor=center] at (-6.1,-2) {\includegraphics[scale = 0.22]{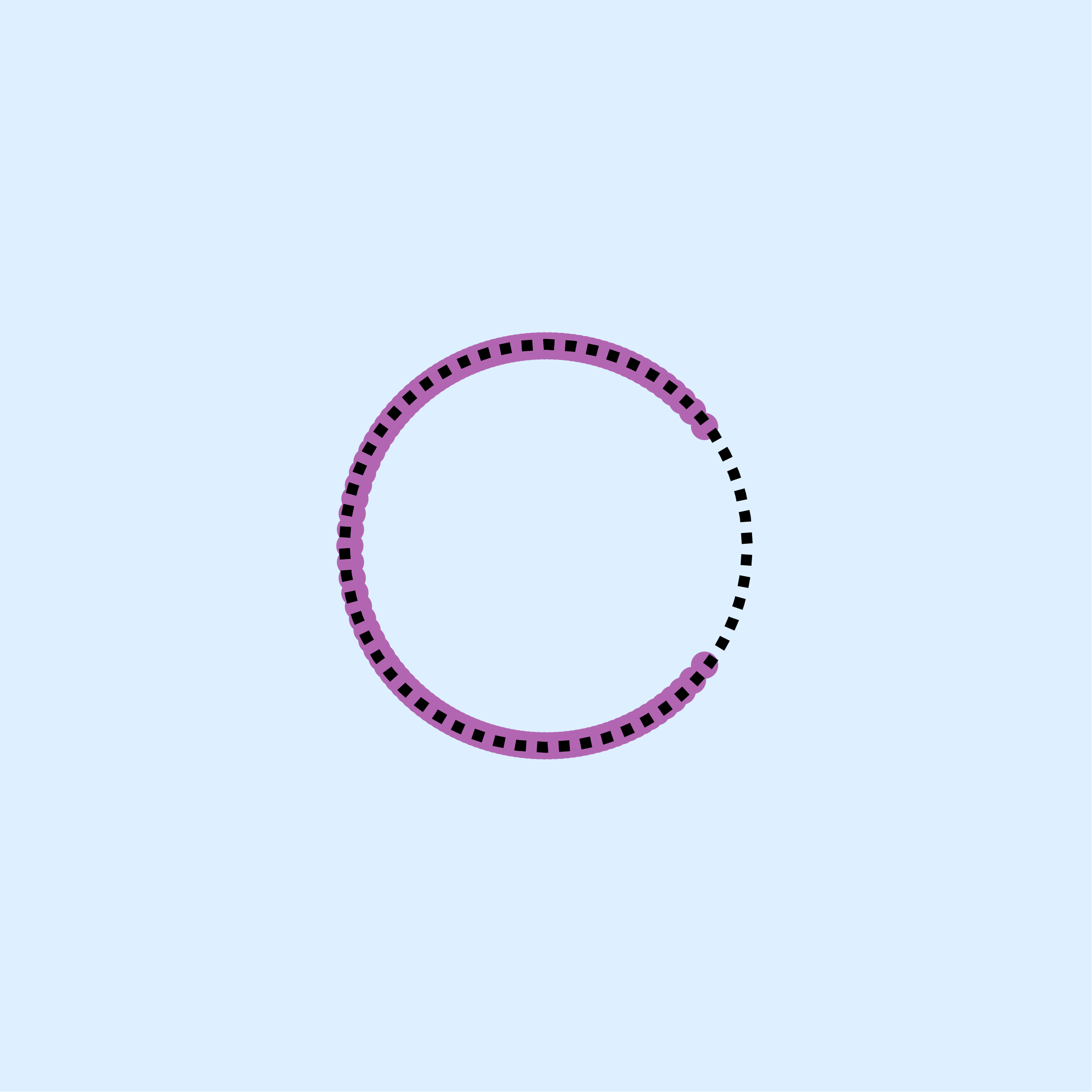}};

\node[anchor=center] at (0,-5.9) {\includegraphics[scale = 0.22]{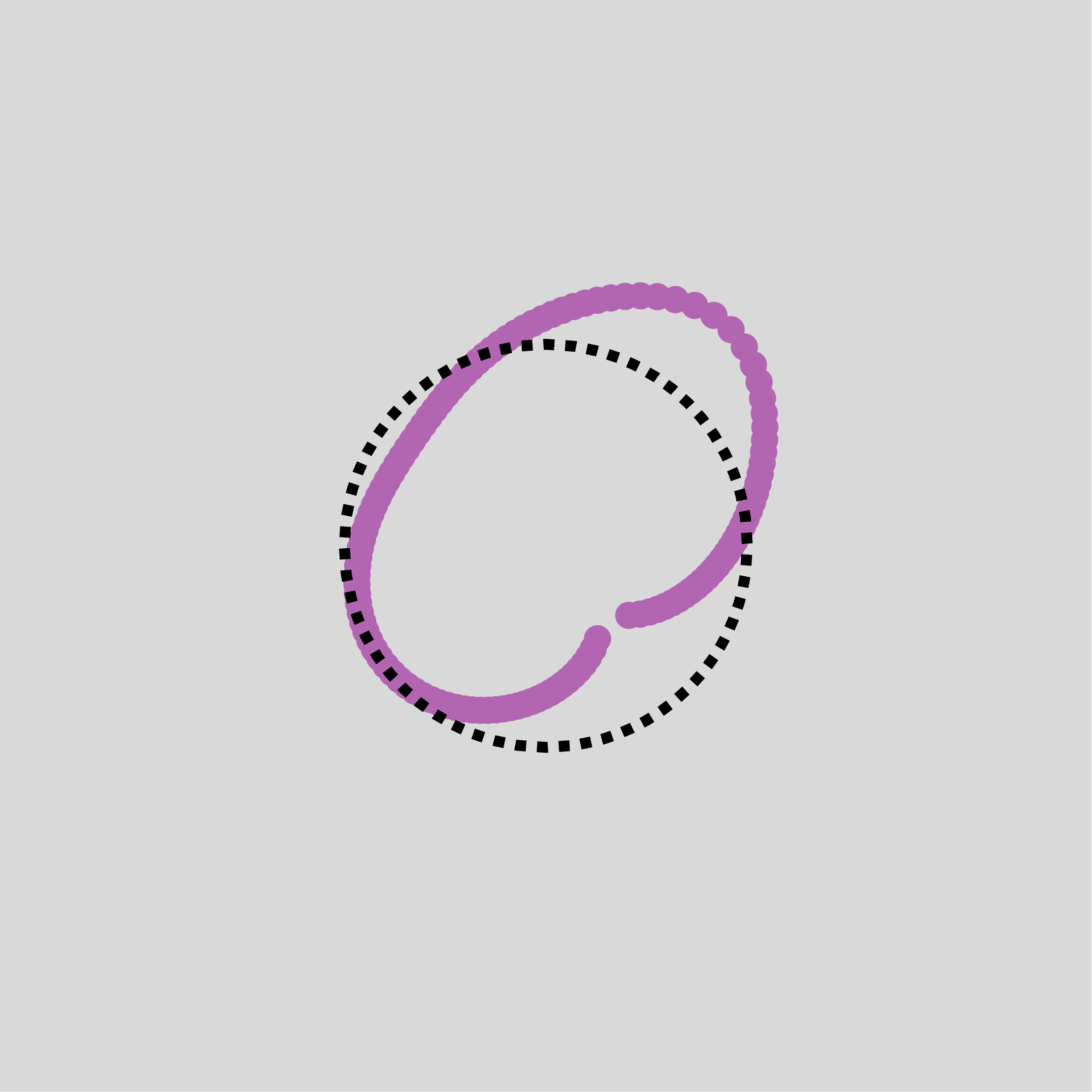}};

\draw[line width = 2pt] (4.545,0.45) rectangle (4.545+3.1,3.55);

\draw[line width = 2pt] (4.545,0.45) rectangle (4.545+3.1,3.55);

\draw[line width = 2pt] (4.545,-0.45) rectangle (4.545+3.1,-3.55);

\def\hcord{6.1};
\def\vcord{-7.9};

\draw[line width = 2pt] (-4.545+\hcord,0.45+\vcord) rectangle (-4.545-3.1+\hcord,3.55+\vcord);

\draw[line width = 2pt] (-4.545,0.45) rectangle (-4.545-3.1,3.55);

\draw[line width = 2pt] (-4.545,-0.45) rectangle (-4.545-3.1,-3.55);

\coordinate (Sample1) at(0.822,0.07);
 
\coordinate (Sample3) at (-0.24,0.50);
 \coordinate (Sample2) at  (0.12,0.7);
 \coordinate (Sample4) at (-1.57,0.07);
 \coordinate (Sample5) at (0.11,-3.1);

\draw[line width = 2pt] (Sample5) to[out = -85, in = 90] (-1.5455,-4.37);

\draw[line width = 2pt] (Sample5) to[out = -85, in = 90] (-1.5455+3.101,-4.37);

 \draw[line width = 2pt] (Sample1) to[out = -90, in = 180] (4.545,-0.45);
  \draw[line width = 2pt] (Sample1) to[out = -90, in = 180] (4.545,-3.55);

  \draw[line width = 2pt] (Sample2) to[out = 90, in = 180] (4.545,0.45);
  \draw[line width = 2pt] (Sample2) to[out = 90, in = 180] (4.545,3.55);

  \draw[line width = 2pt] (Sample3) to[out = 180-45, in = 0] (-4.545,0.45);
  \draw[line width = 2pt] (Sample3) to[out = 180-45, in = 0] (-4.545,3.55);

  \draw[line width = 2pt] (Sample4) to[out = -90, in = 0] (-4.545,-0.45);
  \draw[line width = 2pt] (Sample4) to[out = -90, in = 0] (-4.545,-3.55);
        
    \end{tikzpicture}
    \caption{Representative root plots of the orthogonal polynomial $p_n$ for $n=N =125$, evaluated at five values of the complex coupling $g$ of the unitary matrix model with a quadratic single-trace potential $V(U)\propto g^{-1}(\text{Tr}(U)+\text{Tr}(U^{-1})+\text{Tr}(U^2)+\text{Tr}(U^{-2}))$ (see \eqref{eq:promotedpotential}), with one sample taken from each colored region of the phase diagram. 
    The central figure shows the phase diagram in the complex $g$-plane, while the surrounding panels display the corresponding root distributions in the eigenvalue $z$-plane.
    These colored regions correspond to phase approximations classified by the number of cuts in the associated eigenvalue spectral density. Each root plot is displayed alongside the phase diagram, with the corresponding sample indicated by a black dot and connected to its root plot. Within each panel, the polynomial roots are shown as purple dots, and the unit circle is indicated by a black dashed line. The specific sample values used here are listed in subsection~\ref{subsec:Complicatedmodel}.}
    \label{fig:CMMintroduction}
\end{figure}

Finally, in \textbf{section~\ref{sec:phase diagram}}, we apply the asymptotic bootstrap
estimate method to a numerical exploration of the phase diagram of unitary matrix
models in complex ’t~Hooft coupling space, where positivity is not available.
In particular, we study the Gross-Witten-Wadia model \cite{Gross:1980he,Wadia:1980cp} (see figures~\ref{fig:GWWphasediagram}, \ref{fig:GWWonecutsamples}, \ref{fig:GWWtwocutsamples} and \ref{fig:GWWungappedsamples}), as well as a model with a more complicated potential involving single-trace terms quadratic in the unitary matrix. 
The latter exhibits a richer phase structure, with multiple phases corresponding to distinct cut configurations of the eigenvalue spectral density. Figure \ref{fig:CMMintroduction} provides a preview of representative plots for each phase. This exploration leverages a large $N$ correspondence between the accumulation locus of orthogonal polynomial roots and the support of the planar eigenvalue spectral density. This correspondence has been understood at different levels of rigor in, \textit{e.g.}, \cite{b09,b07,s07,bt11,aam13,hkl14,bt16} for Hermitian matrix models. 
Furthermore, an exhaustive numerical exploration of the complex phase diagram of Hermitian matrix models, based on this correspondence, has already been carried out in \cite{krsst25a} whose approach we will extend here for single-trace unitary matrix models. 

An interesting observation is that, while phase boundaries within a given phase are typically associated with anti-Stokes lines (see black lines in figure \ref{fig:CMMintroduction}) of the instanton action, our numerical results suggest that some Stokes lines can also serve as a useful \emph{proxy} for additional phase boundaries (see dashed lines in figure \ref{fig:CMMintroduction}). 
This behavior is corroborated by the accumulation of zeros of orthogonal polynomials constructed from the moments obtained via the asymptotic bootstrap estimate method. 
Additional appendices provide further details of computations presented in the main text.

\section{An asymptotic bootstrap estimate method}

\label{sec:method}

In this section, we setup, review and generalize the shoestring bootstrap method introduced in \cite{Berenstein:2025itw} to efficiently compute moments (Fourier coefficients) of a rather broad class of measures on $S^1$. These problems and their treatment are a simplified toy model for problems of quantum mechanics on the circle whose solution can also be approached via bootstrap methods \cite{Aikawa:2021eai,Berenstein:2021loy,Tchoumakov:2021mnh}.

The method leverages the asymptotic (in the Fourier index) behavior of the moments and allows for a very efficient linear computation.
We start by setting up the problem on $S^1$ and reviewing the positivity bootstrap approach. We then to introduce the shoestring bootstrap and estimate the associated error bounds, already observed in \cite{Berenstein:2025itw}. Finally, we remark that the bootstrap method is trivially generalizable to more generic complex measures. The resulting method solely relies on asymptotic constrains of the moments and thus should be seen as an asymptotic bootstrap estimate method.

\paragraph{Setup and the positivity bootstrap bounds:}
Let us consider the following problem on $S^1$. Assume that we have a probability measure given by
\begin{equation}\label{eq:circle measure}
\d\mu = \exp [V(\theta)] \d\theta \,,
\end{equation}
where for simplicity we take $V(\theta)$ to be a periodic function given by a cosine expansion, rather than a more general periodic function,
\begin{equation}
    V(\theta)= c_0+\sum_{k=1}^K 2 \frac{c_k}{k} \cos(k \theta) \,,
\end{equation}
with no sine terms. The variable $\theta$ is periodic with period $2\pi$. We begin by assuming that the coefficients $c_k$ are real-valued, as befits a probability distribution. Importantly, this assumption will be relaxed later in the paper. The constant $c_0$ is adjusted so that the probability measure is unit-normalized,
\begin{equation}
    \int_{-\pi}^\pi \d\mu = 1 \,.
\end{equation}
Our goal is to determine the Fourier coefficients, or moments, of the measure $\d\mu$, given by
\begin{equation}
a_n = \int_{-\pi}^\pi \d\mu \exp(i n\theta) \,.
\end{equation}
The simplifying assumption guarantees that $a_n=a_{-n}$, as a consequence of the symmetry $V(\theta)=V(-\theta)$. 
Moreover, one can show that the $a_n$ are real-valued in the case we have described.

The $a_n$ satisfy a recursion relation that arises from the integration-by-parts formula
\begin{equation}
\int_{-\pi}^\pi \partial_\theta( \exp(i n \theta +V(\theta)) \d \theta=0 \,,
\end{equation}
which can be written as the following identity
\begin{equation}\label{eq:deriv rec relations}
-\sum_{k=1}^K c_k a_{n-k} +  n a_n + \sum_{k=1}^K  c_k a_{n+k} =0 \,.
\end{equation}
Together with the equation $a_n=a_{-n}$, this implies that the recursion determines all moments once $a_0=1, a_1, \dots, a_K$ are specified.

The positivity bootstrap program leverages the fact that the measure $\d\mu$ satisfies
\begin{equation}
\int_{-\pi}^\pi \d\mu |f(\theta)|^2 \geq 0
\end{equation}
for any finite (truncated) Fourier series $f(\theta)= \sum_{s=0}^S  b_s \exp(i s \theta)$.
This condition is equivalent to requiring that the $(S+1)\times(S+1)$ Toeplitz matrix
\begin{equation}\label{eq:PSD matrix}
    M_{S}= \begin{pmatrix} 1& a_1 & a_2 & \dots\\
    a_1 &1 & a_1 & \ddots\\
    a_2& a_1 & 1 &\ddots\\
    \vdots & \ddots & \ddots & \ddots
    \end{pmatrix}\succeq 0
\end{equation}
is positive semidefinite. Given the recursion, $M_S$ depends on only $K$ variables. 
Optimization problems of the form $\min a_\ell$ (or $\max a_\ell$) subject to $M_S\succeq 0$ for fixed $S$ are semidefinite programming problems (SDPs) that yield two-sided bounds on the possible values of the $a_\ell$. 
The bounds are nested, meaning that $[\min a_\ell,
\max a_\ell]_{S+1}\subset [\min a_\ell, \max a_\ell]_{S}$.
Moreover, the feasible sets $\{ a_{1}, \dots, a_{K} \mid M_S\succeq 0\}$ are convex and nested. 
When we let $S\to \infty$, it is expected that the $a_\ell$ converge exponentially fast to their true values. 

\paragraph{Cheap and shoestring bootstrap methods and error estimates:}
A cheap bootstrap two-sided bound can be obtained by demanding that the tail of Fourier coefficients satisfies
$|a_{S}|\leq 1, \dots, |a_{S-K+1}|\leq 1$, which is a simple subset of the positivity conditions \eqref{eq:PSD matrix}. 
These inequalities are sufficient to bound the values of $a_{1},\ldots,a_K$ within exponentially small intervals. 
One can then check positivity of $M_{R}$ for some $R<S$, thus finding a point in the feasible region of the SDP problem for some smaller $R$ that is very close to $S$. 

A second, even better method arises from the observation that asymptotically $a_S\to 0$ exponentially fast. 
A very good approximation is therefore to set $a_S=0, \dots, a_{S-K+1}=0$ for sufficiently large $S$. Making use of the recursion \eqref{eq:deriv rec relations}, one can then obtain a very precise estimate for $a_1, \dots, a_K$, with roughly twice as many digits of precision compared to the cheap bootstrap bounds. 
Since for each $S$ this procedure produces a unique value of $a_1, \dots, a_K$ by solving a finite linear system, the solution can be obtained without invoking an SDP solver.
Positivity is then checked a posteriori and does not enter into the determination of the solution.

This simple method was shown to yield exponentially accurate estimates for the moments of the distribution in \cite{Berenstein:2025itw}. 
Here, we give an argument for why this method works so well for this class of distributions on the circle.
The idea is to understand both the bounds and the errors in the estimates as arising from properties of the general solutions of the recursion relation.


From Fourier analysis, we expect that $a_S\to 0$ as $S\to\infty$. We take fast convergence to mean that, asymptotically, the sequence satisfies
$|a_S|\ll |a_{S-1}|\ll \dots $.
Under this assumption, we approximate the recursion by retaining only the leading terms,
\begin{equation}\label{eq:rec approx}
    - c_K a_{S-K} +S a_S\simeq 0 \,,
\end{equation}
which yields the estimate\footnote{This asymptotic analysis has some similarity with the approach in \cite{Huang:2025sua} to study quantum mechanics problems on the circle. }
\begin{equation}
    a_S \simeq \frac{c_K}{S} a_{S-K} \,.
\end{equation}

Repeated application of the approximate recursion \eqref{eq:rec approx} yields an estimate for each congruence class $S \bmod K$. One finds 
\begin{equation}\label{eq:asym decaying aS}
a_S \simeq d_{S\bmod K} \times \left(\frac{c_K}{K}\right)^{\lfloor\frac{S}{K}\rfloor} \frac{1}{\Gamma\left(\frac{S}{K}+1\right)} \,
\end{equation}
where $d_{S\bmod K}$ is some constant that depends on the congruence class $S\bmod K$ but does not grow with $S$. 
The gamma function in the denominator shows that the large Fourier modes $a_S$ decay to zero faster than exponentially for sufficiently large $S$. 

Having argued for the asymptotic decay of the physical Fourier modes, let us now examine the general solution of the recursion relation.
In addition to the decaying solution discussed above, the recursion admits linearly independent solutions that grow rapidly as $S\to\infty$.

The growing solution can be estimated by reversing the asymptotic argument used previously. 
As $S$ increases, the coefficient multiplying $a_S$ in the recursion grows linearly in $S$, making the evolution increasingly sensitive to earlier values.
In the regime $|a_S|\ll |a_{S+K}|\ll \dots$, the recursion can be approximated by
\begin{equation}\label{eq:rec approx 2}
S a_S + c_K a_{S+K} \simeq 0 \, .
\end{equation}
This leads to a growing solution of the form
\begin{equation}
    a^{\mathrm{gr}}_S \simeq d'_{S\bmod K} \times \left(\frac{K}{ |c_K|}\right)^{\frac{S}{K}-1} \Gamma\left(\frac{S}{K}\right) \,,
\end{equation}
where again $d'_{S\bmod K}$ is some constant that depends on the congruence class $S\bmod K$ but does not grow with $S$.

Thus, if the initial data, i.e. a presumed value for a given low Fourier mode, contain an error of size $\delta$, this error is amplified at large $S$ as
\begin{equation}
    |a_S| \sim \delta \, \left(\frac{K}{|c_K|}\right)^{\frac{S}{K}} \Gamma\left(\frac{S}{K}\right).
\end{equation}
In the cheap bootstrap, one only enforces bounds of order one on $a_S$, allowing errors of size
\begin{equation}
\delta_{\text{cheap}} \sim \left(\frac{|c_K|}{K}\right)^{\frac{S}{K}} \frac{1}{\Gamma\left(\frac{S}{K}\right)} \, .
\end{equation}
By contrast, in the shoestring bootstrap one sets $a_S=0$, so that the mismatch at large $S$ is of the same order as the true physical value of $a_S$.
As a result, the induced error in this case is suppressed by an additional factor of $\left(\frac{|c_K|}{K}\right)^{\frac{S}{K}} /\hspace{2pt} \Gamma\left(\frac{S}{K}\right)$.
Hence, the error in the shoestring bootstrap is parametrically the square of the error in the cheap bootstrap. 
This explains why the shoestring method typically yields roughly \emph{twice} as many correct digits in the estimate of $a_1,\dots,a_K$.\footnote{A potential concern is that some of the coefficients $d_{S\bmod K}$ appearing in the asymptotic solutions might vanish.
This can only occur when the potential has an additional discrete symmetry, so that certain Fourier modes vanish identically.
In such cases, setting $a_S=0$ for those values of $S$ is in fact exact, and the above analysis can be restricted to the nonvanishing subsequence, for which the same error estimates hold.}

The asymptotic error estimates also indicate how large $S$ must be for either method to be effective.
Parametrically, one requires that $(|c_K|/K)^{\frac{S}{K}}\ll \Gamma\left(\frac{S}{K}\right)\sim (S/K)^{\frac{S}{K}}$.
Larger values of $|c_K|$ therefore require pushing the truncation to larger $S$ (in typical situations, once $S\gtrapprox|c_K|$), while once this regime is reached the precision improves rapidly with only modest further increases in $S$.
In this regime, the subleading terms neglected in the approximate recursion are also parametrically suppressed, unless there is a large hierarchy among the coefficients $c_k$.

The upshot is that the shoestring bootstrap provides an accurate solution without the need to run a full SDP.
The cheap bootstrap already converges exponentially fast and slightly overestimates the convex region defined by $M_S\succeq 0$, so the SDP can only marginally improve upon it.
By contrast, the shoestring bootstrap method suppresses the growing solution parametrically more strongly, explaining its significantly better numerical performance.
Here, violations of positivity are not expected, and a single positivity check suffices as a consistency test.

\paragraph{Letting go of positivity:}
The analysis above relies only on the recursion relations satisfied by the Fourier coefficients and on their asymptotic behavior at large mode number.
In particular, none of the estimates require the coefficients $c_k$ to be real or the measure $\d\mu$ to be positive.
When the $c_k$ are complex, positivity is lost, but the recursion relations remain unchanged, and the asymptotic analysis continues to apply.

In this more general setting, $a_S,a_{S-1},\dots,a_{S-K+1}=0$ still yields asymptotically fast convergence to the correct solution.
This motivates a method that replaces positivity with asymptotic control, which we refer to as an \textbf{asymptotic bootstrap estimate}.

\begin{figure}[ht]
\begin{center}
\includegraphics[width=10cm]{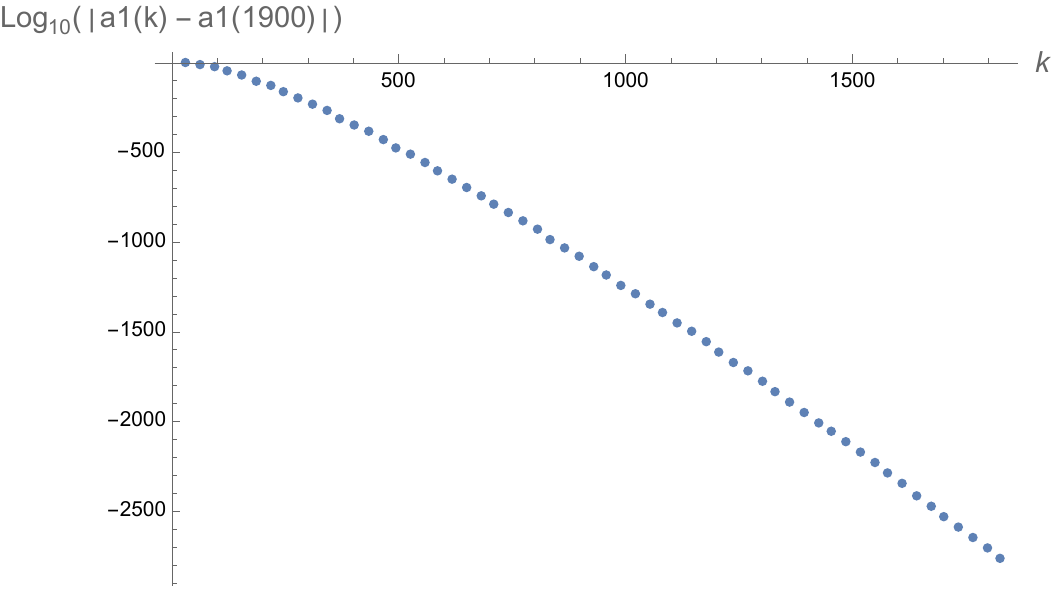}
\caption{Absolute error for the example \eqref{eq:example error}, as a function of the truncation order $k$.}\label{fig:Error1_20_20}
\end{center}
\end{figure}

\begin{figure}[ht]
\begin{center}
\includegraphics[width=10cm]{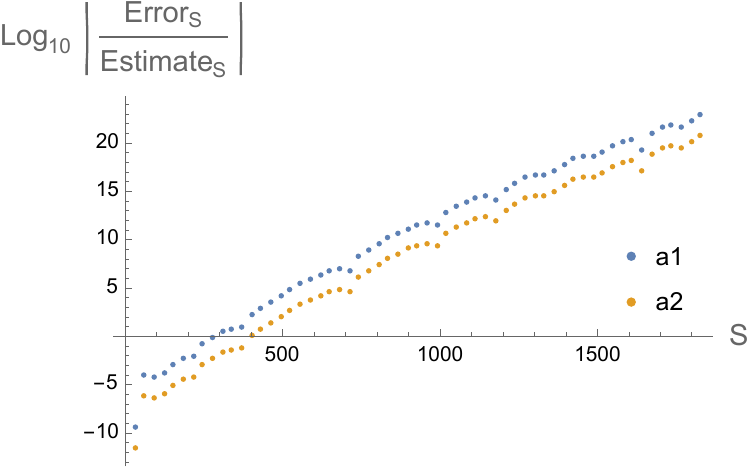}
\caption{Ratio of the numerical error divided by an estimated asymptotic error \eqref{eq:asym error}, shown for both $a_1$ (in blue) and $a_2$ (in yellow).
The errors appear highly correlated and exhibit a mild downward curvature. }\label{fig:error_estimated}
\end{center}
\end{figure}

\paragraph{An example of the estimate and its error:}
As an illustration of the method and its error estimates, we consider the problem of computing the first two Fourier moments $a_1$ and $a_2$ for the normalized measure
\begin{equation}\label{eq:example error}
\int \d\mu= {\cal N}\int \d\theta \exp(20 \cos(\theta)+20\cos(2\theta)) \,,
\end{equation}
where ${\cal N}$ is a constant that ensures unit normalization. 
To obtain a reference value, we solve the recursion relations at a very large truncation order, choosing $S=1902$ and imposing $a_S=a_{S+1}=0$.
The resulting values of $a_1$ and $a_2$ will be used as a benchmark in the comparison below.
We then repeat the computation for truncation orders
$S\in (32,1850)$ and compare the resulting values with the benchmark obtained at $S=1902$.
For a given truncation order $S$, the estimate of the absolute error is $|a_1(S)-a_1(1902)|$.
To quantify the rate of convergence, we plot the base-10 logarithm of this error as a function of $S$, as shown in figure~\ref{fig:Error1_20_20}.  
From this plot, we infer that the computation achieves better than 2800 digits of precision. 
We now compare the numerical results with the asymptotic error estimate discussed
previously, which is parametrically given by
\begin{equation}\label{eq:asym error}
    \delta \sim  \left(\frac{|c_K|}{K} \right)^{2S/K}\frac{1}{\Gamma\left(\frac S K\right)^2} \,,
\end{equation}
up to a constant of order one. The ratio of the numerical error to the asymptotic error estimate \eqref{eq:asym error} is plotted in figure~\ref{fig:error_estimated}.
Note that the asymptotic error estimate differs from the numerical error by more than an overall constant, showing a systematic drift.
The estimate deviates by roughly $20$ digits over a range of $2000$ truncation orders (about 1\% error in the number of digits per step). That is, the prediction of roughly $2800$ digits of precision is only off by the last $20$ digits. 
Up to this small correction, our method indeed achieves the expected high level of precision. 

The reason for this discrepancy lies in the fact that the asymptotic error analysis above assumes that subleading terms in the recursion are parametrically suppressed.
Using \eqref{eq:asym decaying aS}, we have that
$a_{S+1}/{a_S}\simeq (C_K/K)^{1/K}/(S/K)^{1/K}\simeq S^{-1/K}$ 
so that the neglected terms are suppressed only by fractional inverse powers of
the truncation order.
As a result, the onset of the true asymptotic regime requires large values of $S$, especially for larger $K$.

Thus, the observed drift between the numerical error and the asymptotic estimate reflects pre-asymptotic effects rather than a breakdown of the method.
Within this interpretation, the numerical results are fully consistent with the theoretical estimate, which correctly captures the overall scale of the error and the resulting level of precision.
Additional details of the numerical evaluations and a better estimate of the error at finite $n$ which has not yet reached the asymptotic regime can be found in the appendix \ref{sec:numerics_rec}.

In the following sections, we illustrate this approach by applying it to unitary matrix models, where it can be tested numerically for other observables.

\section{Wilson loops and non-perturbative effects in unitary matrix models}
\label{sec:Wilson loops and instantons}

In this section we test the effectiveness of the asymptotic bootstrap estimate on a simple but nontrivial example: a unitary matrix integral,
\begin{equation}
    Z_N(\boldsymbol{t}) = \frac{1}{\text{Vol}(U(N))} \int_{U(N)} \d U \, \exp\!\left(V(U)\right) \,,
    \label{eq:unitmatrix1}
\end{equation}
where $\d U$ is the Haar measure on $U(N)$ and $V(U)$ is a single-trace potential.
We consider potentials of the form
\begin{equation}
   V(U) =  \sum_{k=1}^{K}\frac{t_k}{k}\left(\Tr(U^k)+\Tr(U^{-k})\right),
   \label{eq:firstpotential}
\end{equation}
with $K \in \mathbb{Z}_{\geq 1}$ and couplings $\boldsymbol{t} \in \mathbb{C}^K$.

Our main observables are (unnormalized) Wilson loop expectation values, 
\begin{equation} \label{eq:unitmatrix2}
    \widehat W_n(N) = \left\langle \Tr(U^n)\right\rangle \,,
\end{equation}
for $n \in \mathbb{Z}_{\geq1}$.
We approach their non-perturbative computation using the machinery of orthogonal polynomials, which can be implemented numerically via the asymptotic bootstrap estimate introduced in section~\ref{sec:method}, and explained in more detail in the next subsection.
The normalized Wilson loop observables are denoted by $W_n(N)=\frac1N \widehat W_n(N)$.
For example, $\widehat W_0=N$ and $W_0=1$, illustrating that the normalized and unnormalized observables have different large-$N$ behavior.

There are two natural large-$N$ limits of interest.
In the strong-coupling limit, the couplings $t_k$ are held fixed as $N \to \infty$.
In the ’t~Hooft limit, one instead keeps the rescaled couplings $\tilde t_k \equiv t_k/N$ fixed as $N \to \infty$.
Accordingly, in the ’t~Hooft limit the normalized Wilson loops $W_n(N)$ are expected to admit a finite large-$N$ limit, while in the strong-coupling limit it is the unnormalized observables $\widehat W_n(N)$, with $n\neq 0$, that remain
finite.

This distinction is already visible at the perturbative level.
As we will see in the next subsection, in the strong-coupling limit the unnormalized Wilson loop expectation values take the simple form $\widehat W_n = t_n$ to all orders in perturbation theory.
Dividing by $N$, this implies $W_n = \tilde t_n$, illustrating that in the ’t~Hooft limit it is the normalized Wilson loops that have a well-defined large-$N$ limit when the rescaled couplings $\tilde t_k$ are held fixed.
Non-perturbative instanton corrections to these relations will be analyzed in subsection~\ref{subsec:instantons}. 
See appendix~\ref{app:pert Wilson} for an alternative perturbative derivation of these relations.

The results presented in subsection~\ref{subsec:instantons} mostly pertain to the strong-coupling limit, where the potential experienced by a single eigenvalue is held fixed. 
By contrast, the numerical exploration of the phase diagram in complex coupling space presented in section~\ref{sec:phase diagram} will pertain to the ’t~Hooft limit of the model. 

To test the asymptotic bootstrap estimate in practice, we now turn to explicit unitary matrix models that allow for both analytical and numerical control. 
As a nontrivial cross-check, we compare the bootstrap estimates with analytical, large-$N$ instanton computations.
More precisely, we resort to the recently uncovered instanton expansion of ungapped unitary matrix integrals~\cite{Eniceicu:2023cxn,cmt24} and apply it to the models studied here. In appendix \ref{app:LefschetzThimbleAnalysis}, we provide a complementary Lefschetz thimble analysis of unitary matrix models, approaching the instanton expansion from a resurgent perspective and providing further details and support of the computations carried in subsection \ref{subsec:instantons}.
As test cases, we consider the Gross-Witten-Wadia (GWW) model,
\begin{align}\label{eq:GWW}
V(U) = t \left( \Tr(U) + \Tr(U^{-1}) \right),
\end{align}
for which only the first coupling is nonvanishing, as well as a slightly more complicated model with
\begin{equation}\label{eq:potential}
V(U) = 10 \left( \Tr(U) + \Tr(U^{-1}) + \Tr(U^2) + \Tr(U^{-2}) \right),
\end{equation}
corresponding to $t_1=10$ and $t_2=20$.
These models will be used both to analyze non-perturbative instanton effects and to explore numerically the phase structure in complex coupling space.

\subsection{Asymptotic bootstrap estimate of the moments and orthogonal polynomials}
\label{subsec:bootstrap and ortho polys}

It is convenient to recast the matrix integral \eqref{eq:unitmatrix1} in terms of its eigenvalues. 
Diagonalizing the unitary matrix yields
\begin{align}\label{eq:ZN diagonalize}
Z_N(\boldsymbol{t}) &= \frac{1}{N!} \left[\prod_{n=1}^{N}\int_{S^1} \frac{\rmd z_n}{2\pi i  z_n}\exp\left(V(z_n)\right)\right] \Delta(\boldsymbol{z})\Delta(\boldsymbol{z^{-1}}) \,,
\end{align}
where $\Delta(\boldsymbol{z})=\prod_{i<j}(z_i-z_j)=\det z_i^j$ is the well-known Vandermonde determinant. 
After diagonalization, the model reduces to a system of $N$ eigenvalues on the unit circle, each subject to the same potential $V(z)$ and interacting through the Vandermonde determinant.
Thus, each eigenvalue is weighted by a measure of the form \eqref{eq:circle measure} studied in the previous section.

A standard technique to study matrix integrals of this type is to introduce monic orthogonal polynomials $\{p_n(z)\}_{n\geq 0}$ with respect to the measure \eqref{eq:circle measure}.
These are orthogonal with respect to the bilinear form
\begin{align}\label{eq:form 1}
\langle f | g \rangle = \int_{S^1}\frac{\d z}{2\pi i z} \, \e^{V(z)} g(z) f(z^{-1}) \,.
\end{align}
They satisfy
\begin{align}
\langle p_m | p_n \rangle = h_m \, \delta_{m,n} \,.
\end{align}
The Vandermonde determinant $\Delta(\boldsymbol{z}) \equiv \det(z_j^{\,k})$ can equivalently be written in terms of the orthogonal polynomials as
$\Delta(\boldsymbol{z}) = \det\!\bigl(p_k(z_j)\bigr)$.
As a consequence, the partition function evaluates to
$Z_N = \prod_{n=0}^{N-1} h_n$.
Furthermore, the expectation value of a Wilson loop can be expressed in terms of the orthogonal polynomials as
\begin{align}\label{eq:Wn from otho polys}
W_{-n}=W_n = \frac{1}{N}\langle \Tr(U^n)\rangle
= \frac{1}{N} \sum_{k=0}^{N-1} \frac{\langle p_k \,|\, z^n p_k\rangle}{h_k}\, .
\end{align}
The reason for the equality is that the measure is invariant under $\theta\to -\theta$, or equivalently $z\to 1/z$ and this also translates to the Wison loops.
This is special about the type of potential we chose. 

In practice, the orthogonal polynomials can be computed by a Gram-Schmidt orthogonalization of the monomials $\{1,z,z^2,\dots\}$ with respect to the bilinear form \eqref{eq:form 1}.
Equivalently, this amounts to orthogonalization with respect to the Toeplitz moment matrix $M_S$ in \eqref{eq:PSD matrix}.
Writing
\[
p_n(z)=z^n+\sum_{j=0}^{n-1} c_{n,j} z^j ,
\]
the orthogonality conditions $\langle p_n|z^m\rangle=0$ for $m<n$ determine the coefficients $c_{n,j}$ through a linear system involving $M_S$.
Having computed the moments to very high precision using the asymptotic bootstrap estimate method described in section~\ref{sec:method}, we can thus obtain the orthogonal polynomials reliably by a standard Gram-Schmidt procedure.\footnote{We attach \texttt{Mathematica} code that implements the asymptotic bootstrap estimate method, constructs the orthogonal polynomials, and reproduces the results presented in the following sections.}
This procedure applies equally well when the potential has complex couplings.

Finally, having computed the orthogonal polynomials with high precision up to some large fixed degree $N_{\max}$, we can study their zeros in order to probe the structure of the spectral density.
More specifically, we use the general fact that
\begin{align}
\left\{ z^* \mid p_n(z^*)=0 \right\} \subset \text{supp}(\rho) \,,
\end{align}
where $\rho$ denotes the spectral density.
This relation is well understood for Hermitian matrix integrals (see, for instance,~\cite{b09,b07,s07,bt11,aam13,hkl14,bt16} for rigorous arguments and \cite{krsst25a} for numerical phase diagram explorations, leveraging this relation), and we \emph{assume} that it continues to hold for unitary matrix integrals.
In section~\ref{sec:phase diagram}, we use this correspondence to explore the topology of the support of the spectral density and to map out the phase diagram of unitary matrix integrals in complexified coupling space.

\subsection{The strong coupling limit}

In this subsection, we briefly describe how our numerical results based on the asymptotic bootstrap estimate method correctly reproduce the expected large-$N$ results as we approach the strong coupling limit of the model. 
We further highlight the precision of the method and motivate the analysis of non-perturbative effects that follows.
For concreteness, we focus on the model \eqref{eq:potential}. 

For this model, the strong-coupling solution predicts $\widehat W_1=10$, $\widehat W_2=20$, and $\widehat W_n=0$ $(n\geq 3)$, corresponding to an eigenvalue density of
\begin{equation}
    \rho(\theta) = \frac1{2\pi}[N+20 \cos(\theta)+40 \cos(2\theta)] \,,
\end{equation}
since the unnormalized Wilson loops $\widehat W_n$ are the Fourier coefficients of $\rho(\theta)$. 
Requiring positivity of the density implies that $N\gtrsim 42$, obtained by evaluating the density at its minima, which occur at $\theta=\pm (\pi -\tan ^{-1}\left(3 \sqrt{7}\right))$.\footnote{This is a simple bound in the spirit of \cite{Anderson:2016rcw}.} This indicates the onset of the strong coupling phase. 

\begin{figure}[ht]
\begin{center}
\includegraphics[width=10cm]{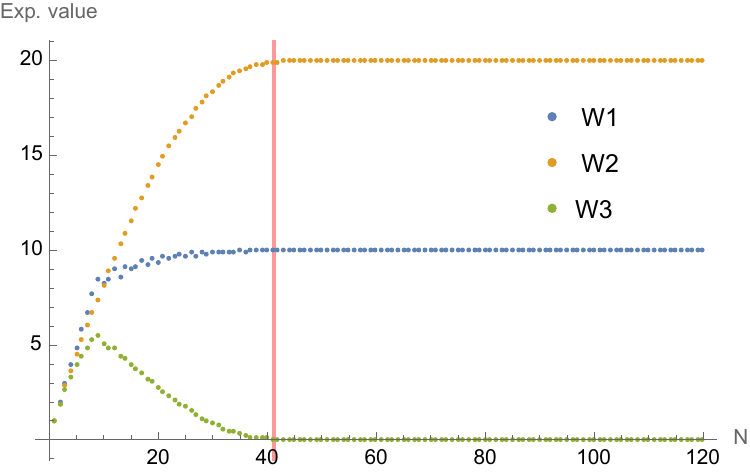}
\caption{Plot of expectation values of (unnormalized) Wilson loops for the potential \eqref{eq:potential} at increasing values of $N$. The vertical line indicates the expected transition to strong coupling. We see that at large enough $N$,
$\widehat W_1\simeq 10,\widehat W_2\simeq 20, \widehat W_3\simeq 0 $ as expected. The convergence to these values occurs very rapidly once $N$ surpasses the estimated transition value indicted in red.}\label{fig:Vevs123N}
\end{center}
\end{figure}

Using the asymptotic bootstrap estimate method together with the calculation of the associated orthogonal polynomials, we compute the Wilson loops $\widehat W_n(N)$ numerically. 
The results are shown in figure~\ref{fig:Vevs123N}. As $N$ increases past the expected threshold, the numerical values rapidly converge to the strong-coupling predictions. In particular, $\widehat W_1$ and $\widehat W_2$ approach their expected limits, while higher Wilson loops are strongly suppressed. 
For example, $\widehat W_3$ is computed from \eqref{eq:Wn from otho polys} as a sum over $N$ contributions, each of order one. The vanishing of $\widehat W_3$ thus results from large cancellations among these terms. Note that the fact that this cancellation is reproduced with high numerical precision provides a strong test of the stability and accuracy of the method.

\begin{figure}[ht]
\begin{center}
\includegraphics[width=10cm]{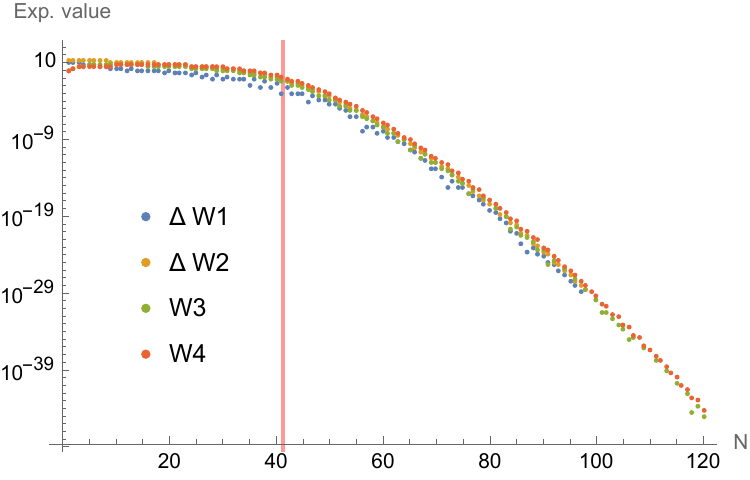}
\caption{Plot of the residual between the numerical finite-$N$ result versus the expected large-$N$ result, for increasing values of $N$. The expected transition to strong coupling is indicated in red.}\label{fig:LogVevs1234N}
\end{center}
\end{figure}

To better quantify the convergence, we study the deviation of the numerical computation for $\widehat W_n(N)$ from its large $N$ exact value. As shown in figure~\ref{fig:LogVevs1234N}, these deviations decay much faster than any power of $\frac 1N$. Again, note that we obtain cancellations in these expectation values sensitive to $40$ decimal places, from sums with terms of order one. 
This fast suppression is consistent with the expectation that perturbation theory around the strong-coupling solution is exact, and that all corrections are non-perturbative in $N$.
This is in agreement with the exact computation at $N=\infty$ of the matrix integral \eqref{eq:unitmatrix1} (see, for instance, \cite[appendix C]{Eniceicu:2023cxn}). 
An alternative derivation of this fact is explained in appendix \ref{app:strong_couplng_results}.

\begin{figure}[ht]
\begin{center}
\includegraphics[width=10cm]{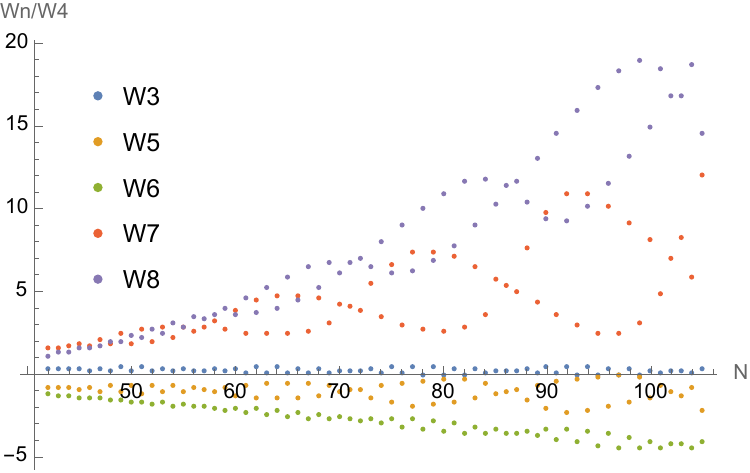}
\caption{Ratios of different Wilson loops $\widehat W_n$ relative to $\widehat W_4$ as a function of $N$. Notice the rapid oscillatory behaviors, suggesting interference between different instanton configurations.}\label{fig:Ratios35678}
\end{center}
\end{figure}

Next, assuming that the large $N$ corrections arise from non-perturbative large $N$ instantons in the matrix model, it is useful to consider ratios of Wilson loop expectation values at the same $N$.
If the leading large-$N$ behavior is governed by a common instanton action, these ratios largely cancel the dominant exponential factor, making residual features, such as oscillations, more visible as shown in figure~\ref{fig:Ratios35678}. 
We first note that the ratios are now of order one, consistent with the na\"\i ve expectation that the leading exponential suppression cancels. However, the ratios show pronounced oscillations whose amplitude increases with $N$. In particular, $\widehat W_3$ changes sign multiple times over the range shown, and $\widehat W_5$ and $\widehat W_7$ are on the verge of changing signs for larger values of $N$ beyond those plotted. 
This behavior suggests that more than one instanton configuration contributes to the large-$N$ expansion, leading to a nontrivial interference pattern between multiple contributions. 
The oscillations occur in increments of $N$ very close one, as seen in the appearance of a double envelope for each $\widehat W_n$. In addition, there is a longer period of order $N\simeq 26$, roughly aligned with the crossings of these double envelopes. The goal of the next section is to show that a detailed instanton analysis can account for this behavior.

\subsection{Eigenvalue instantons and saddle-point expansions}
\label{subsec:instantons}

We now turn to an analytical approach for computing the (unnormalized) Wilson loop expectation values \eqref{eq:unitmatrix2} in the ungapped phase. These predictions are derived via appropriate saddle-point expansions of the unitary matrix integral \eqref{eq:unitmatrix1}, evaluated around eigenvalue configurations that feature tunneling of both regular eigenvalues \cite{d90a,d93,m08,msw08} and anti-eigenvalues \cite{mss22}. Accounting for all such configurations leads to an instanton expansion for the unitary matrix integral which has been shown to match (to leading order in \cite{Eniceicu:2023cxn} and to all orders in \cite{cmt24}) the Fredholm determinant expansion derived in \cite{m22} (see also \cite{o99,bo99} for the original derivation of the Fredholm determinant expansion in the context of fermionic determinants), whose direct relation with the giant graviton expansion \cite{ai19,afim19,afim20a,afim20b,afimy20,fim21,i21,gl24} has been cleared out in \cite{e23}.

In what follows, we will review this instanton expansion and understand how one can extract analytical predictions out of it by performing suitable saddle-point expansions. In appendix \ref{app:LefschetzThimbleAnalysis}, we present an alternative and rather illuminating argument that motivates this expansion by applying simple Lefschetz–thimble considerations (akin to those in \cite{msw08,mss22}) to the unitary matrix integrals. Finally, we match our analytical predictions with the asymptotic bootstrap estimate results, derived via the orthogonal polynomial formula \eqref{eq:Wn from otho polys}, thereby providing a highly non-trivial cross-check.

\paragraph{Spectral geometry of unitary matrix integrals:} We are interested in addressing the large $N$ expansion of the matrix integral \eqref{eq:unitmatrix1} in the ungapped phase. In this phase, the perturbative large $N$ expansion is famously trivial, yielding (see, for instance \cite[appendix C]{Eniceicu:2023cxn}, following a result from \cite{diaconis1994eigenvalues}) 
\begin{equation}
    Z^{(0)}_N(\boldsymbol{t}) = \prod_{k=1}^K\exp\left(\frac{t_k^2}{k}\right) \,.
    \label{eq:Instantons3}
\end{equation}

To properly formulate the instanton expansion of \cite{cmt24} and to extract further analytical predictions via saddle-point techniques, it is convenient to rewrite it in a form that is
holomorphic in the eigenvalues and closely analogous to a Hermitian matrix
integral.
Following, for example, \cite{Eniceicu:2023cxn}, this is achieved by expressing the unitary
Vandermonde factor $|\Delta(\boldsymbol z)|^2$ in terms of $\Delta(\boldsymbol z)^2$. On the unit circle $\bar{z}=z^{-1}$ and thus 
\begin{align}
|\Delta(\boldsymbol z)|^2
=
\Delta(\boldsymbol z)\,\Delta(\boldsymbol z^{-1})
=
(-1)^{\frac12 N(N-1)}
\left(\prod_{i=1}^N z_i^{-(N-1)}\right)
\Delta(\boldsymbol z)^2 \,.
\end{align}
As a result, the partition function may be rewritten as
\begin{align}
     Z_N(\boldsymbol{t}) &= (-1)^{\frac{1}{2}N(N-1)}\frac{1}{N!} \left[\prod_{n=1}^{N}\int_{S^1} \frac{\rmd z_n}{2\pi i}z_n^{-N}\exp\left(V(z_n)\right)\right] \Delta^2(\boldsymbol{z})\nonumber \\& = (-1)^{\frac{1}{2}N(N-1)}\frac{1}{N!} \left[\prod_{n=1}^{N}\int_{S^1} \frac{\rmd z_n}{2\pi i }\exp\left(V(z_n)-N\log(z_n)\right)\right]\Delta^2(\boldsymbol{z}) \,.
     \label{eq:Instantons2}
\end{align}
In this form, the unitary matrix integral can be analyzed using the familiar Lefschetz-thimble and instanton framework (see \cite{msw08,mss22}) and we do so in appendix \ref{app:LefschetzThimbleAnalysis}, providing alternative motivation for the instanton expansion we will shortly present.

\begin{figure}
    \centering
    \begin{tikzpicture}

\def\vspace{2.5};

\draw[line width = 0pt,fill = darktangerine,fill opacity=0.2] (-4,2.5)--(-0.82,2.5)to [out = -50, in = 90] (-0.5,1.6)to [out = -90, in = 180] (0,1.6-0.5)to [out = 0, in = -90] (0.5,1.6)to [out = 90, in = 180+50] (0.82,2.5)--(4,2.5)--(6,2.5+1.2)-- (-2,2.5+1.2)-- cycle;

\draw[fill = LightBlue,fill opacity=0.2,line width = 0pt] (-4,0) -- (4,0) -- (6,1.2) --(0.66,1.2)to[out = 90+50, in = -90](0.5,1.6)to[out = 90, in   = 0] (0,1.6+0.5)to[out = 180, in   = 90] (-0.5,1.6)to[out = -90, in =90-50](-0.66,1.2) -- (-2,1.2)-- cycle;

\draw[line width = 0pt,fill = darktangerine,fill opacity=0.2] (2.5+0.26,0.6+1.91) to[out = -60, in = 90] (3,1.6)to[out = 90, in = 180+60](3.24,0.6+1.91)--cycle;

\draw[line width = 2pt] (-4,0+\vspace) -- (4,0+\vspace) -- (6,1.2+\vspace) -- (-2,1.2+\vspace) -- cycle;

 \draw[line width = 2pt] (1.5,0.75) to[out = 180-20, in   = -90] (0.5,1.6);

 \draw[line width = 2pt] (-1.5,0.75) to[out = 20, in   = -90] (-0.5,1.6);

\draw[line width = 2pt,color = ForestGreen,dashed] (0.5+0.02,1.619)to[out = 180-20, in   = 20] (-0.5-0.02,1.619);

\draw[line width = 2pt] (-0.66,1.2)--(-2,1.2)--(-4,0) -- (4,0) -- (6,1.2) --(0.66,1.2);

\draw[line width = 2pt] (0.5,1.6)to[out = 90, in   = 0] (0,1.6+0.5)to[out = 180, in   = 90] (-0.5,1.6)to[out = -90, in =90-50](-0.66,1.2);

\draw[line width = 2pt] (1.5,0.75+2.2) to[out = 180+20, in   = 90] (0.5,1.6);

\draw[line width = 2pt, color = ForestGreen] (0.5,1.6)to[out = 180+20, in   = -20] (-0.5,1.6);

\draw[line width = 2pt] (-1.5,0.75+2.2) to[out =-20, in   = 90] (-0.5,1.6);

\draw[line width = 2pt] (-0.5,1.6)to [out = -90, in = 180] (0,1.6-0.5)to [out = 0, in = -90] (0.5,1.6);

\draw[line width = 2pt] (2.5,0.6) to[out = 20, in = -90] (3,1.6)to[out = -90, in = 180-20] (3.5,0.6);

\draw[line width = 2pt] (2.5-0.5,0.6+2.5) to[out = -20, in = 90] (3,1.6)to[out = 90, in = 180+20] (3.5+0.5,0.6+2.5);


\filldraw[color=ForestGreen, line width=2pt](0.5,1.6) circle (0.05);

\filldraw[color=ForestGreen, line width=2pt](-0.5,1.6) circle (0.05);

\filldraw[color=cornellred, line width=2pt](3,1.6) circle (0.05);


   \node[Blue] at (6.5,0.5){$y^+(z)$};   
   \node[Orange] at (6.5,2.8){$y^-(z)$}; 
    \end{tikzpicture}
    \caption{Illustration of a typical spectral curve arising in the ungapped phase of a unitary matrix model. The eigenvalue distribution is shown as a green line, and a saddle-point is marked by a red dot. The first branch of the curve is shaded in blue, while the second one appears in orange.}
    \label{fig:Instantons4}
\end{figure}
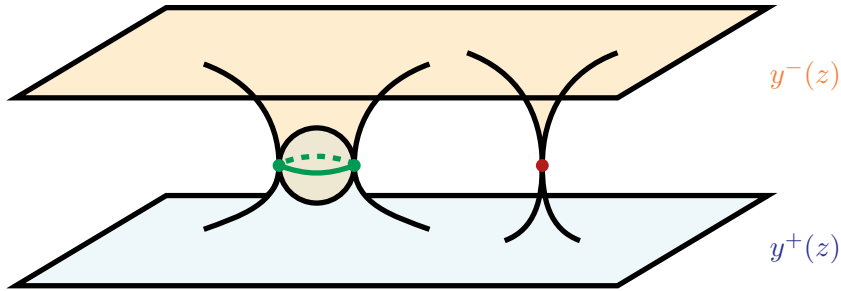

In order to address instanton contributions to the matrix integral \eqref{eq:Instantons2}, we have to compute the effective potential
\begin{equation}
    V_{\text{eff}}(z) = -\frac{V(z)}{N} +\log(z) -\frac{2}{N}\left\langle\text{Tr}\left(\log(z-U)\right)\right\rangle \,.
    \label{eq:effpotential}
\end{equation}
One can show that (see, for instance, \cite[appendix C]{Eniceicu:2023cxn}) 
\begin{equation}
    \left\langle\text{Tr}\left(\log\left(z-U\right)\right)\right\rangle = \begin{lcases} &
          -\sum_{k=1}^K\frac{t_k}{k} z^{-k} + N\log(z)  \hspace{1pt} , \hspace{1pt} |z| >1 \\& -\sum_{k=1}^K\frac{t_k}{k}z^{k}  \hspace{1pt} , \hspace{1pt} |z| <1
    \end{lcases} \,.
    \label{eq:resolvent}
\end{equation}
Using the equations \eqref{eq:firstpotential} and \eqref{eq:resolvent}, we can rewrite the effective potential as
\begin{equation}
    V_{\text{eff}}(z) = \begin{lcases} & V^+_{\text{eff}}(z) = 
        -\log(z) -\frac{1}{N}\sum_{k=1}^K\frac{t_k}{k}\left(z^k-z^{-k}\right) \hspace{1pt} , \hspace{1pt} |z| >1 \\ & V^-_{\text{eff}}(z) = \log(z) +\frac{1}{N}\sum_{k=1}^K\frac{t_k}{k}\left(z^k-z^{-k}\right) \hspace{1pt} , \hspace{1pt} |z| <1
    \end{lcases}.
    \label{eq:Instantons13}
\end{equation}
Analytically continuing the branches $V^\pm_{\text{eff}}(z)$ separately gives rise to two distinct effective potentials. The spectral curve is formally defined as the Riemann surface upon which the derivative of the effective potential is defined. In the ungapped phase, the spectral curve is disconnected, featuring two connected components that intersect at the unit circle. The connected components read
\begin{equation}
    y^\pm(z) = \frac{\rmd V_{\text{eff}}^\pm}{\rmd z}(z) \,.
\end{equation}
This spectral curve is schematically displayed in figure \ref{fig:Instantons4}.
We will henceforth refer to $y^+(z)$ as the first branch and $y^-(z)$ as the second branch of the spectral curve.

Following \cite{mss22}, eigenvalues tunneling through the first branch will be called regular eigenvalues, while those tunneling through the second branch will be referred to as anti-eigenvalues. 
This conventional choice differs from that in \cite{Eniceicu:2023cxn}, where the second branch was disconnected and defined as the complex domain reached via the analytic extension of the functions $V^\pm_{\text{eff}}(z)$ into inside/outside the unit circle, respectively. 
Instead, here the first branch is given by the analytic continuation of $V^{+}_{\text{eff}}(z)$ into the entire complex plane.\footnote{One way to see this more clearly is as follows. Consider the unitary matrix integral in the one-cut, gapped phase. Here, the effective potential $V^{+}_{\text{eff}}(z)$ inside the unit circle is well-defined and trivially the analytic continuation (without crossing any branch cuts) of the function outside the circle. Our convention for the first branch is then the one consistent from the limit of this phase to the ungapped phase.}
This choice will be reflected in the eigenvalue content underlying the instanton expansion of \eqref{eq:unitmatrix1}. Indeed, while in \cite{Eniceicu:2023cxn,cmt24} the components of this expansion are obtained by considering exclusively the tunneling of anti-eigenvalues, ours will be obtained by considering the tunneling of pairs consisting of one eigenvalue and one anti-eigenvalue (see appendix \ref{app:LefschetzThimbleAnalysis} for more details). 

This picture is physically more sound, since an anti-eigenvalue should be thought of as an eigenvalue “hole”, increasing by one the number of eigenvalues remaining in the distribution from which it departed \cite{mss22}.\footnote{This interpretation is akin to the Dirac sea picture and we refer the reader to \cite{mss22} for details on this analogy.} Thus, the tunneling of a pair consisting of one eigenvalue and one anti-eigenvalue necessarily leaves unchanged the number of eigenvalues remaining on the unit circle distribution. This is precisely what one expects for tunneling in the ungapped phase, since any decrease in the number of eigenvalues would cause the cut to reopen, leading back to a gapped phase.

In order to simplify the notation in what follows, we fix $y(z) = y^+(z)$ and $V_{\text{eff}}(z) = V_{\text{eff}}^+(z)$.

\paragraph{The instanton expansion and saddle-point analysis:} 
Let us now write the instanton expansion formula for the generic unitary matrix model \eqref{eq:unitmatrix1}. The expansion reads \cite{m22,lr23,cmt24}
\begin{equation}
    Z_N(\boldsymbol{t}) = Z^{(0)}_N(\boldsymbol{t}) + \sum_{n=1}^{+\infty}Z^{(n)}_N(\boldsymbol{t}) \,,
    \label{eq:Instantons7}
\end{equation}
where 
\begin{align}
    Z^{(n)}_N(\boldsymbol{t}) &= \frac{Z_N^{(0)}(\boldsymbol{t})}{\left(n!\right)^2}\left[\prod_{m=1}^n \oint_{\mathcal{C}}\frac{\rmd z_m}{2\pi}\oint_{\bar{\mathcal{C}}}\frac{\rmd \bar{z}_m}{2\pi} \exp\left(-N\left(V_{\text{eff}}(z_m)-V_{\text{eff}}(\bar{z}_m)\right)\right)\right] \times \nonumber \\ &\quad\Delta^2(\boldsymbol{z})\Delta^2(\boldsymbol{\bar{z}})\prod_{m=1}^{n}\prod_{n=1}^{n}\frac{1}{(z_m-\bar{z}_{n})^2} \,,
    \label{eq:expansionNcontribution}
\end{align}
and $\mathcal{C},\bar{\mathcal{C}}\subset \mathbb{C}$ are circular contours with radius smaller than one and greater than one, respectively. The expansion \eqref{eq:Instantons7} is convergent \cite{m22} and should be understood as the partition function's non-perturbative completion, generated by accounting for regular instantons and their ghost partners \cite{Eniceicu:2023cxn,cmt24}. In particular, $Z^{(n)}_N(\boldsymbol{t})$ should be understood as the contribution of the instanton configuration associated with a background in which $n$ eigenvalues and $n$ anti-eigenvalues have tunneled to saddles located inside and outside the unit circle, respectively. See appendix \ref{app:LefschetzThimbleAnalysis} for more details on this.

In order to extract analytical predictions from \eqref{eq:Instantons7}, we need to expand the integrals \eqref{eq:expansionNcontribution} around saddle-points. This can be tricky and is best understood in concrete examples. We will consider the GWW model \eqref{eq:GWW} as a pedagogical example (already worked out in \cite{Eniceicu:2023cxn}) and then move on to the model \eqref{eq:potential}.

The GWW model has two saddles located at 
\begin{align}
    & z_1^\star =\frac{1}{2 \tilde{t}}\left(-1-\sqrt{1-4 \tilde{t}^{\hspace{1pt}2}}\right) \label{eq:GWWs1} \,,\\ 
    & z_2^\star =\frac{1}{2 \tilde{t}}\left(-1+\sqrt{1-4 \tilde{t}^{\hspace{1pt}2}}\right)\label{eq:GWWs2} \,,
\end{align}
where
\begin{equation}
    \tilde{t} = \frac{t}{N} \,.
\end{equation}
The ungapped phase occurs for $ \tilde{t} < 1/2$ in which case $z_1^\star$ will take place outside the unit circle while $z_2^\star$ will take place inside. In order to evaluate \eqref{eq:expansionNcontribution} analytically, we need to perform a saddle-point expansion of the integrals over $\mathcal{C}$ and $\bar{\mathcal{C}}$ around the saddles $z_2^\star$ and $z_1^\star$, respectively.

More concretely, we begin by deforming the contours $\mathcal{C}$ and $\bar{\mathcal{C}}$ into the steepest-descent and steepest-ascent contours associated with the saddles $z_2^\star$ and $z_1^\star$, respectively (see appendix \ref{app:LefschetzThimbleAnalysis} for definitions and further details).\footnote{Appendix \ref{app:LefschetzThimbleAnalysis} provides a complementary argument that does not involve such a contour deformation.}
We can then rewrite \eqref{eq:expansionNcontribution} as
\begin{align}
    Z^{(n)}_N(t) = &\frac{Z_N^{(0)}(t)}{\left(n!\right)^2}\left[\prod_{m=1}^n \int_{\mathcal{C}_2^\star}\frac{\rmd z_m}{2\pi}\int_{\bar{\mathcal{C}}^\star_1}\frac{\rmd \bar{z}_m}{2\pi} \exp\left(-N\left(V_{\text{eff}}(z_m)-V_{\text{eff}}(\bar{z}_m)\right)\right)\right] \times \nonumber \\ &\Delta^2(\boldsymbol{z})\Delta^2(\boldsymbol{\bar{z}})\prod_{m=1}^{n}\prod_{\bar{m}=1}^{n}\frac{1}{(z_m-\bar{z}_{\bar{m}})^2} \,,
    \label{eq:GWWInstanton}
\end{align}
where $C_2^\star \subset \mathbb{C}$ ($\bar{C}_1^\star\subset \mathbb{C}$) is the steepest-descent (steepest-ascent) contour associated with the saddle $z_2^\star$ ($z_1^\star$). We then perform the change of variables
\begin{align}
    & z_m = z_2^\star +\sqrt{\frac{1}{N}} \frac{x_m}{\sqrt{V''_{\text{eff}}(z_2^\star)}} \,,\\
    & \bar{z}_m = z_1^\star +\sqrt{\frac{1}{N}} \frac{\bar{x}_m }{\sqrt{-V''_{\text{eff}}}(z_1^\star)} \,,
\end{align}
where the new variables take values in $\mathbb{R}$, leading, after a sequence of Taylor expansions, to Gaussian integrals that can be evaluated straightforwardly. For instance, we can write
\begin{align}
    Z_N^{(1)}(t) = & \frac{Z_N^{(0)}(t)}{\sqrt{-V_{\text{eff}}''(z_1^\star)V_{\text{eff}}''(z_2^\star)}} \frac{e^{-2 A(t) N}}{(z_2^\star-z_1^\star)^2}\frac{1}{N}\int_{\mathbb{R}}\frac{\rmd x}{2\pi} \int_{\mathbb{R}}\frac{\rmd \bar{x}}{2\pi}e^{-\frac{x^2}{2}-\frac{\bar{x}^2}{2}}  \left(1+\mathcal{O}\left(\frac{1}{N}\right)\right) \,,
\end{align}
where
\begin{equation}
    A(t) = \frac{ V_{\text{eff}}(z_2^\star)-V_{\text{eff}}(z_1^\star)}{2} =  -\sqrt{1-4\tilde{t}^{\hspace{1pt}2}} + \text{arccosh}\left(\frac{1}{2\tilde{t}}\right) \,.
    \label{eq:GWWA}
\end{equation}
Performing the Gaussian integrations yields the following expansion,
\begin{equation}
    \frac{Z_N^{(1)}(t)}{Z_N^{(0)}(t)} = e^{-2A(t)N}\sum_{m=1}^{+\infty} C_m(t) N^{-m} \,,
    \label{eq:GWWSaddlePointExpansion}
\end{equation}
where the first few coefficients read
\begin{align}
    & C_1(t) = -\frac{\tilde{t}^{\hspace{1pt}2}}{2 \pi  \left(1-4 \tilde{t}^{\hspace{1pt}2}\right)^{3/2} }\,,\label{eq:GWWLeadingCoeff}\\
    &C_2(t) = -\frac{\tilde{t}^{\hspace{1pt}2} \left(18 \tilde{t}^{\hspace{1pt}2}+13\right)}{12 \pi  \left(4 \tilde{t}^{\hspace{1pt}2}-1\right)^3 }\,,\label{eq:GWWSubLeadingCoeff}\\
    & C_3(t) =-\frac{\tilde{t}^{\hspace{1pt}2} \left(1188 \tilde{t}^{\hspace{1pt}4}+2484 \tilde{t}^{\hspace{1pt}2}+241\right)}{144 \pi  \left(1-4 \tilde{t}^{\hspace{1pt}2}\right)^{9/2} }\,.\label{eq:GWWSubSubLeadingCoeff}
\end{align}
The instanton action \eqref{eq:GWWA} and the leading coefficient \eqref{eq:GWWLeadingCoeff} agree exactly with the results of \cite{Eniceicu:2023cxn}.\footnote{This agreement follows after translating to the conventions of \cite{Eniceicu:2023cxn}, which amounts to the redefinition $t \mapsto N/(2t)$.}
This concludes our saddle-point analysis of the GWW model.

Now, we consider the unitary matrix model \eqref{eq:potential}. Using the equation \eqref{eq:Instantons13}, we can write the effective potential as
\begin{equation}
    V_{\text{eff}}(z) = -\log(z) -\frac{10}{N}\left(z-\frac{1}{z}+z^2-\frac{1}{z^2}\right) \,,
    \label{eq:Instantons8}
\end{equation}
and the associated saddles read
\begin{align}
       & z_1^\star = \frac{1}{40} \left(-\sqrt{10} \sqrt{-2 N-\sqrt{825-20 N}-75}+\sqrt{825-20 N}-5\right) \,, \\ 
        & z_2^\star = \frac{1}{40} \left(\sqrt{10} \sqrt{-2 N+\sqrt{825-20 N}-75}-\sqrt{825-20 N}-5\right) \,, \\
       & z_3^\star = \frac{1}{40} \left(-\sqrt{10} \sqrt{-2 N+\sqrt{825-20 N}-75}-\sqrt{825-20 N}-5\right) \,, \\
       & z_4^\star = \frac{1}{40} \left(\sqrt{10} \sqrt{-2 N-\sqrt{825-20 N}-75}+\sqrt{825-20 N}-5\right) \,.
    \label{eq:Instantons12}
\end{align}
The ungapped phase occurs for $N > 165/4$ in which case the saddle points $z_1^\star$ and $z_3^\star$ lie outside the unit circle, while the saddle points $z_2^\star$ and $z_4^\star$ lie inside.

As in the previous case, we want to evaluate the integrals \eqref{eq:expansionNcontribution} analytically by resorting to a saddle-point expansion of the integrals over $\mathcal{C}$ and $\bar{\mathcal{C}}$ around saddles taking place inside and outside the unit circle respectively. We start by deforming the circular contours $\mathcal{C}$ and $\bar{\mathcal{C}}$ into $\mathcal{C}_2^\star+\mathcal{C}_4^\star$ and $\bar{\mathcal{C}}_1^\star+\bar{\mathcal{C}}_3^\star$, respectively. Figure \ref{fig:deformation} shows how this deformation of contours reproduces the \emph{sum} of the relevant steepest ascent/descent contours.\footnote{Again, appendix \ref{app:LefschetzThimbleAnalysis} provides a complementary argument that does not involve such a contour deformation.}
We can then rewrite \eqref{eq:expansionNcontribution} as
\begin{figure}
    \centering
    \begin{tikzpicture}[scale = 0.75]


\filldraw[decorate, decoration={snake, segment length=5pt, amplitude=3pt},color=ForestGreen, fill=Green!5, line width=2pt](0,0) circle (3);





\draw[color=Orange, line width=2pt](-1.5*0.9,-4.5*0.9) to[out = 180-20, in = 0](-1.5*0.9-7,-2);
\draw[color=Orange, line width=2pt](-1.5*0.9,4.5*0.9) to[out = 180+20, in = 0](-1.5*0.9-7,2);
\draw[color=Orange, line width=2pt](-1.5*0.9,-4.5*0.9) to[out = -20, in = 180](-1.5*0.9+9,-2);
\draw[color=Orange, line width=2pt](-1.5*0.9,4.5*0.9) to[out = 20, in = 180](-1.5*0.9+9,2);

\draw[blue,rotate = 18,line width = 2pt] (0+0.08,1-0.08) ellipse (0.7 and 0.9);
\draw[blue,rotate = -18,line width = 2pt] (0+0.08,-1+0.08) ellipse (0.7 and 0.9);



\draw[color= black, line width=2pt](0,0) circle (3.5);

\draw[color= black, line width=2pt](0,0) circle (2.5);




\filldraw[color= cornellred, line width=2pt](-1.5*0.9,4.5*0.9) circle (0.1);
\filldraw[color= cornellred, line width=2pt](-1.5*0.38,4.5*0.38) circle (0.1);
\filldraw[color= cornellred,  line width=2pt](-1.5*0.9,-4.5*0.9) circle (0.1);
\filldraw[color= cornellred,  line width=2pt](-1.5*0.38,-4.5*0.38) circle (0.1);

\filldraw[color= black, line width=2pt](0,0) circle (0.1);


\draw[line width = 2pt,->] (-3.5-0.2,0) -- (-8.2,0);
\draw[line width = 2pt,->] (-3.4,0+1.4) to[out = 180-20, in = -80] (-5.5,2.2);
\draw[line width = 2pt,->] (-2.67,0+2.5) to[out = 180-40, in = -90+30] (-3,3.2);
\draw[line width = 2pt,->] (0,3.5+0.15) -- (0,4.1);

\draw[line width = 2pt,->] (-3.4,-0-1.4) to[out = 180+20, in = 80] (-5.5,-2.2);
\draw[line width = 2pt,->] (-2.67,0-2.5) to[out = 180+40, in = 90-30] (-3,-3.2);
\draw[line width = 2pt,->] (0,-3.5-0.15) -- (0,-4.1);

\draw[line width = 2pt,->] (3.5+0.2,0) -- (8.2,0);
\draw[line width = 2pt,->] (3.4,0+1.4) to[out = +20, in = 180+80] (5.5,2.2);
\draw[line width = 2pt,->] (2.67,0+2.5) to[out = 40, in = -90-20] (3,3.3);

\draw[line width = 2pt,->] (3.4,0-1.4) to[out = -20, in = -180-80] (5.5,-2.2);
\draw[line width = 2pt,->] (2.67,0-2.5) to[out = -40, in = 90+20] (3,-3.3);

\draw[line width = 2pt,->] (-2.5+0.2,0) -- (-0.8,0);
\draw[line width = 2pt,->] (-2.5+0.4+0.25,-1.4) to[out = 40, in = 180](-1.1,-1.2);
\draw[line width = 2pt,->] (1,-2.5+0.43) to[out = 90+30, in = -30] (0.5,-1.5);
\draw[line width = 2pt,->] (2-0.05,-1.2) to[out = 90+60, in = 0] (0.7,-0.82);

\draw[line width = 2pt,->] (2.5-0.2,0) -- (0.6,0);

\draw[line width = 2pt,->] (-2.5+0.4+0.25,1.4) to[out = -40, in = -180](-1.1,1.2);
\draw[line width = 2pt,->] (1,2.5-0.43) to[out = -90-30, in = 30] (0.5,1.5);
\draw[line width = 2pt,->] (2-0.05,1.2) to[out = -90-60, in = 0] (0.7,0.82);

\node[color=blue] at(-1.4,-0.55){\scalebox{1.3}{$\mathcal{C}_4^\star$}};
\node[color=blue] at(-1.4,0.55){\scalebox{1.3}{$\mathcal{C}_2^\star$}};
\node[color=Orange] at(0,-4.95){\scalebox{1.3}{$\bar{\mathcal{C}}_3^\star$}};
\node[color=Orange] at(0,4.9){\scalebox{1.3}{$\bar{\mathcal{C}}_1^\star$}};

\node[color=cornellred] at(-0.3,1.2){\scalebox{1.3}{$z_2^\star$}};
\node[color=cornellred] at(-0.3,-1.2){\scalebox{1.3}{$z_4^\star$}};
\node[color=cornellred] at(-1.4,4.75-0.05){\scalebox{1.3}{$z_1^\star$}};
\node[color=cornellred] at(-1.4,-4.75+0.05){\scalebox{1.3}{$z_3^\star$}};

\node[color=black] at(2,0.5){\scalebox{1.3}{$\mathcal{C}$}};

\node[color=black] at(3.95,0.5){\scalebox{1.3}{$\bar{\mathcal{C}}$}};

    \end{tikzpicture}
    \caption{Pictorial representation of the deformation (indicated by arrows) of the circular contours $\mathcal{C}$ and $\bar{\mathcal{C}}$ (shown in black) into the sum of steepest-descent contours $\mathcal{C}_2^\star + \mathcal{C}_4^\star$ (shown in blue) and the sum of steepest-ascent contours $\bar{\mathcal{C}}_1^\star + \bar{\mathcal{C}}_3^\star$ (shown in orange), respectively. The saddle points are marked by red dots, while the unit-circle eigenvalue distribution is represented by the green wavy curve.}
    \label{fig:deformation}
\end{figure}
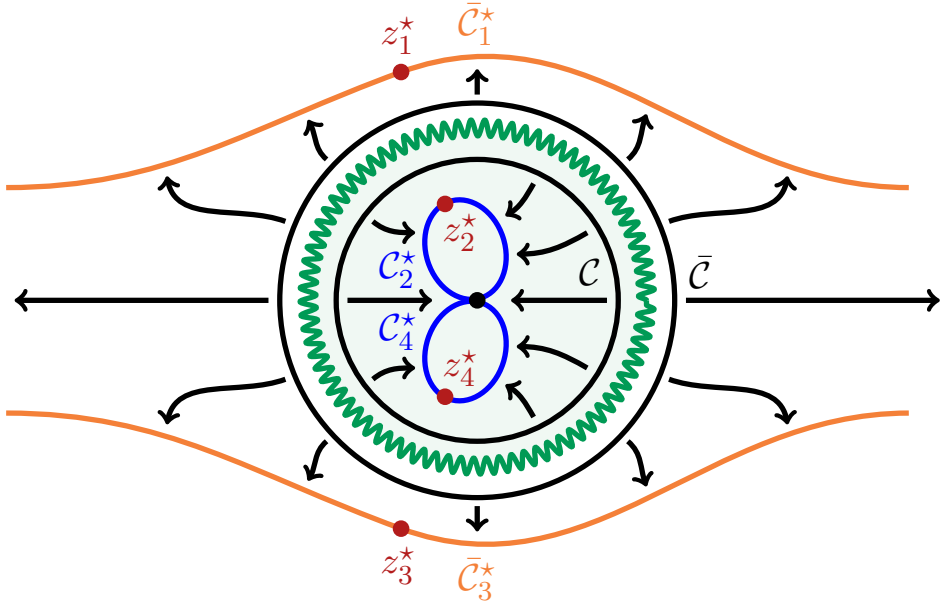
\begin{align}
     Z^{(n)}_N = &\frac{Z_N^{(0)}}{\left(n!\right)^2}\left[\prod_{m=1}^n \int_{\mathcal{C}_2^\star+\mathcal{C}_4^\star}\frac{\rmd z_m}{2\pi}\int_{\bar{\mathcal{C}}_1^\star+\bar{\mathcal{C}}_3^\star}\frac{\rmd \bar{z}_m}{2\pi} \exp\left(-N\left(V_{\text{eff}}(z_m)-V_{\text{eff}}(\bar{z}_m)\right)\right)\right] \times \nonumber \\ &  \Delta^2(\boldsymbol{z})\Delta^2(\boldsymbol{\bar{z}}) \prod_{m=1}^{n}\prod_{\bar{m}=1}^{n}\frac{1}{(z_m-\bar{z}_{\bar{m}})^2} \,.
\end{align}

As in the GWW model, we focus on the leading instanton contribution to \eqref{eq:Instantons7}, obtained by setting $n = 1$ in the expression above. Due to linearity in the integration domain, we can write 
\begin{equation}
    \frac{Z_N^{(1)}}{Z^{(0)}_N} = \text{I}_{2,1}(N)+\text{I}_{2,3}(N)+\text{I}_{4,1}(N)+\text{I}_{4,3}(N) \,,
    \label{eq:Instantons9}
\end{equation}
where
\begin{equation}
    \text{I}_{i,j}(N) = \int_{\mathcal{C}_i^\star}\frac{\rmd z}{2\pi}\int_{\bar{\mathcal{C}}_j^\star}\frac{\rmd \bar{z}}{2\pi} \exp\left(-N\left(V_{\text{eff}}(z)-V_{\text{eff}}(\bar{z})\right)\right)\frac{1}{(z-\bar{z})^2} \,.
\end{equation}
for $i = 2,4$ and $j = 1,3$. Following the procedure leading up to \eqref{eq:GWWSaddlePointExpansion}, we can perform a saddle-point expansion of the integrals above, yielding
\begin{align}
I_{i,j}(N) = \exp\left(-2 A(z_i^\star,z_j^\star)N\right) \sum_{m=1}^{+\infty} C_m(z_i^\star,z_j^\star) \, N^{-m} \,,
\label{eq:Instantons11}
\end{align}
where 
\begin{equation}
    A(z,w) = \frac{V_{\text{eff}}(z)-V_{\text{eff}}(w)}{2} 
    \label{eq:Acomplicated}
\end{equation}
is the instanton action. The coefficients $C_m(z,w)$ can be computed systematically, and we provide the first three in appendix \ref{appendix:coefficients}.

\paragraph{Wilson loop expectation values:} To derive an analytical prediction for the (unnormalized) Wilson loop expectation value \eqref{eq:unitmatrix2} for some $n \in \mathbb{N}$ it is useful to perturb the potential \eqref{eq:firstpotential} by introducing a source term of the form
\begin{equation}
    V_{\text{source}}(z) = g\left(z^n+\frac{1}{z^n}\right) \,,
\end{equation}
for some small $g>0$. Due to the invariance of the Haar measure under Hermitian conjugation, one can write
\begin{equation}
   \widehat{W}_n(N) = \frac{1}{2}\frac{\partial}{\partial g}\log\left( Z_N(\boldsymbol{t})\right)\Bigg\vert_{g=0} \,.
    \label{eq:Instantons17}
\end{equation}
Therefore, using the equation \eqref{eq:Instantons3}, we can write 
\begin{equation}
    \widehat{W}_{n}(N) = \widehat{W}_{n}^{(0)}(N) +  \widehat{W}_{n}^{(1)}(N) + \cdots \,,
    \label{eq:Instantons10}
\end{equation}
where the dots denote higher-instanton corrections and
\begin{align}
  & \widehat{W}_{n}^{(0)}(N)= \frac{1}{2}\frac{1}{Z_N^{(0)}(\boldsymbol{t})}\frac{\partial}{\partial g} Z^{(0)}_N(\boldsymbol{t}) \Bigg\vert_{g=0} = t_n \,,\label{eq:Wilson loop pert}
  \\ & \widehat{W}_{n}^{(1)}(N)=  \frac{1}{2}\frac{\partial}{\partial g}\left(\frac{Z_N^{(1)}(\boldsymbol{t})}{Z^{(0)}_N(\boldsymbol{t})}\right)\Bigg\vert_{g=0} \,.
  \label{eq:one-instanton}
\end{align}
The right-hand side of the equation above has the natural form of a saddle-point expansion (see equations \eqref{eq:GWWSaddlePointExpansion} and \eqref{eq:Instantons9}).

Let us now use the analytical saddle-point predictions obtained above to test the non-perturbative predictive power of the asymptotic bootstrap estimate method introduced in section \ref{sec:method}. More concretely, we consider the ratio
\begin{equation}
    Q_n(N) = \frac{\widehat{W}_{n}^{(1)}(N)}{W_{n}^{\text{bootstrap}}(N)-\widehat{W}_{n}^{(0)}(N)} \,.
    \label{eq:ratio}
\end{equation}
where $W_{n}^{\text{bootstrap}}(N)$ denotes the numerical prediction (for a fixed $N$) obtained directly from the formula \eqref{eq:Wn from otho polys} for which the intervening orthogonal polynomials are computed using the asymptotic bootstrap estimate method. If we assume that our method produces results with a very large accuracy (see section \ref{sec:method} for a discussion of the error estimates), it is natural to expect that:

\begin{itemize}
    \item The ratio \eqref{eq:ratio} converges to $1$ as $N$ increases. This is natural, as we expect subleading corrections to the saddle-point expansion \eqref{eq:one-instanton} to decay polynomially fast and subleading instanton corrections to decay exponentially fast, as $N$ increases.

    \item  The ratio \eqref{eq:ratio} approaches 1 as we consider more terms in the saddle-point expansion. 
\end{itemize}
In what follows, we check whether these features are observed in our working examples, namely the GWW model \eqref{eq:GWW} and the model \eqref{eq:potential}. This provides a highly non-trivial cross-check of our method.

\begin{figure}[ht]
  \centering

  \begin{subfigure}[t]{0.45\textwidth}
    \centering
    \includegraphics[width=\linewidth]{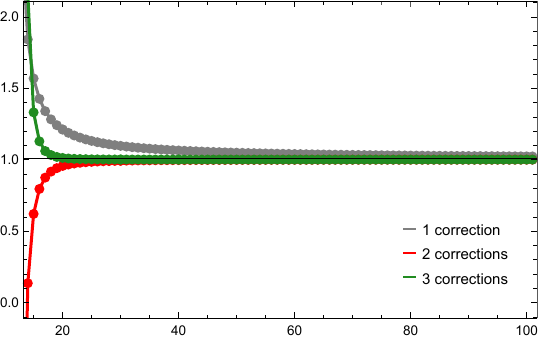}
    \caption{$n = 1$}
  \end{subfigure}
  \hfill
  \begin{subfigure}[t]{0.45\textwidth}
    \centering
    \includegraphics[width=\linewidth]{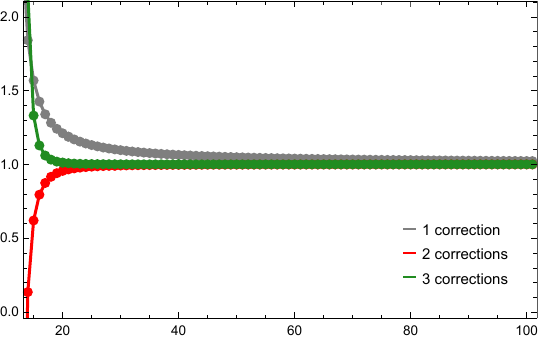}
    \caption{$n = 2$}
  \end{subfigure}

  \vspace{1em}

  \begin{subfigure}[t]{0.45\textwidth}
    \centering
    \includegraphics[width=\linewidth]{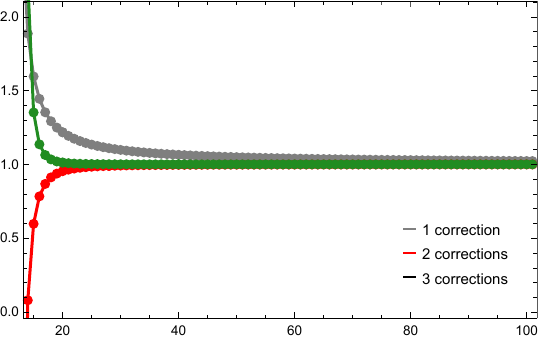}
    \caption{$n = 3$}
  \end{subfigure}
  \hfill
  \begin{subfigure}[t]{0.45\textwidth}
    \centering
    \includegraphics[width=\linewidth]{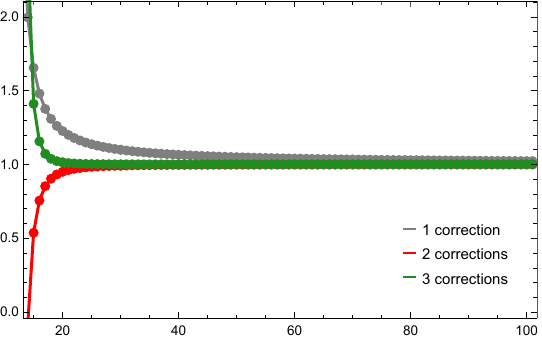}
    \caption{$n = 4$}
  \end{subfigure}

  \caption{Plot of the ratio \eqref{eq:ratio}, sampled for the values $10 \le N \le 100$ in the ungapped phase of GWW model \eqref{eq:GWW} for $t = 5$ and $n = 1,2,3,4$.}
    \label{fig:Instantons10}
\end{figure}

In figure \ref{fig:Instantons10}, we plot the ratio \eqref{eq:ratio} for the GWW model \eqref{eq:GWW} along the range $10 \le N \le 100$ for $t = 5$ (notice that we are only picking values in the ungapped phase) and $n = 1,2,3,4$. We present a series of plots (in various colors) illustrating the effect of including successive correction terms in the saddle-point expansion \eqref{eq:GWWSaddlePointExpansion}. The gray line corresponds to the case where only the leading correction \eqref{eq:GWWLeadingCoeff} is included. The red line incorporates both the leading and subleading \eqref{eq:GWWSubLeadingCoeff} corrections. Finally, the green line represents the inclusion of the third correction term \eqref{eq:GWWSubSubLeadingCoeff}. As expected, the ratio seems to converge to $1$ as we increase $N$ (the first point above). Moreover, the accuracy of the match improves markedly with each additional correction term (the second point above).

\begin{figure}[ht]
  \centering

\begin{subfigure}[t]{0.45\textwidth}
    \centering
    \includegraphics[width=\linewidth]{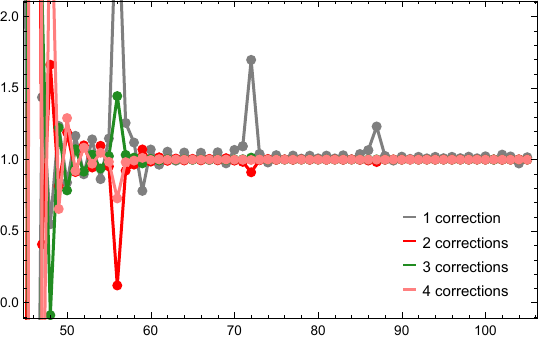}
    \caption{$n = 1$}
  \end{subfigure}
  \hfill
  \begin{subfigure}[t]{0.45\textwidth}
    \centering
    \includegraphics[width=\linewidth]{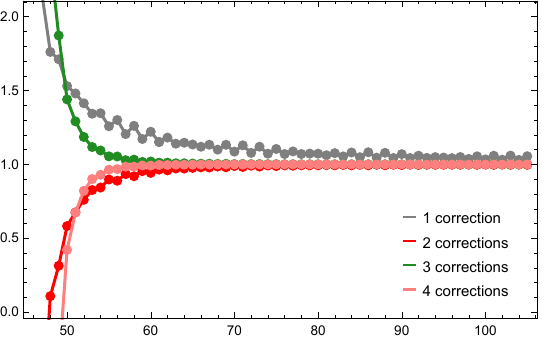}
    \caption{$n = 2$}
  \end{subfigure}

  \vspace{1em}
  \begin{subfigure}[t]{0.45\textwidth}
    \centering
    \includegraphics[width=\linewidth]{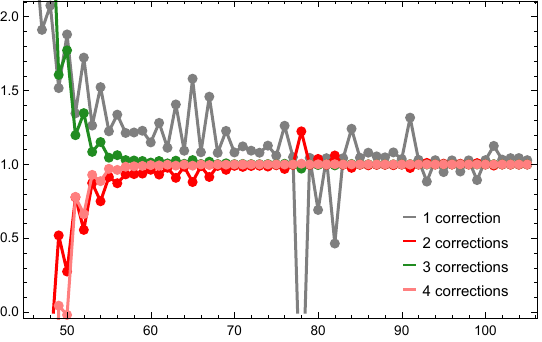}
    \caption{$n = 3$}
  \end{subfigure}
  \hfill
  \begin{subfigure}[t]{0.45\textwidth}
    \centering
    \includegraphics[width=\linewidth]{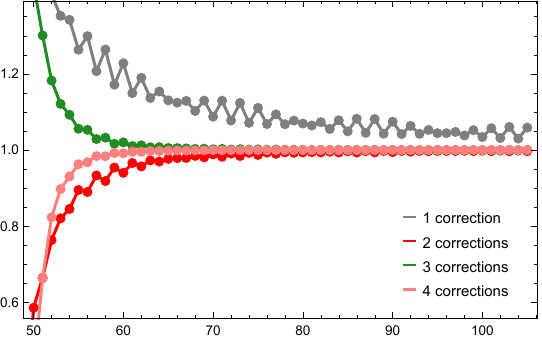}
    \caption{$n = 4$}
  \end{subfigure}

  \vspace{1em}

  \begin{subfigure}[t]{0.45\textwidth}
    \centering
    \includegraphics[width=\linewidth]{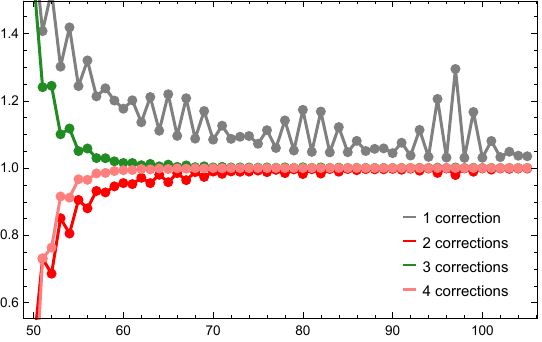}
    \caption{$n = 5$}
  \end{subfigure}
  \hfill
  \begin{subfigure}[t]{0.45\textwidth}
    \centering
    \includegraphics[width=\linewidth]{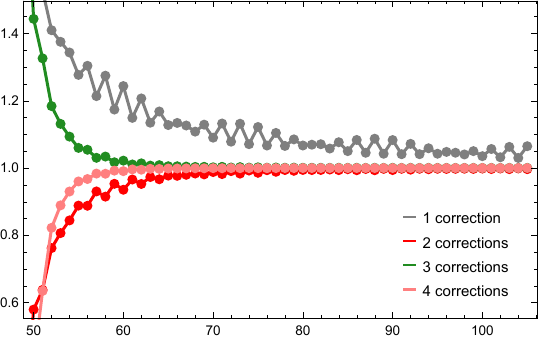}
    \caption{$n = 6$}
  \end{subfigure}

  \caption{Plot of the ratio \eqref{eq:ratio}, sampled for the values $42 \le N \le 105$ in the ungapped phase of the unitary matrix model \eqref{eq:potential} for $n = 1,2,3,4,5,6$. It is interesting to notice that the first correction has large oscillations. This is entirely due to the denominator, which as seen in figure \ref{fig:Ratios35678} can have large oscillations near zero. With subsequent perturbative corrections we see that the agreement becomes excellent. This means that the instanton interference phases and corrections capture exactly the numerical answer.}
  
    \label{fig:Instantons3}
\end{figure}

In figure \ref{fig:Instantons3}, we plot the ratio \eqref{eq:ratio} for the model \eqref{eq:potential} along the range $42 \le N \le 105$ (again, only picking values in the ungapped phase) for $n = 1,2,3,4,5,6$. We present a series of plots (in various colors) illustrating the effect of including successive correction terms in the saddle-point expansion \eqref{eq:Instantons11}. The color coding remains unchanged save for the addition of the pink line, accounting for the inclusion of the fourth correction term in \eqref{eq:Instantons11}. As expected, the ratio seems to converge to $1$ as we increase $N$ (the first point above). Moreover, the accuracy of the match improves markedly with each additional correction term (the second point above).

To obtain a more concrete (less visual) sense of the second point, we collect in table \ref{tab:errortable2} samples of the error
\begin{equation}
    \delta(N) = |Q_n(N)-1|
    \label{eq:error}
\end{equation}
for the GWW model over the range $60\le N \le 64$ for $t = 5$ and $n = 2$, organized according to the number of successive correction terms accounted for in the saddle-point expansion \eqref{eq:GWWSaddlePointExpansion}. Similarly, in table \ref{tab:errortable} we collect samples of the error \eqref{eq:error} for the model \eqref{eq:potential} over the range $82 \le N \le 86$ for $n = 3$, organized in the same fashion. In both cases, we observe very good agreement, which clearly improves with the inclusion of successive correction terms, as expected.

\begin{table}[h]
    \centering
    \begin{tabular}{c|c|c|c|c|c}
         $\delta(N)$& $N = 60$& $N = 61$& $N = 62$& $N = 63$& $N = 64$  \\
       \hline
       1 correction  & $3.84\times10^{-2}$& $3.77\times10^{-2}$& $3.71\times10^{-2}$&$3.64\times10^{-2}$& $3.58\times10^{-2}$ \\
       \hline
       2 corrections  & $1.10\times10^{-3}$& $1.06\times10^{-3}$& $1.02\times10^{-3}$&$9.82\times10^{-4}$& $9.47\times10^{-4}$  \\
       \hline
       3 corrections  & $3.00\times10^{-5}$& $2.82\times10^{-5}$& $2.66\times10^{-5}$&$2.51\times10^{-5}$& $2.37\times10^{-5}$ 
    \end{tabular}
    \caption{Samples of the error \eqref{eq:error} for the GWW model \eqref{eq:GWW} over the range $60 \le N \le 64$ for $t = 5$ and $n = 2$, evaluated while incorporating various correction terms in the saddle-point expansion \eqref{eq:GWWSaddlePointExpansion}.}
    \label{tab:errortable2}
\end{table}

\begin{table}[h]
    \centering
    \begin{tabular}{c|c|c|c|c|c}
         $\delta(N)$& $N = 82$& $N = 83$& $N = 84$& $N = 85$& $N = 86$  \\
       \hline
       1 correction  & $5.36\times10^{-1}$& $4.20\times10^{-2}$& $2.41\times10^{-1}$&$4.52\times10^{-2}$& $7.90\times10^{-2}$ \\
       \hline
       2 corrections  & $6.00\times10^{-2}$& $2.74\times10^{-3}$& $2.36\times10^{-2}$&$2.90\times10^{-3}$& $6.46\times10^{-3}$  \\
       \hline
       3 corrections  & $6.79\times10^{-3}$& $2.72\times10^{-4}$& $2.49\times10^{-3}$&$2.70\times10^{-4}$& $6.30\times10^{-4}$ \\
       \hline
       4 corrections  & $9.24\times10^{-4}$& $3.48\times10^{-5}$& $3.16\times10^{-4}$&$3.24\times10^{-5}$& $7.45\times10^{-5}$
    \end{tabular}
    \caption{Samples of the error \eqref{eq:error} for the model \eqref{eq:potential} over the range $82 \le N \le 86$ for $n = 3$, evaluated while incorporating various correction terms in the saddle-point expansion \eqref{eq:Instantons11}.}
    \label{tab:errortable}
\end{table}

\section{Phase diagram in the coupling complex plane}
\label{sec:phase diagram}

Within the context of Hermitian matrix models, it is well known that in the ’t Hooft limit, the roots of orthogonal polynomials accumulate along the support of the eigenvalue spectral density \cite{b09,b07,s07,bt11,aam13,hkl14,bt16} (see \cite{krsst25a} for detailed numerical explorations of this correspondence for the quartic and cubic Hermitian matrix models). 

In what follows, we will assume that this result extends to the generic single-trace unitary matrix model \eqref{eq:unitmatrix1}, beyond the ungapped phase. This assumption can be motivated analytically (albeit heuristically) by the fact that the model may be recast as an approximate Hermitian matrix model (see equation \eqref{eq:Instantons2}) and, more importantly, is empirically supported by the numerical explorations presented later in this section.

As described in subsection \ref{subsec:bootstrap and ortho polys}, the asymptotic bootstrap estimate method introduced in section \ref{sec:method} provides an efficient numerical procedure for computing orthogonal polynomials in unitary matrix models (further validated by the instanton analysis of subsection \ref{subsec:instantons}). We may therefore use this method to compute orthogonal polynomials for large values of $N$ and plot the associated root distributions, with the goal of investigating the topology of the eigenvalue spectral density as a function of the matrix model couplings. As concretely understood in \cite{krsst25a}, sharp changes in this topology signal phase transitions, which are most naturally understood as occurring in the complex coupling plane or phase diagram.\footnote{Computing the roots of orthogonal polynomials is not the only numerical approach to probing the topology of the eigenvalue spectral density. For instance, one may alternatively solve the equations of motion for the eigenvalue integration variables in \eqref{eq:Instantons2} at finite but large $N$. Such an approach was explored in \cite{bdv16}, providing numerical evidence for multi-cut phases of the GWW model, and in \cite{krsst25a} for the quartic and cubic Hermitian matrix models.}
Thus, a numerical exploration of the phase diagram provides a valuable opportunity to test the effectiveness of the asymptotic bootstrap estimate method in regimes where the measure is complex and positivity is absent. 

In what follows, we present a numerical exploration of the phase diagram for the GWW model \eqref{eq:GWW} in subsection \ref{subsec:GWW} and for the model \eqref{eq:potential} in subsection \ref{subsec:Complicatedmodel}. In the case of the GWW model, our results agree with previous analyses, including those of \cite{cgkt22,m08,m89}. 
For more complicated potentials such as \eqref{eq:potential}, few closed-form results are available in the literature, and to the best of our knowledge the results presented here are new.

In order to conduct efficient numerical explorations of the phase diagram, we must develop a ``smart'' criterion for selecting sampling points, rather than relying on a sufficiently fine (but bounded) grid of complex coupling values in the hope of capturing all phases. Our chosen criterion exploits exact analytical predictions (or suitable approximations) for the phase boundaries, allowing us to determine \emph{a priori} which regions of parameter space should be sampled in order to probe distinct topologies of the eigenvalue spectral density. 
Concretely, this criterion is based on the structure of anti-Stokes lines associated with a given phase and, somewhat unexpectedly, also by the corresponding Stokes lines. 
Provided that distinct topologies are indeed observed numerically in the predicted regions, this exploration supports the validity and practical usefulness of the asymptotic bootstrap estimate method.

\subsection{The Gross-Witten-Wadia model}
\label{subsec:GWW}

The GWW model instanton action in the ungapped phase reads (see equation \eqref{eq:GWWA})
\begin{equation}
    A_{\text{ungapped}}(g) = -\sqrt{1-\frac{4}{g^2}}+ \text{arccosh}\left(\frac{g}{2}\right) \,,
    \label{eq:niceA}
\end{equation}
where 
\begin{equation}\label{eq:def inverse coupling g}
    g=\tilde{t}^{-1} \,.
\end{equation}
In this section, we parametrize the couplings in terms of $g$, yielding a visually clearer phase diagram in which nontrivial phases occupy compact, bounded regions of the complex $g$-plane.

The real part of the instanton action captures the amount of energy required to transport an eigenvalue (anti-eigenvalue) from the eigenvalue distribution on the unit circle to the saddle $z_2^\star$ ($z_1^\star$). Consequently, for large real values of $A(g)$, configurations with eigenvalues localized at the saddles are expected to be exponentially suppressed and thus subdominant. This situation changes drastically when 
\begin{equation}
    \text{Re}\left( A_{\text{ungapped}}(g)\right) =0 \,,
    \label{eq:antistokes}
\end{equation}
as beyond this point carrying eigenvalues from the unit circle distribution to the saddles no longer costs energy. It is therefore not surprising that the locus satisfying the condition above marks the boundary between two distinct phases of the matrix model, distinguished by the emergence of cuts around the saddle locations in the former phase. 

This locus consists of a set of lines, commonly referred to as \emph{anti-Stokes lines}, which may terminate either at singularities of the effective potential (such as $\infty$) or at critical points (double-scaling points). We then expect that the phase boundary separating the the ungapped phase from other phases takes place at the anti-Stokes line of the instanton action \eqref{eq:niceA} \cite{ps92}. 

Following \cite{m89,m08,cgkt22}, we know that the instanton action associated with the phase featuring a cut centered around $\theta = 0$ reads
\begin{equation}
    A_{\text{one-cut}}^{\theta = 0}(g) = \sqrt{\frac{2}{g}\left(\frac{2}{g}-1\right)}-\frac{1}{2}\text{arccosh}\left(\frac{4}{g}-1\right) \,.
    \label{eq:A1cutr}
\end{equation}
while the instanton action associated with the phase featuring a cut centered around $\theta = \pi$ reads
\begin{equation}
    A_{\text{one-cut}}^{\theta = \pi}(g) = A_{\text{one-cut}}^{\theta = 0}(-g) \,.
    \label{eq:A1cutl}
\end{equation}

\begin{figure}
    \centering
    \includegraphics[width=0.5\linewidth]{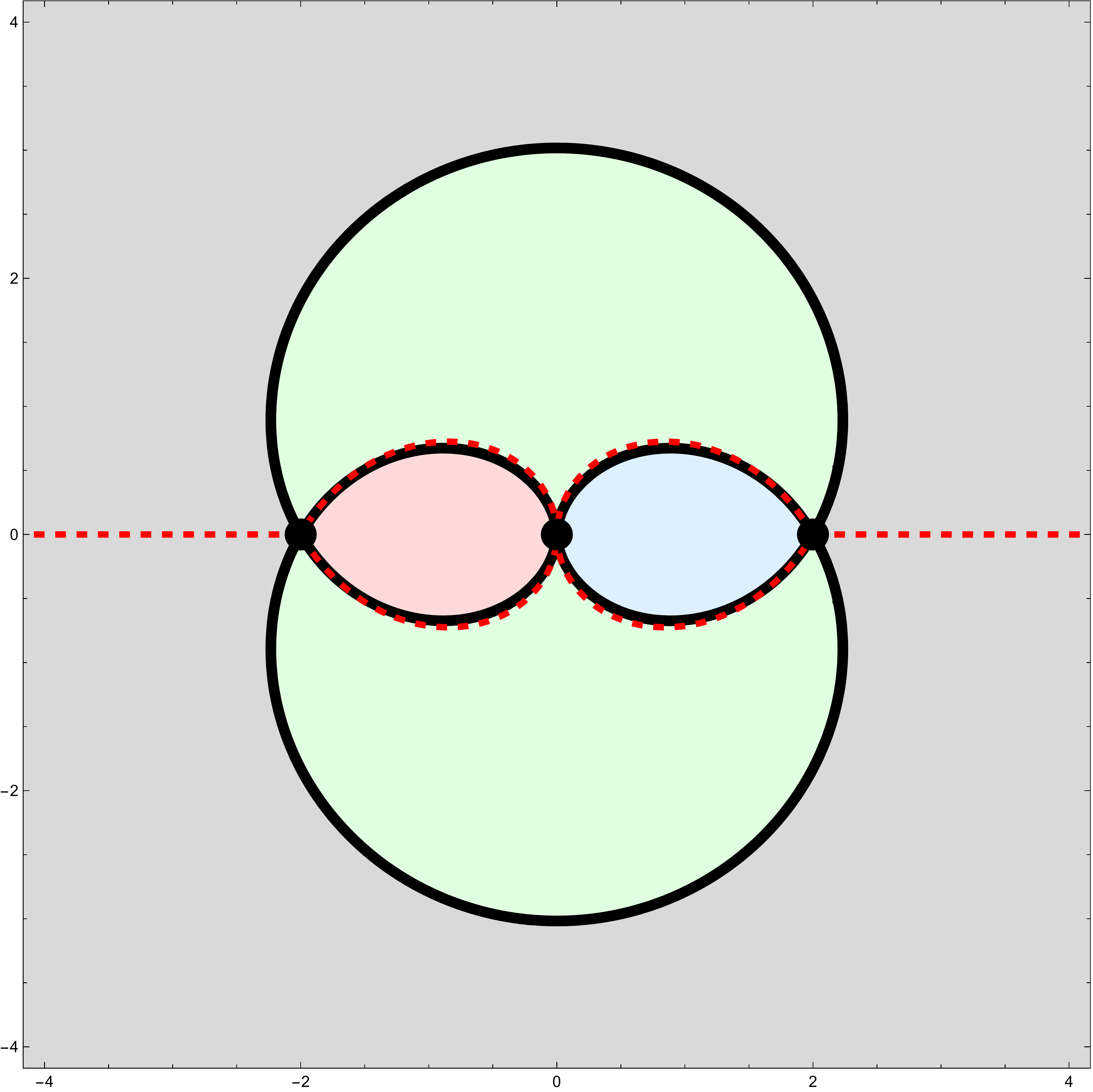}
    \caption{Phase diagram of the GWW model in the complex $g$ plane. The anti-Stokes lines corresponding to the instanton actions \eqref{eq:niceA}, \eqref{eq:A1cutr}, and \eqref{eq:A1cutl} are shown as solid black lines, terminating at the critical points $g = \pm 2$ and the singular point $g = 0$, indicated by black dots. The Stokes lines associated with the instanton action \eqref{eq:niceA} are shown as red dashed lines. The different phases are shaded in light gray, green, red and blue.}
    \label{fig:GWWphasediagram}
\end{figure}

Taking the discussion above into account, we expect the anti-Stokes lines of the instanton actions \eqref{eq:niceA}, \eqref{eq:A1cutr}, and \eqref{eq:A1cutl} to frame the phase boundaries in the GWW matrix model phase diagram. In figure \ref{fig:GWWphasediagram}, we plot these lines, highlighting the regions corresponding to the different expected phases of the model with distinct colors. We expect that each phase will exhibit a distinct topology of the eigenvalue spectral density, and we will confirm this by examining the large $N$ root distribution of orthogonal polynomials, computed using the asymptotic bootstrap estimate method for values of $g$ sampled in each phase.

In this figure, we also plotted the \emph{Stokes lines} associated with the instanton action \eqref{eq:niceA} which can be defined as the locus of points such that (compare with equation \eqref{eq:antistokes})
\begin{equation}
    \text{Im}\left(A_{\text{ungapped}}(g)\right) = 0.
\end{equation}
These mark the points at which the Stokes phenomena occurs, where a discontinuity in the asymptotic expansion of the exact solution takes place \cite{abs18}. In the context of unitary (and Hermitian) matrix models, this manifests itself as a topology change of the Lefschetz thimbles associated with the eigenvalues on the saddles or distributions (cuts) \cite{msw08,mss22}. In particular, the Stokes line shown in figure \ref{fig:GWWphasediagram} along the positive real axis is precisely the line on which we identified the Lefschetz thimble topology change illustrated in figure \ref{fig:Instantons1} of appendix \ref{app:LefschetzThimbleAnalysis}.

Interestingly, the complex Stokes lines shown in figure \ref{fig:GWWphasediagram} provide a very good approximation to the anti-Stokes lines associated with the one-cut phases. In the next subsection, when we turn to the model \eqref{eq:potential}, we will exploit this observation to construct an approximate prediction for the phase boundaries, thereby facilitating its numerical exploration.

It is also interesting to note that the phase diagram shown in figure \ref{fig:GWWphasediagram} bears a striking resemblance to that of the cubic Hermitian matrix model studied in \cite{krsst25a}. In this comparison, the ungapped phase of the unitary matrix model appears to correspond to the analogue of the trivalent phase in the Hermitian counterpart.

Having obtained an exact prediction for the phase diagram of the GWW model (figure \ref{fig:GWWphasediagram}), we now turn to the asymptotic bootstrap estimate method introduced in section \ref{sec:method} in order to generate large $N$ orthogonal polynomials. We then study their root distributions for values of $g$ sampled across the various phases shown in figure \ref{fig:GWWphasediagram}. 

Note that the phase diagram for unitary matrix integrals of the form \eqref{eq:firstpotential} is symmetric under complex conjugation $t_k \mapsto t_k^*$ (or of their reciprocals $g_k$ used in the present section). This follows from the fact that the unitary matrix integral at coupling $t_k^*$ is the complex conjugate of the integral at coupling $t_k$. 
As a result, the moments, Toeplitz matrices, and orthogonal polynomials at $t_k^*$ are complex conjugates of those at $t_k$, implying that the topology of the phase structure is invariant under reflection across the real $t_k$-axis. 
This symmetry is reflected in the phase diagrams and in the corresponding root plots of the two models studied in this paper.  
In addition, the phase structure of the GWW model is symmetric under a $\pi$ rotation of the coupling $g$, since the transformation $g \to -g$ can be compensated by a $\pi$ rotation of the eigenvalue variable $z$ in the GWW potential.

\paragraph{Phase diagram exploration:} In order to realize the correspondence between the root distribution of orthogonal polynomials and the support of the eigenvalue spectral density, it is necessary to consider very large values of $N$. For our practical purposes, $N = 125$ will suffice. Additionally, we will need to compute orthogonal polynomials $p_n$ for very large values of $n$, since a large number of roots is required to accurately reproduce the eigenvalue spectral density (the cuts) and to resolve its topology changes across all phases. In what follows, we fix $n = N$. 

\begin{figure}
    \centering
    \begin{tikzpicture}
        \node at (0,0) {\includegraphics[scale = 0.30]{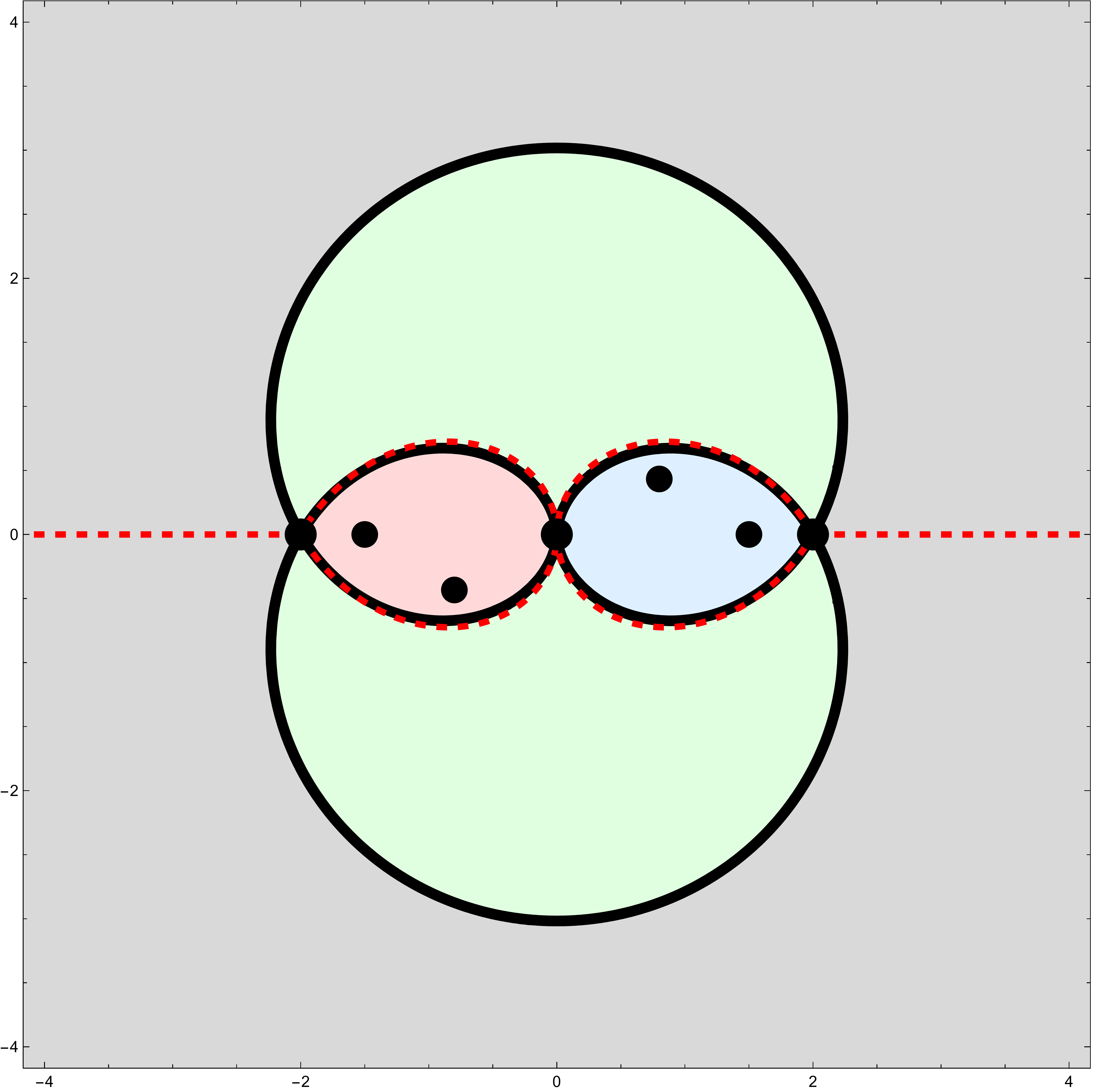}};

\node[anchor=center] at (6.1,-2) {\includegraphics[scale = 0.22]{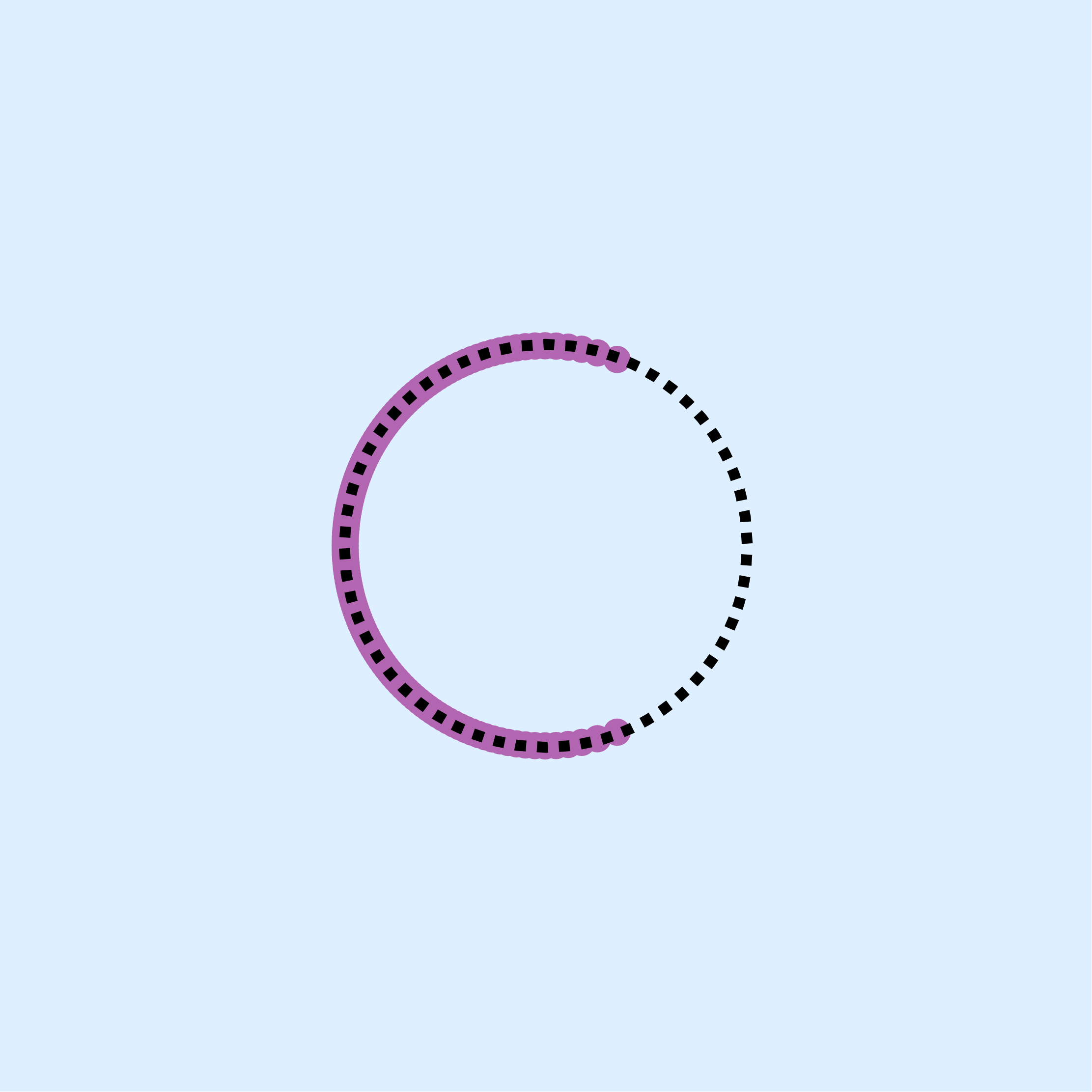}};

\node[anchor=center] at (6.1,2) {\includegraphics[scale = 0.22]{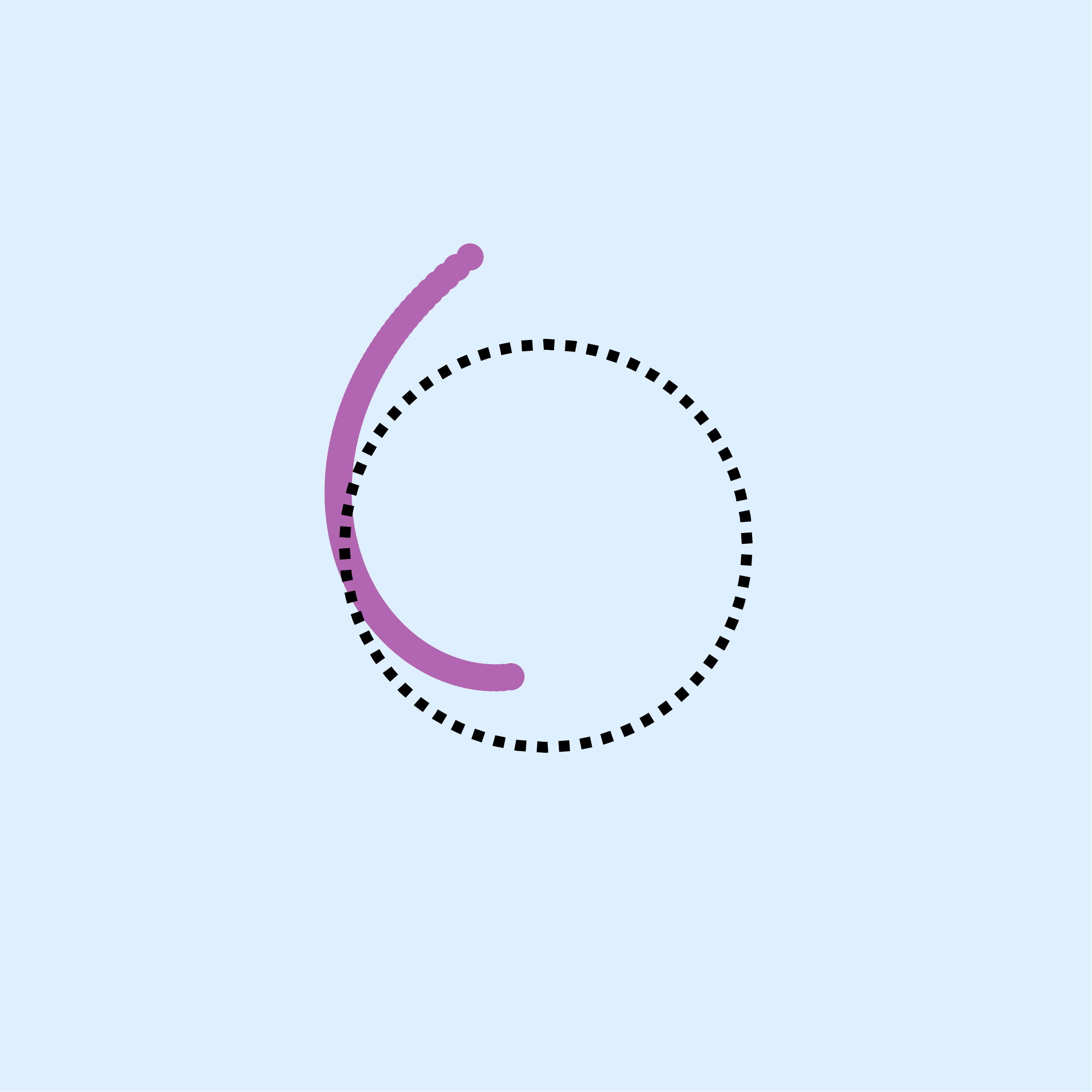}};

\node[anchor=center] at (-6.1,2) {\includegraphics[scale = 0.22]{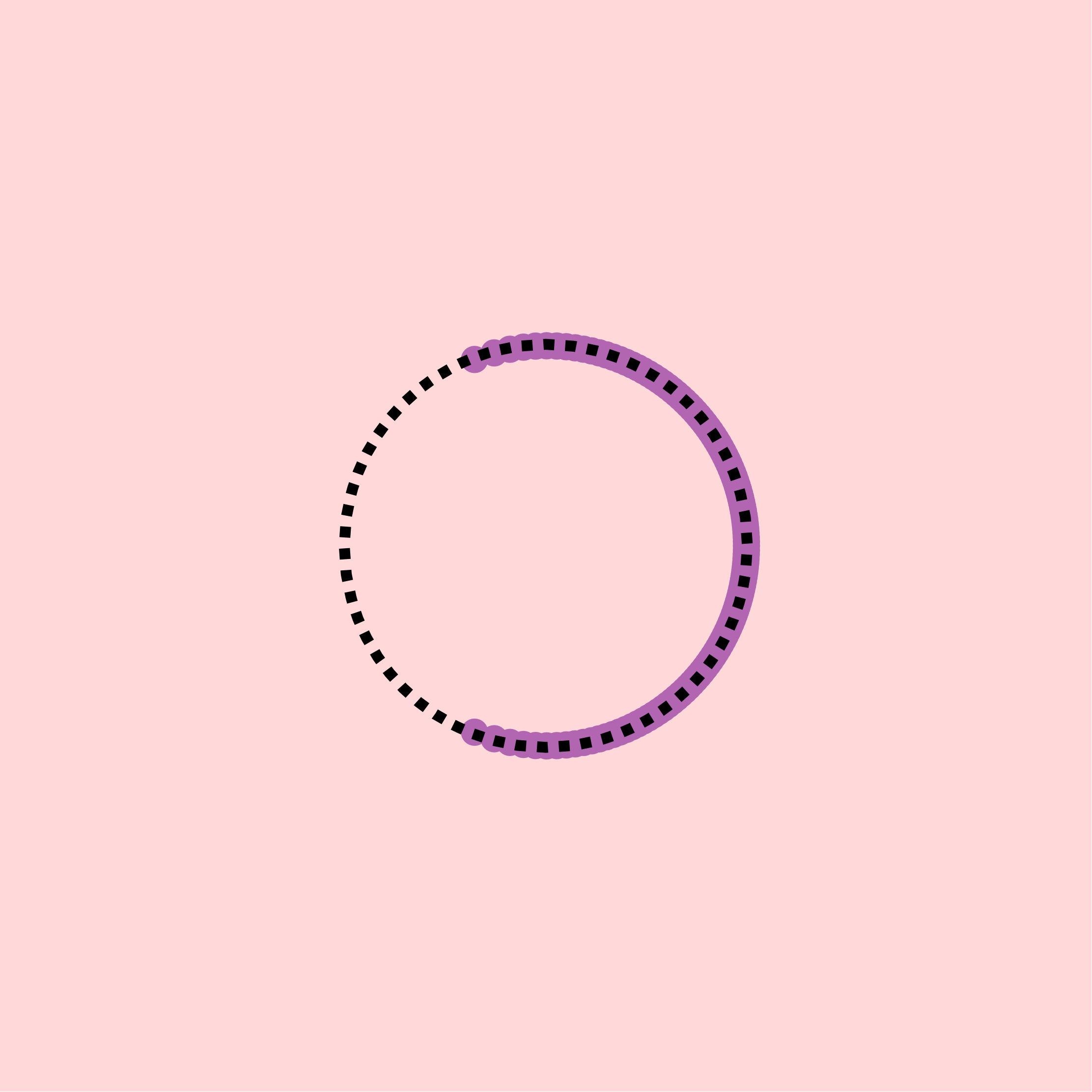}};

\node[anchor=center] at (-6.1,-2) {\includegraphics[scale = 0.22]{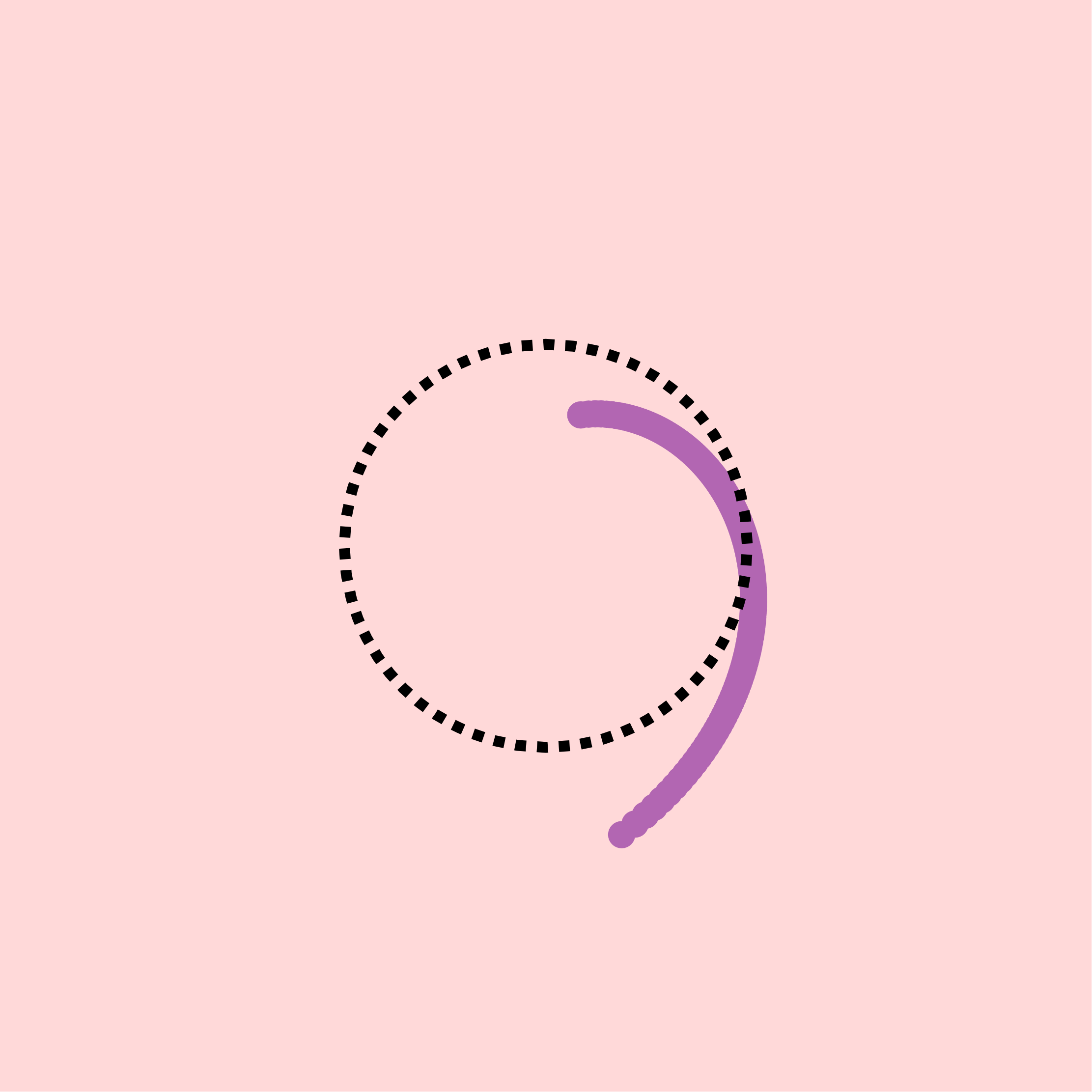}};

\draw[line width = 2pt] (4.545,0.45) rectangle (4.545+3.1,3.55);

\draw[line width = 2pt] (4.545,-0.45) rectangle (4.545+3.1,-3.55);

\draw[line width = 2pt] (-4.545,0.45) rectangle (-4.545-3.1,3.55);

\draw[line width = 2pt] (-4.545,-0.45) rectangle (-4.545-3.1,-3.55);

 \coordinate (Sample1) at (1.38,0.07);
 \coordinate (Sample2) at (0.75,0.45);
 \coordinate (Sample3) at (-1.23,0.05);
 \coordinate (Sample4) at (-0.648,-0.3);

 \draw[line width = 2pt] (Sample1) to[out = -90, in = 180] (4.545,-0.45);
  \draw[line width = 2pt] (Sample1) to[out = -90, in = 180] (4.545,-3.55);

  \draw[line width = 2pt] (Sample2) to[out = 45, in = 180] (4.545,0.45);
  \draw[line width = 2pt] (Sample2) to[out = 45, in = 180] (4.545,3.55);

  \draw[line width = 2pt] (Sample3) to[out = 90, in = 0] (-4.545,0.45);
  \draw[line width = 2pt] (Sample3) to[out = 90, in = 0] (-4.545,3.55);

  \draw[line width = 2pt] (Sample4) to[out = 180+45, in = 0] (-4.545,-0.45);
  \draw[line width = 2pt] (Sample4) to[out = 180+45, in = 0] (-4.545,-3.55);
        
    \end{tikzpicture}
    \caption{Root plots of the orthogonal polynomial $p_n$ for $n = N = 125$, evaluated at four values of $g$ sampled from the light red and blue phases of figure \ref{fig:GWWphasediagram}. Each root plot is displayed alongside the phase diagram, with the corresponding sample indicated by a black dot and connected to its root plot. Within each panel, the roots are shown as purple dots, and the unit circle is indicated by a black dashed line. The sample values are $g = \pm0.80\pm0.43 i$ and $g = \pm 1.5$.}
    \label{fig:GWWonecutsamples}
\end{figure}

\begin{figure}
    \centering
    \begin{tikzpicture}
        \node at (0,0) {\includegraphics[scale = 0.30]{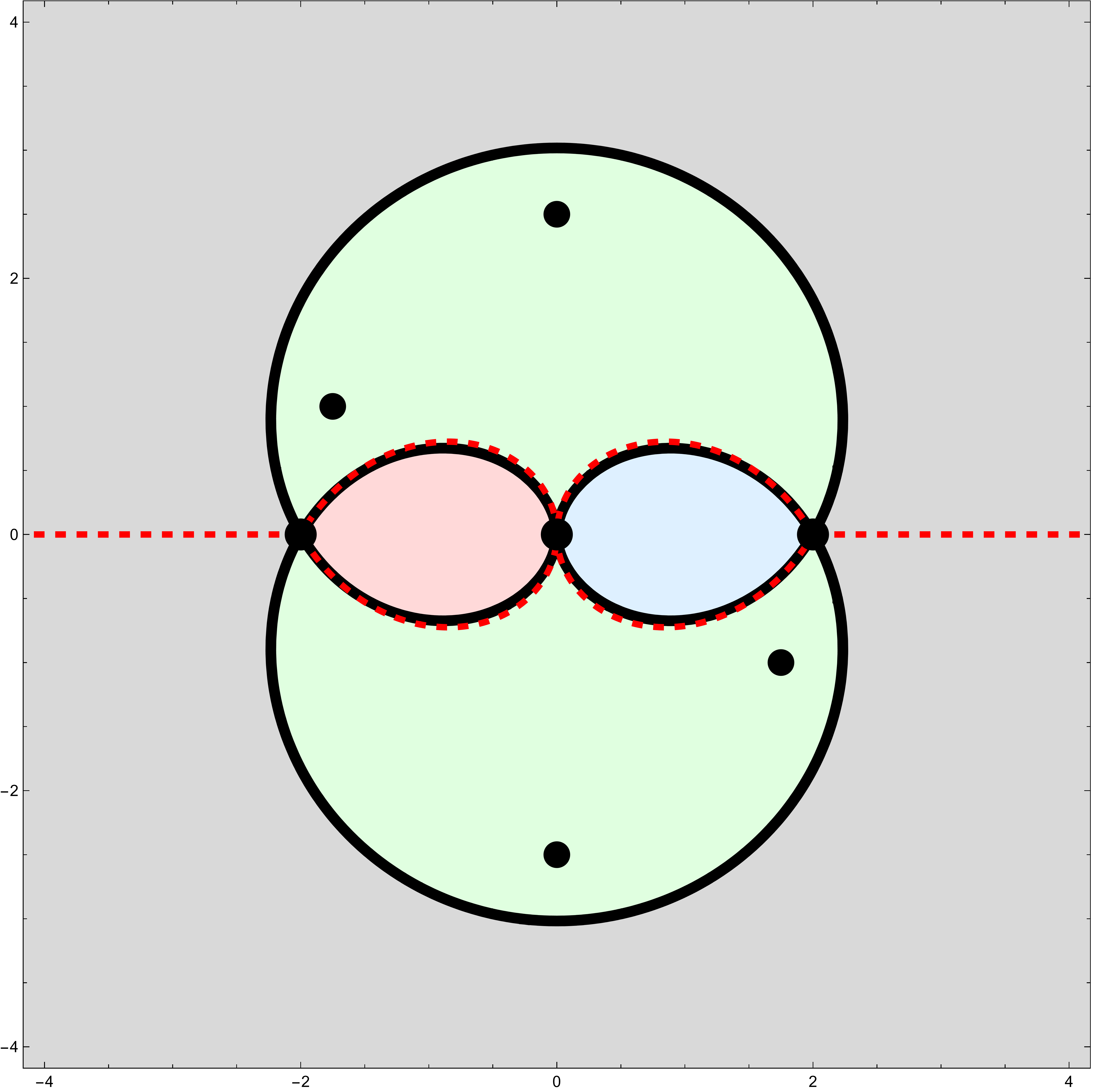}};

\node[anchor=center] at (6.1,2) {\includegraphics[scale = 0.22]{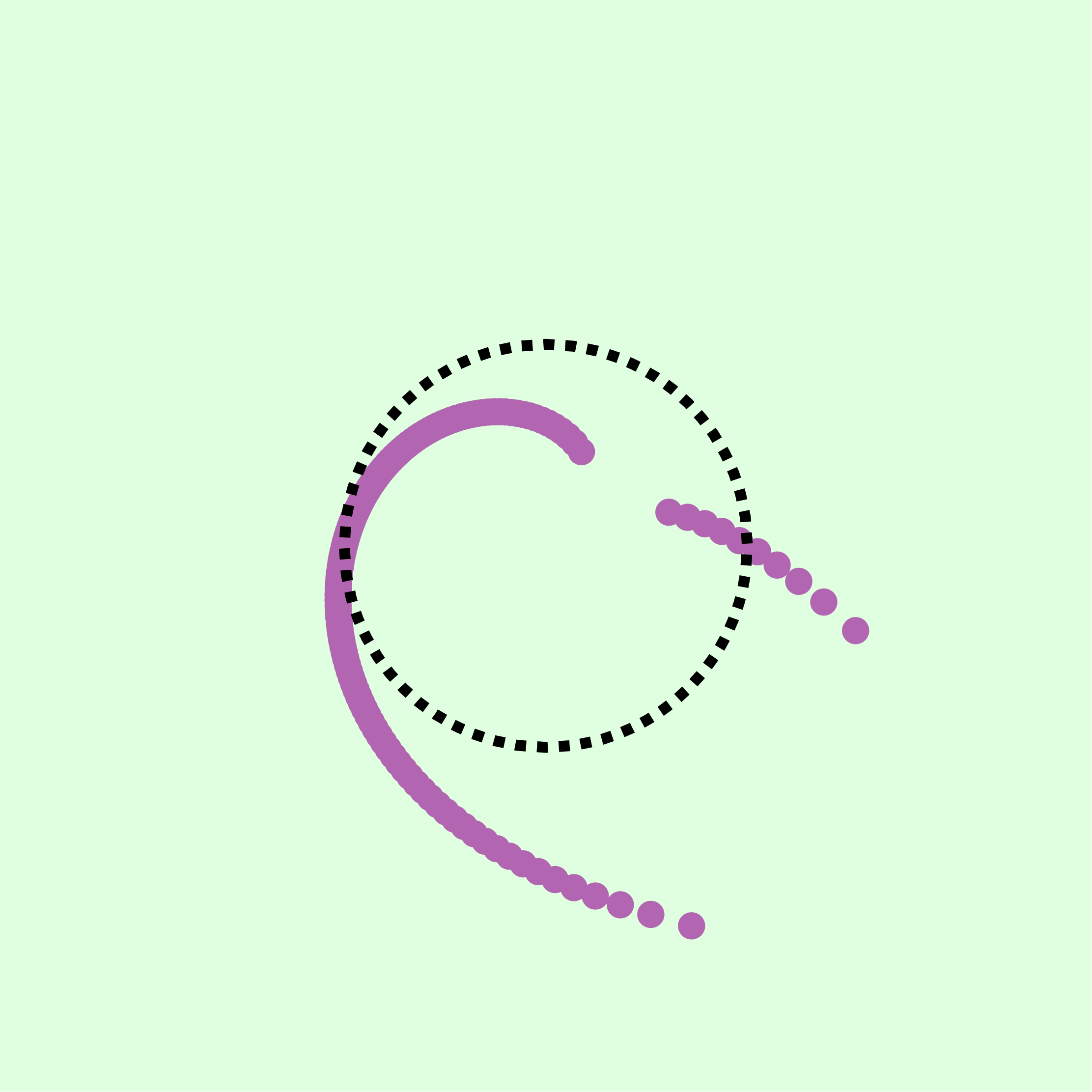}};

\node[anchor=center] at (6.1,-2) {\includegraphics[scale = 0.22]{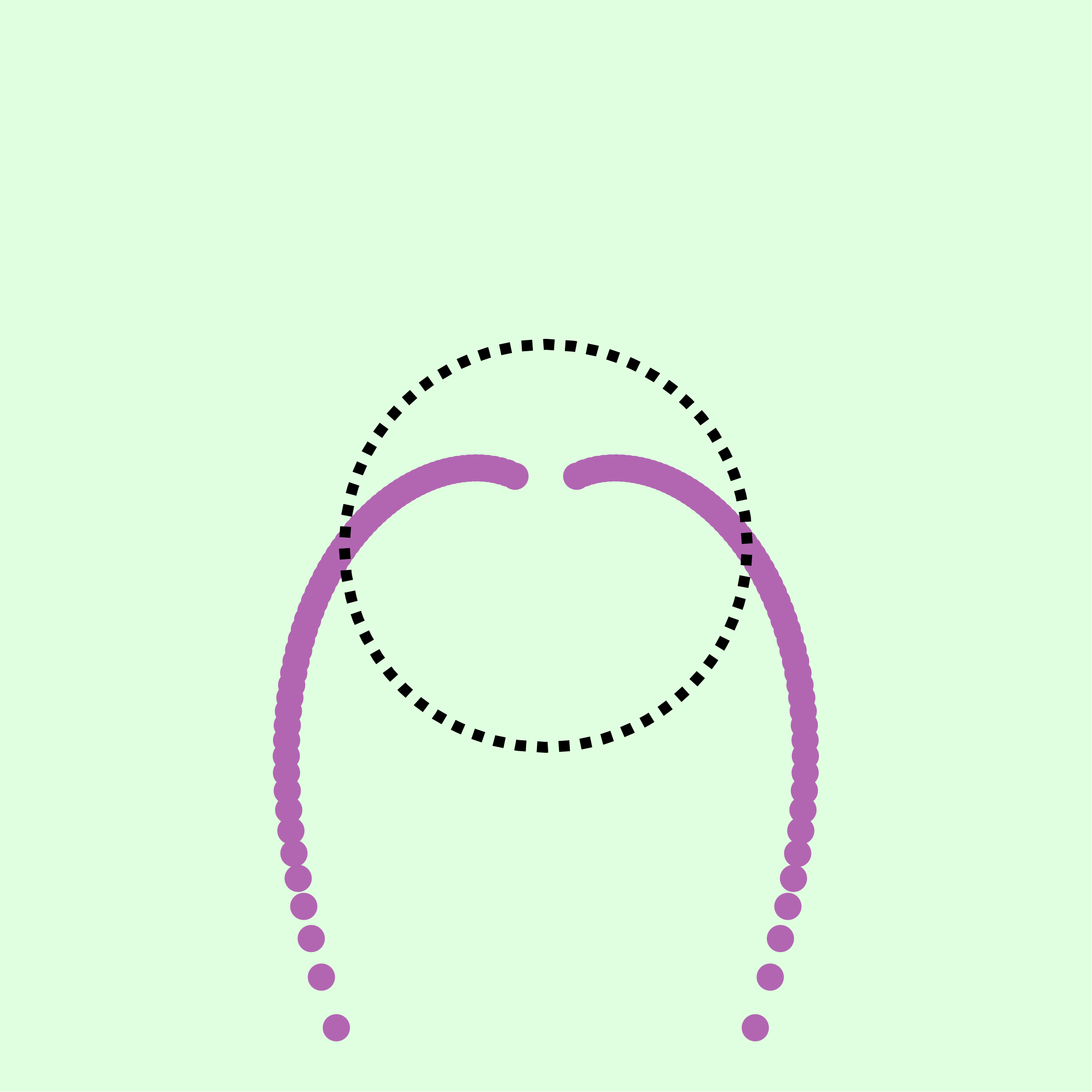}};

\node[anchor=center] at (-6.1,2) {\includegraphics[scale = 0.22]{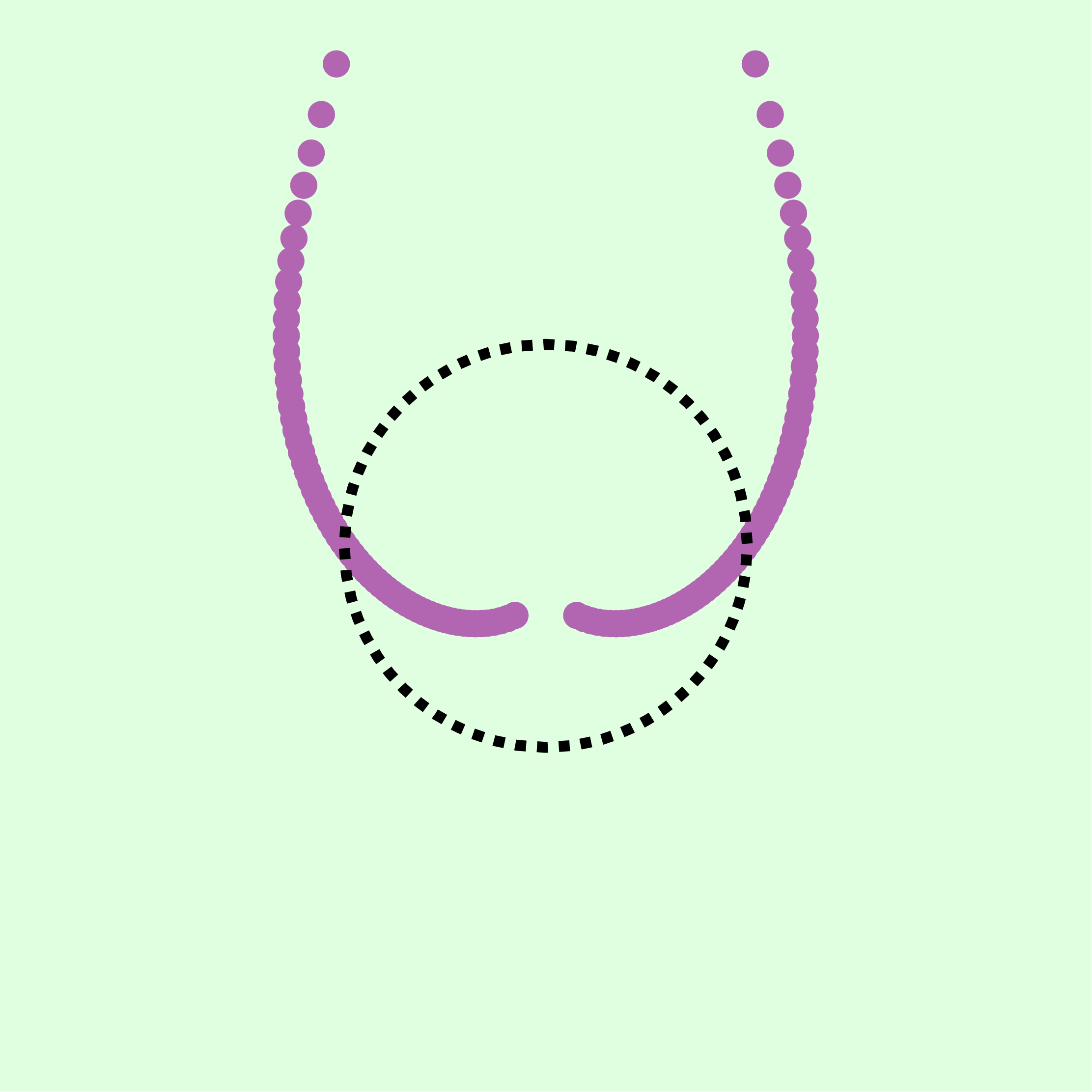}};

\node[anchor=center] at (-6.1,-2) {\includegraphics[scale = 0.22]{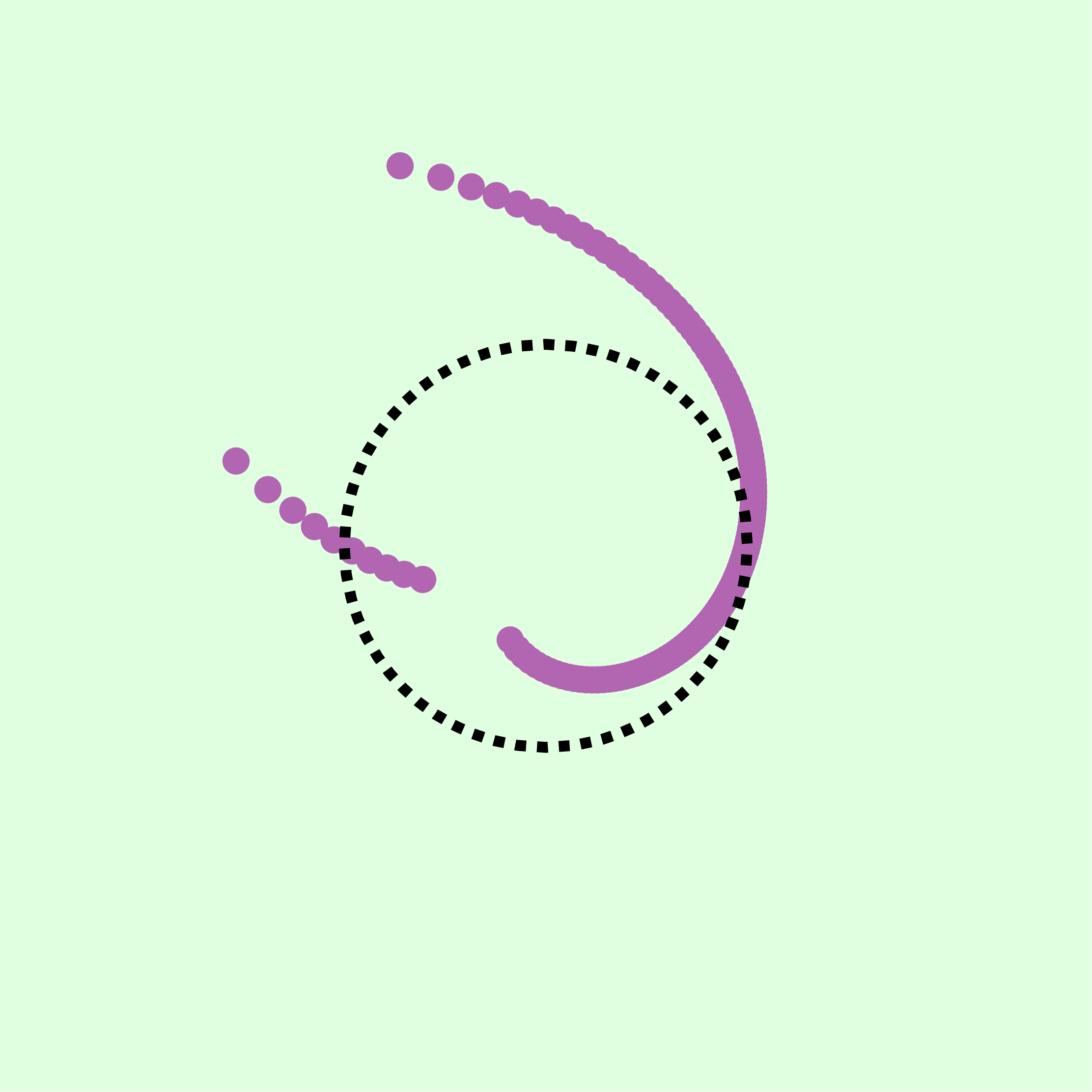}};

\draw[line width = 2pt] (4.545,0.45) rectangle (4.545+3.1,3.55);

\draw[line width = 2pt] (4.545,-0.45) rectangle (4.545+3.1,-3.55);

\draw[line width = 2pt] (-4.545,0.45) rectangle (-4.545-3.1,3.55);

\draw[line width = 2pt] (-4.545,-0.45) rectangle (-4.545-3.1,-3.55);

\coordinate (Sample1) at(0,-2.1);
 \coordinate (Sample2) at  (1.57,-0.8);
 \coordinate (Sample3) at (0,2.25);
 \coordinate (Sample4) at (-1.49,0.95);

 \draw[line width = 2pt] (Sample1) to[out = 0, in = 180] (4.545,-0.45);
  \draw[line width = 2pt] (Sample1) to[out = 0, in = 180] (4.545,-3.55);

  \draw[line width = 2pt] (Sample2) to[out = 0, in = 180] (4.545,0.45);
  \draw[line width = 2pt] (Sample2) to[out = 0, in = 180] (4.545,3.55);

  \draw[line width = 2pt] (Sample3) to[out = 180, in = 0] (-4.545,0.45);
  \draw[line width = 2pt] (Sample3) to[out = 180, in = 0] (-4.545,3.55);

  \draw[line width = 2pt] (Sample4) to[out = 180, in = 0] (-4.545,-0.45);
  \draw[line width = 2pt] (Sample4) to[out = 180, in = 0] (-4.545,-3.55);
        
    \end{tikzpicture}
    \caption{Root plots of the orthogonal polynomial $p_n$ for $n = N = 125$, evaluated at four values of $g$ sampled from the light green phase of figure \ref{fig:GWWphasediagram}. Each root plot is displayed alongside the phase diagram, with the corresponding sample indicated by a black dot and connected to its root plot. Within each panel, the roots are shown as purple dots, and the unit circle is indicated by a black dashed line. The sample values are $g = \pm 1.8\mp 1.0 i$ and $g = \pm 2.5 i$.}
    \label{fig:GWWtwocutsamples}
\end{figure}

\begin{figure}
    \centering
    \begin{tikzpicture}
        \node at (0,0) {\includegraphics[scale = 0.30]{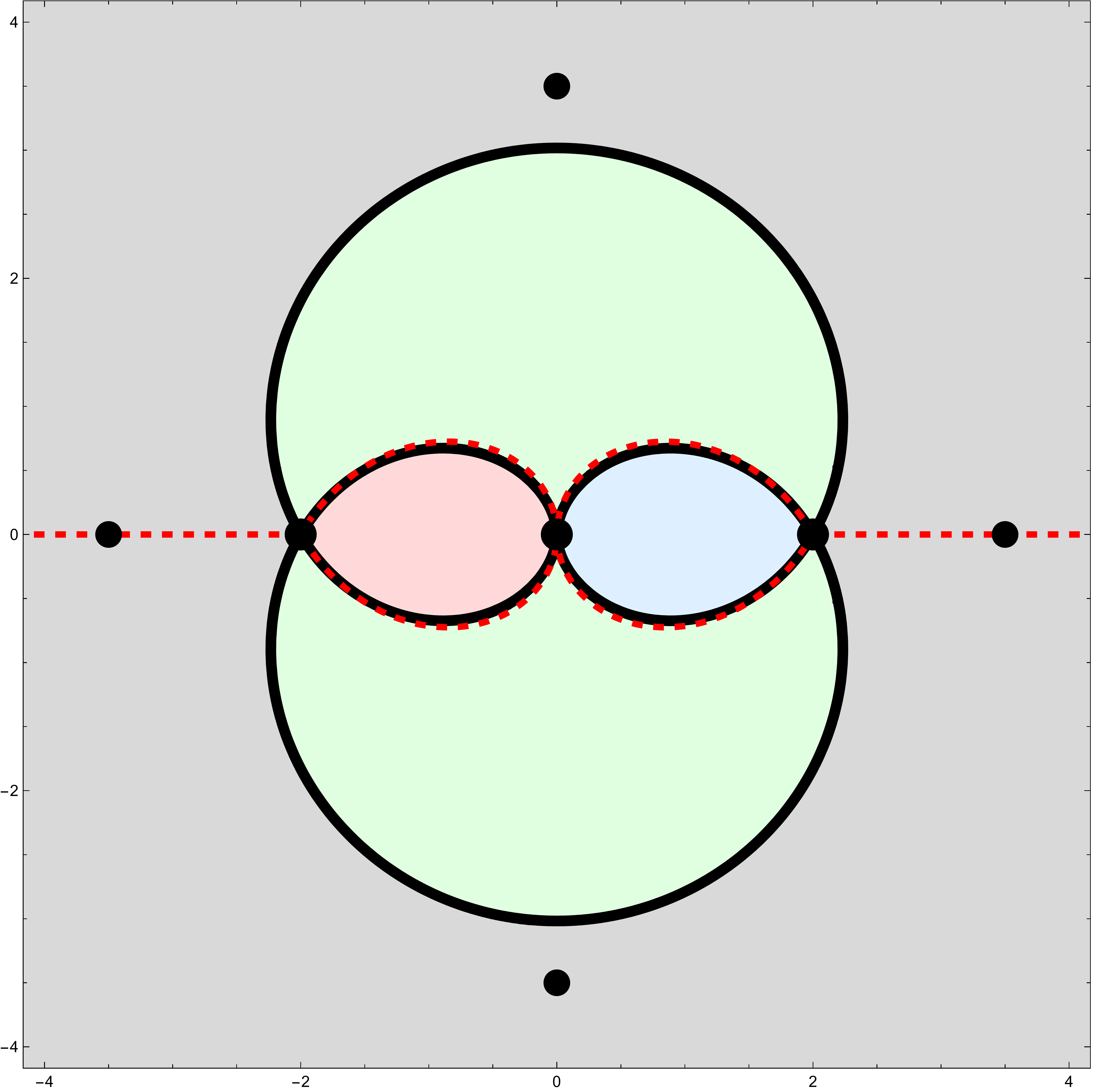}};

\node[anchor=center] at (6.1,2) {\includegraphics[scale = 0.22]{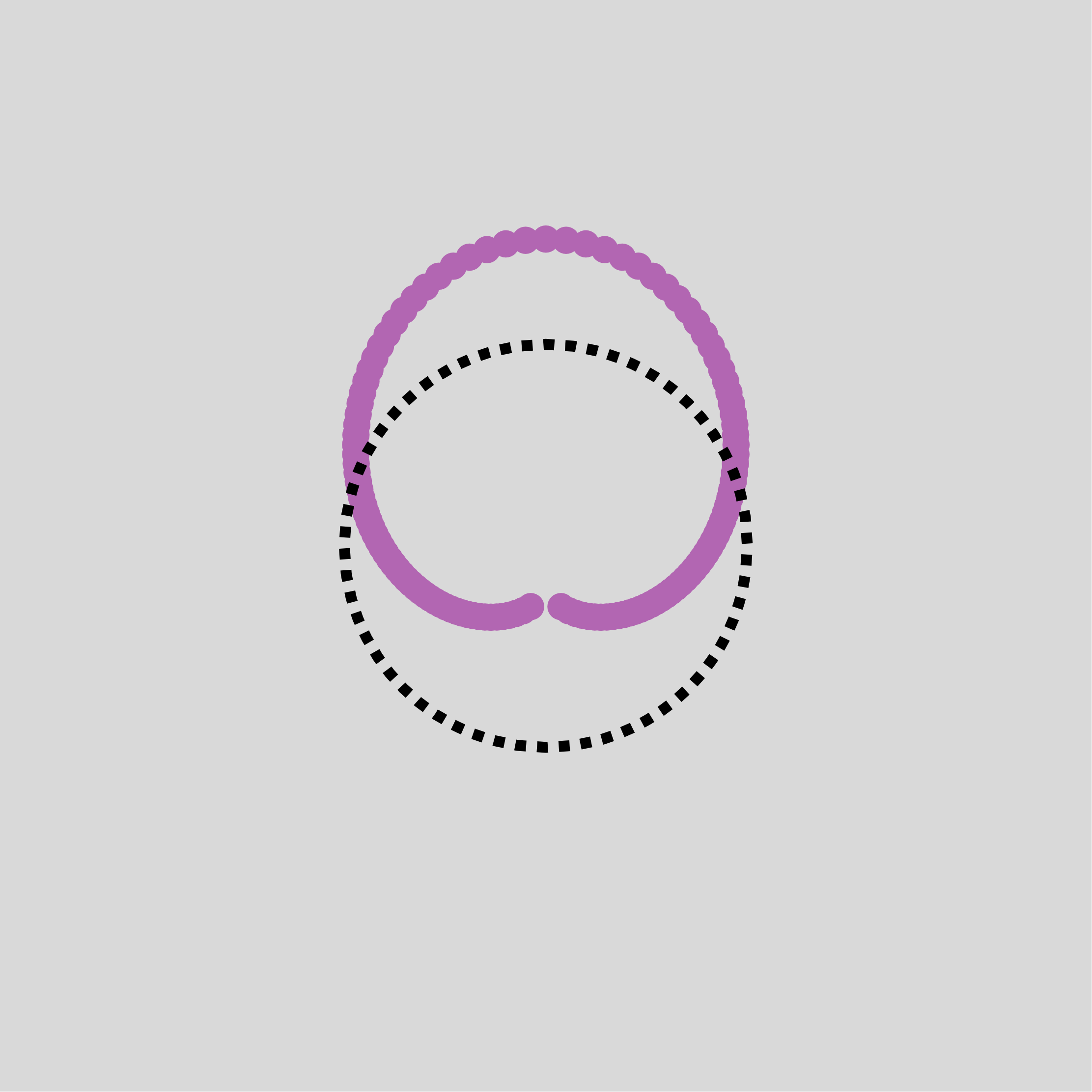}};

\node[anchor=center] at (6.1,-2) {\includegraphics[scale = 0.22]{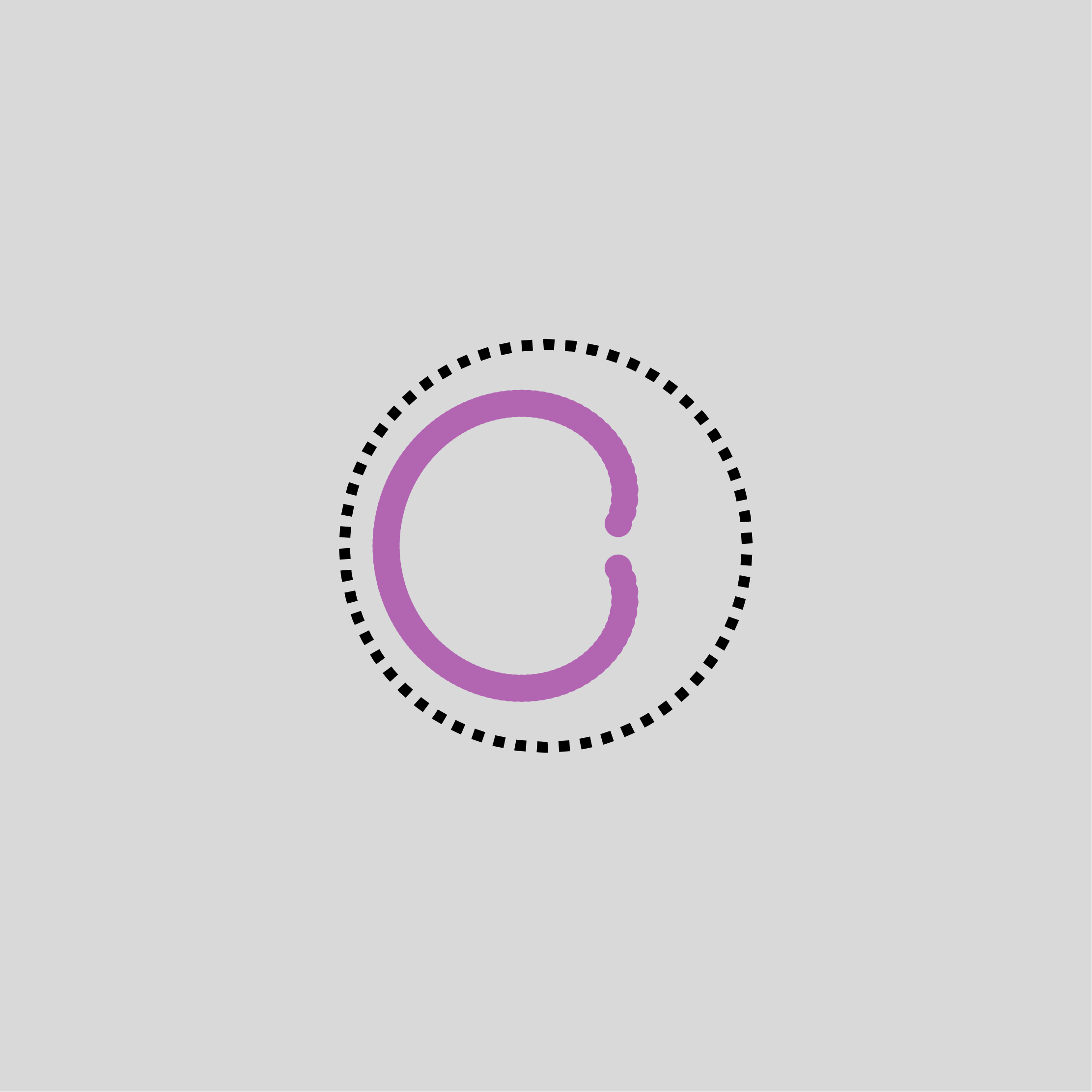}};

\node[anchor=center] at (-6.1,-2) {\includegraphics[scale = 0.22]{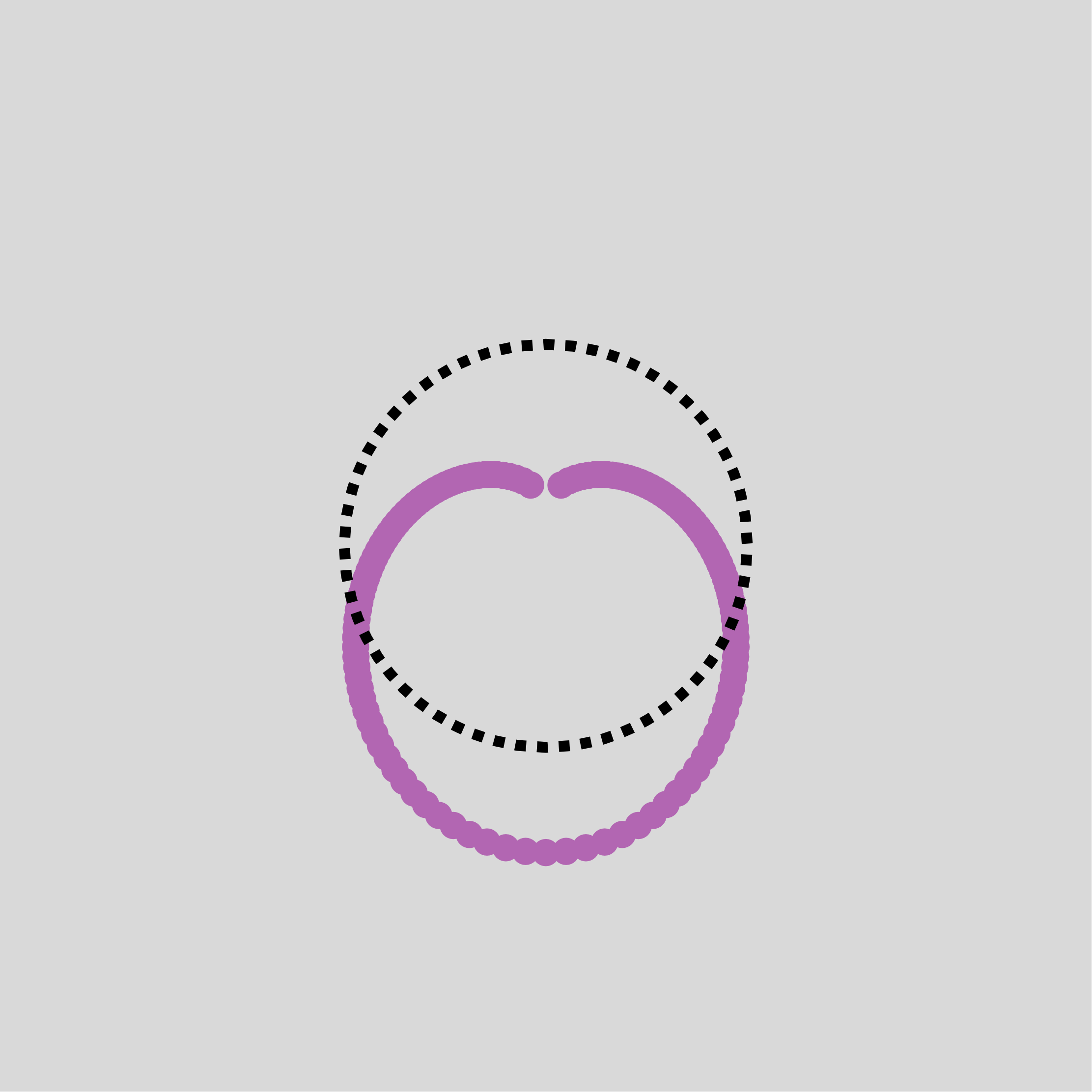}};

\node[anchor=center] at (-6.1,2) {\includegraphics[scale = 0.22]{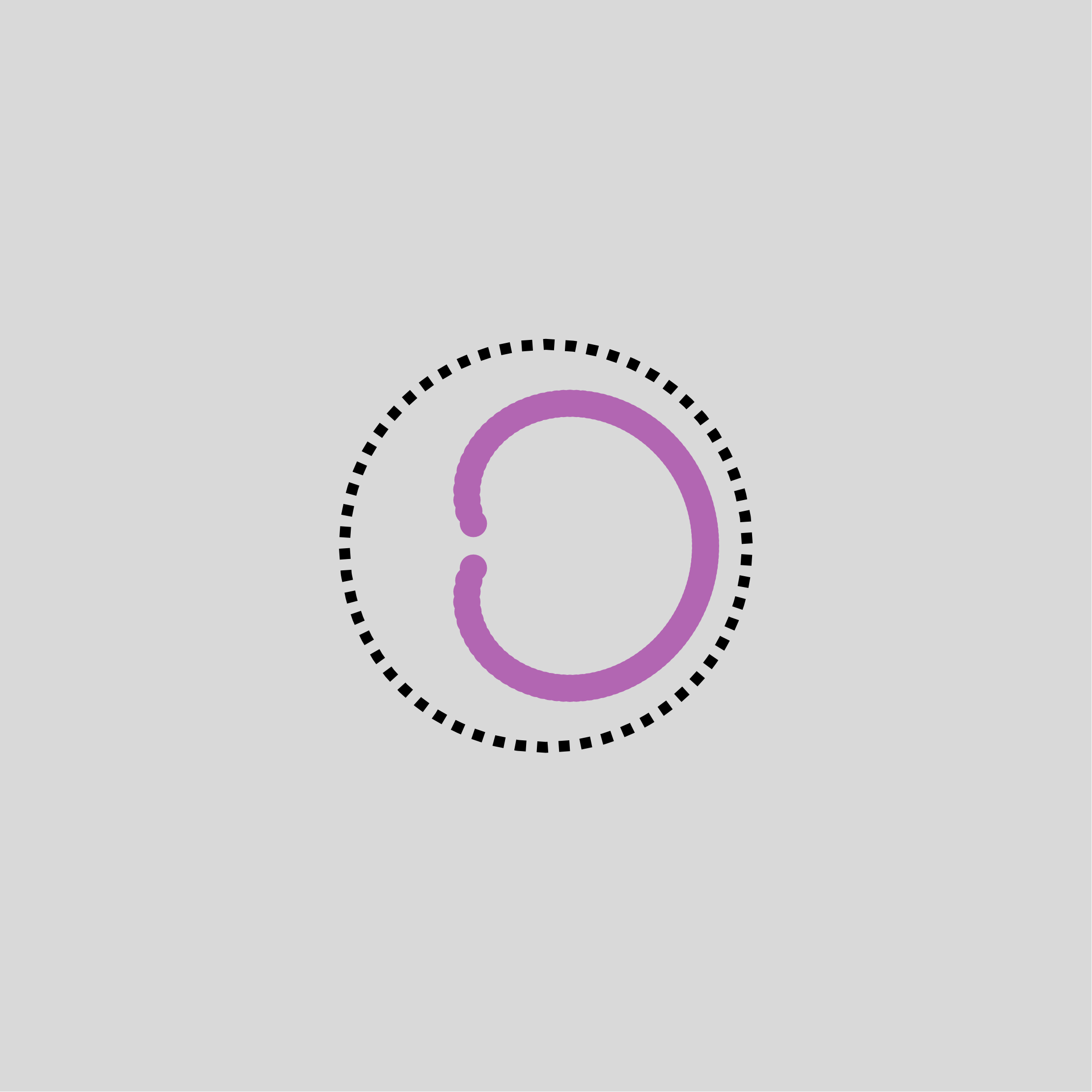}};

\draw[line width = 2pt] (4.545,0.45) rectangle (4.545+3.1,3.55);

\draw[line width = 2pt] (4.545,-0.45) rectangle (4.545+3.1,-3.55);

\draw[line width = 2pt] (-4.545,0.45) rectangle (-4.545-3.1,3.55);

\draw[line width = 2pt] (-4.545,-0.45) rectangle (-4.545-3.1,-3.55);

 \coordinate (Sample1) at (3.12,0.07);
 \coordinate (Sample2) at (0.07,3.1);
 \coordinate (Sample3) at (-2.97,0);
 \coordinate (Sample4) at (0.05,-3);

 \draw[line width = 2pt] (Sample1) to[out = -90, in = 180] (4.545,-0.45);
  \draw[line width = 2pt] (Sample1) to[out = -90, in = 180] (4.545,-3.55);

  \draw[line width = 2pt] (Sample2) to[out = 45, in = 180] (4.545,0.45);
  \draw[line width = 2pt] (Sample2) to[out = 45, in = 180] (4.545,3.55);

  \draw[line width = 2pt] (Sample3) to[out = 90, in = 0] (-4.545,0.45);
  \draw[line width = 2pt] (Sample3) to[out = 90, in = 0] (-4.545,3.55);

  \draw[line width = 2pt] (Sample4) to[out = 180+45, in = 0] (-4.545,-0.45);
  \draw[line width = 2pt] (Sample4) to[out = 180+45, in = 0] (-4.545,-3.55);
        
    \end{tikzpicture}
    \caption{Root plots of the orthogonal polynomial $p_n$ for $n = N = 125$, evaluated at four values of $g$ sampled from the light gray phase of figure \ref{fig:GWWphasediagram}. Each root plot is displayed alongside the phase diagram, with the corresponding sample indicated by a black dot and connected to its root plot. Within each panel, the roots are shown as purple dots, and the unit circle is indicated by a black dashed line. The sample values are $g = \pm3.5 i$ and $g = \pm 3.5$. }
    \label{fig:GWWungappedsamples}
\end{figure}

In figure \ref{fig:GWWonecutsamples}, we plot the root distribution of $p_n$ for four distinct values of $g$ (two purely real and two complex) sampled from the light red and blue phases shown in figure \ref{fig:GWWphasediagram}. It is evident that the light red (light blue) phase corresponds to a one-cut phase in which the eigenvalue spectral density is centered around $z=1$ ($-1$) for real values of $g$. This agrees well with previous results \cite{m08,cgkt22}. In figure \ref{fig:GWWtwocutsamples}, we plot the root distribution of $p_n$ for four distinct values of $g$ (two purely imaginary and two complex) sampled from the light green phase shown in figure \ref{fig:GWWphasediagram}. It is clear that this corresponds to a two-cut phase, agreeing with \cite{cgkt22}. Finally, in figure \ref{fig:GWWungappedsamples}, we show the root distribution of $p_n$ for four distinct values of $g$ (two purely real and two purely imaginary) sampled from the light gray phase shown in figure \ref{fig:GWWphasediagram}. As expected, this phase appears to correspond to the ungapped phase. 

These plots show that the root distributions of the orthogonal polynomials accurately capture the eigenvalue spectral density topology changes across the phases of the GWW model \eqref{eq:GWW}. This provides strong supporting evidence that the asymptotic bootstrap estimate method, together with the computation of the orthogonal polynomials, performs well for complex measures, as anticipated in section \ref{sec:method}. We now turn to the model \eqref{eq:potential} and undertake a similar numerical exploration of its phase diagram.

\subsection{A more complicated model with quadratic single-trace potential}
\label{subsec:Complicatedmodel}

We will now promote the potential \eqref{eq:potential} to
\begin{equation}
    V(U) =  \frac{10}{g}\left(\text{Tr}\left(U\right)+\text{Tr}\left(U^{-1}\right)+\text{Tr}\left(U^2\right)+\text{Tr}\left(U^{-2}\right)\right)
    \label{eq:promotedpotential}
\end{equation}
in order to take the ’t Hooft limit. Using the equation \eqref{eq:Instantons13}, we can write the corresponding effective potential as
\begin{equation}
    V_{\text{eff}}(z) = -\log(z) - \frac{10}{g}\left(z-\frac{1}{z} +z^2 -\frac{1}{z^2}\right)
    \label{eq:Phase1}
\end{equation}
whose saddles read
\begin{align}
     &z_1^\star = \frac{1}{40} \left(-\sqrt{10} \sqrt{-2 g+\sqrt{825-20 g}-75}-\sqrt{825-20 g}-5\right)\\ &z_2^\star =
       \frac{1}{40} \left(\sqrt{10} \sqrt{-2 g+\sqrt{825-20 g}-75}-\sqrt{825-20 g}-5\right)\\  &z_3^\star =\frac{1}{40} \left(-\sqrt{10} \sqrt{-2 g-\sqrt{825-20 g}-75}+\sqrt{825-20 g}-5\right)\\ &z_4^\star =\frac{1}{40} \left(\sqrt{10} \sqrt{-2 g-\sqrt{825-20 g}-75}+\sqrt{825-20 g}-5\right).
        \label{eq:Instantons16}
\end{align}

We begin by deriving analytical predictions for the potential phase boundaries, enabling a more efficient numerical exploration of the phase diagram in what follows. Analogously to what we did in subsection \ref{subsec:GWW}, we will derive these predictions by computing the relevant anti-Stokes lines. The novelty here lies on the fact that we have more than one pair of saddles and as such, one has to take into account multiple instanton actions. More precisely, we will consider the anti-Stokes lines associated with the instanton actions (recall formula \eqref{eq:Acomplicated})
\begin{align}
     &  A_1 = A(z_3^\star,z_4^\star) =\frac{V_{\text{heff}}(z_3^\star) -V_{\text{heff}}(z_4^\star)}{2} \label{eq:A1}\\ 
        &A_2 = A(z_1^\star,z_2^\star) =\frac{V_{\text{heff}}(z_1^\star) -V_{\text{heff}}(z_2^\star)}{2}.\label{eq:A2} 
\end{align}
We considered these instanton actions because they were included in the saddle-point expansion \eqref{eq:Instantons11} and therefore must be taken into account in a complete transseries description of the unitary matrix model partition function. Interestingly, the anti-Stokes lines associated with the remaining instanton actions
\begin{align}
    A_3 = A(z_3^\star,z_2^\star) = \frac{V_{\text{heff}}(z_3^\star) -V_{\text{heff}}(z_2^\star)}{2} \label{eq:A3}\\ 
   A_4 = A(z_1^\star,z_4^\star) = \frac{V_{\text{heff}}(z_1^\star) -V_{\text{heff}}(z_4^\star)}{2} \label{eq:A4}
\end{align}
do not appear to signal a sharp change in the topology of the orthogonal polynomial root distributions in the numerical exploration that follows, and we therefore do not consider them here.

We expect the matrix model to exhibit several additional phases beyond the ungapped phase considered so far. These phases are anticipated to feature between one and four cuts, corresponding to the number of saddles present in the model. Unfortunately, no closed-form expressions are currently known for the corresponding instanton actions in these phases, as their computation requires the solution of a highly non-trivial Riemann–Hilbert problem \cite{m89,cgkt22}. In order to proceed with the numerical exploration, we seek to circumvent this difficulty by constructing analytical approximations to the anti-Stokes lines associated with these instanton actions. This approach is motivated by the observation that some of the Stokes lines associated with the ungapped-phase instanton action \eqref{eq:niceA} provide an excellent approximation, particularly in the vicinity of critical points, to the remaining anti-Stokes lines of the GWW model (see figure \ref{fig:GWWphasediagram}). 

We emphasize that we are not claiming to have determined the exact shape of the phase diagram, nor that our Stokes line approximation method is fully accurate. Indeed, figure \ref{fig:GWWphasediagram} shows that some of the Stokes lines do not approximate anti-Stokes lines. Rather, we draw on empirical evidence observed in the GWW model, where the phase diagram is fully understood, to formulate a practical criterion for selecting sample values of $g$ in a ``smart'' way when probing different eigenvalue spectral density topologies.

Thus, we further consider the Stokes lines associated with the instanton actions \eqref{eq:A1} and \eqref{eq:A2} as well as \eqref{eq:A3} and \eqref{eq:A4}.\footnote{These instanton actions are multi-valued, with branch points located at the critical and singular points. Consequently, the associated Stokes lines may differ from one sheet to another, and in what follows we often consider multiple Stokes lines, living on different sheets, associated with the same instanton action. This choice is primarily empirically motivated, as these lines appear to approximately bound new phases in the numerical analysis presented below. Related considerations emphasizing the relevance of the multi-sheeted structure of instanton actions were already discussed in \cite{krsst25a}.} Moreover, we consider the Stokes lines associated with the instanton action
\begin{equation}
    A_5 = A(z_1^\star,z_3^\star) =\frac{V_{\text{heff}}(z_1^\star) -V_{\text{heff}}(z_3^\star)}{2}
    \label{eq:A5}
\end{equation}
which corresponds to the tunneling of eigenvalues and anti-eigenvalues between the saddles located outside the unit circle. 
While the physical relevance of (anti) Stokes lines of these saddle-to-saddle tunneling was already noticed for Hermitian matrix models in \cite{sst23}, it is still surprising that some of these Stokes lines can approximate phase boundaries in the present case. 

\begin{figure}
    \centering
    \includegraphics[width=0.5\linewidth]{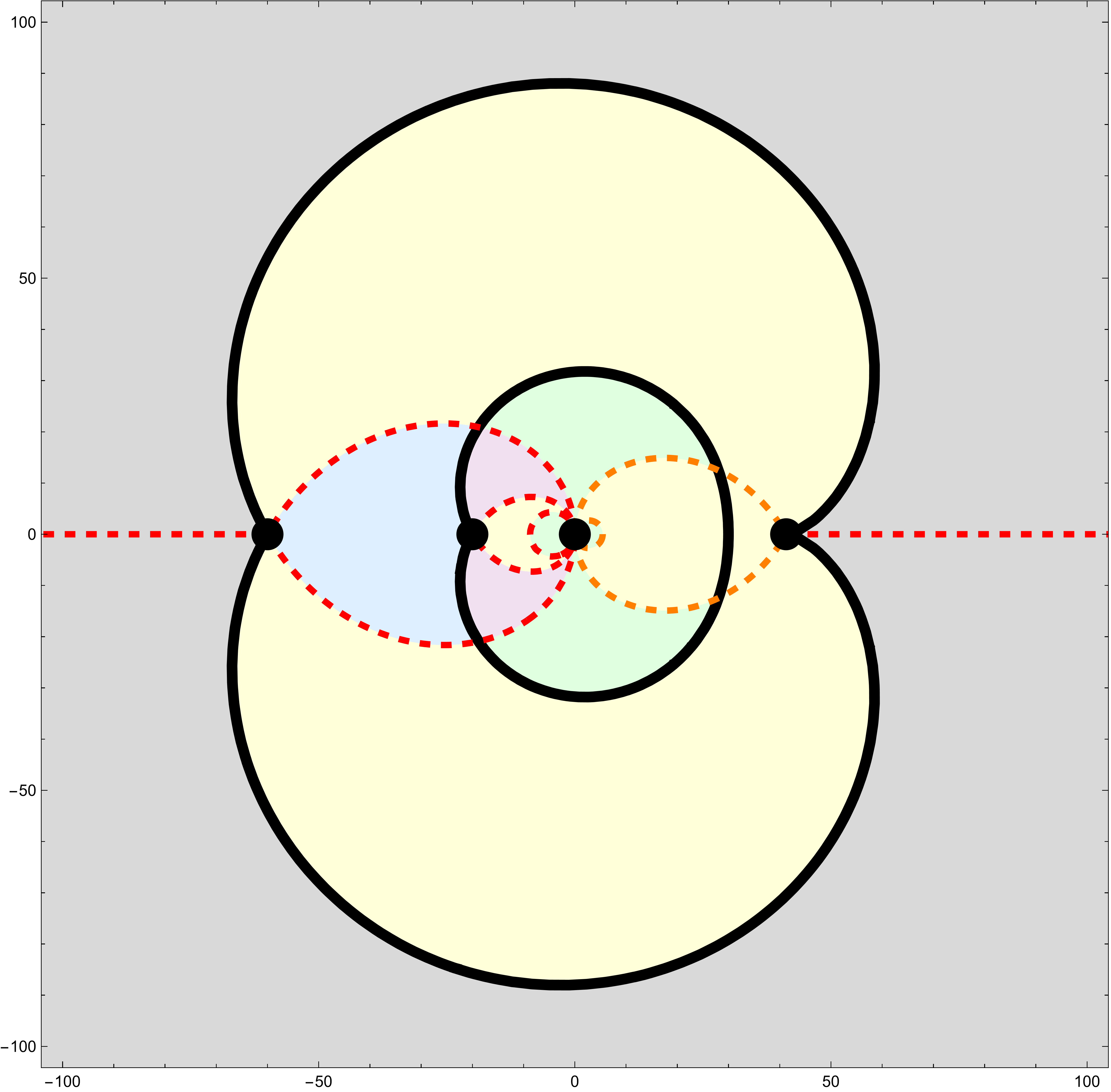}
    \caption{Approximate phase diagram of the model \eqref{eq:promotedpotential} (complex $g$ plane). The anti-Stokes lines corresponding to the instanton actions \eqref{eq:A1} and \eqref{eq:A2} are shown as solid black lines. The critical points $g = -60,-20,165/4$ and the singular point $g = 0$ are indicated by black dots. The Stokes lines associated with the instanton actions \eqref{eq:A1}, \eqref{eq:A2}, \eqref{eq:A3} and \eqref{eq:A4} are shown as red dashed lines while the Stokes lines associated with \eqref{eq:A5} are shown as orange dashed lines. The approximate regions corresponding to the different phases are shaded in light gray, green, yellow, purple, and blue.}
    \label{fig:CMMphasediagram}
\end{figure}

In figure \ref{fig:CMMphasediagram}, we display the relevant Stokes and anti-Stokes lines, and highlight in distinct colors the approximate regions in which different topologies of the eigenvalue spectral density are observed in the following numerical exploration. As in subsection \ref{subsec:GWW}, we examine the large $N$ root distributions of orthogonal polynomials, computed using the asymptotic bootstrap estimate method, for values of $g$ sampled within each region. As in figure \ref{fig:GWWphasediagram}, we do not expect the Stokes lines (shown as dashed curves) to coincide exactly with the phase boundaries. Rather, they should provide a sufficiently accurate approximation (particularly in the vicinity of the critical points) to guide the identification of new phases.

\paragraph{Phase diagram exploration:} Just like in the numerical exploration of the GWW model in subsection \ref{subsec:GWW}, we will need to compute orthogonal polynomials $p_n$ for very large values of $n$, since a large number of roots is required to accurately reproduce the eigenvalue spectral density (the cuts) and to resolve its topology changes across all phases. In what follows, we fix $n =N=125$ just as before. 

\begin{figure}
    \centering
    \begin{tikzpicture}
        \node at (0,0) {\includegraphics[scale = 0.30]{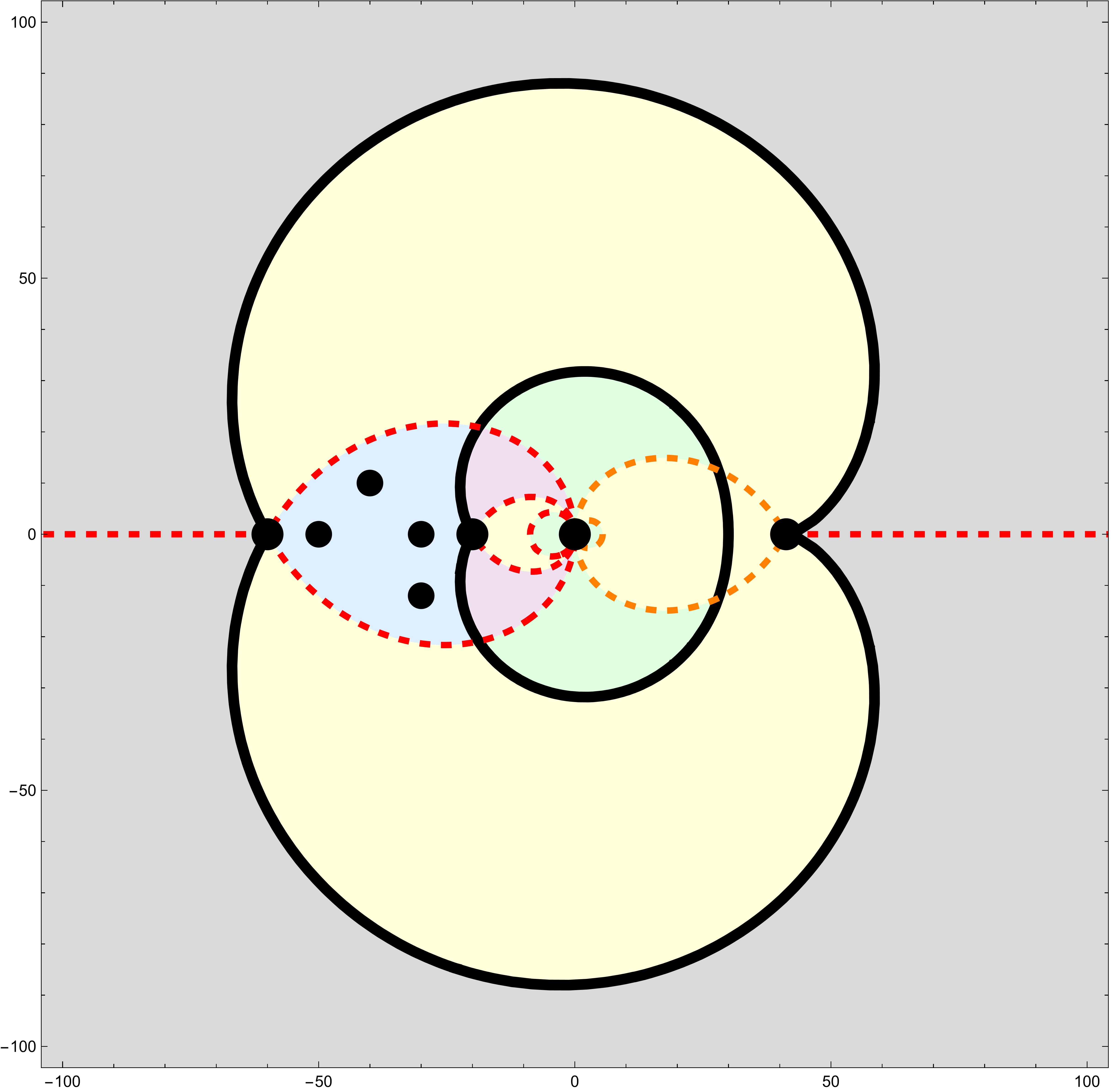}};

\node[anchor=center] at (6.1,2) {\includegraphics[scale = 0.22]{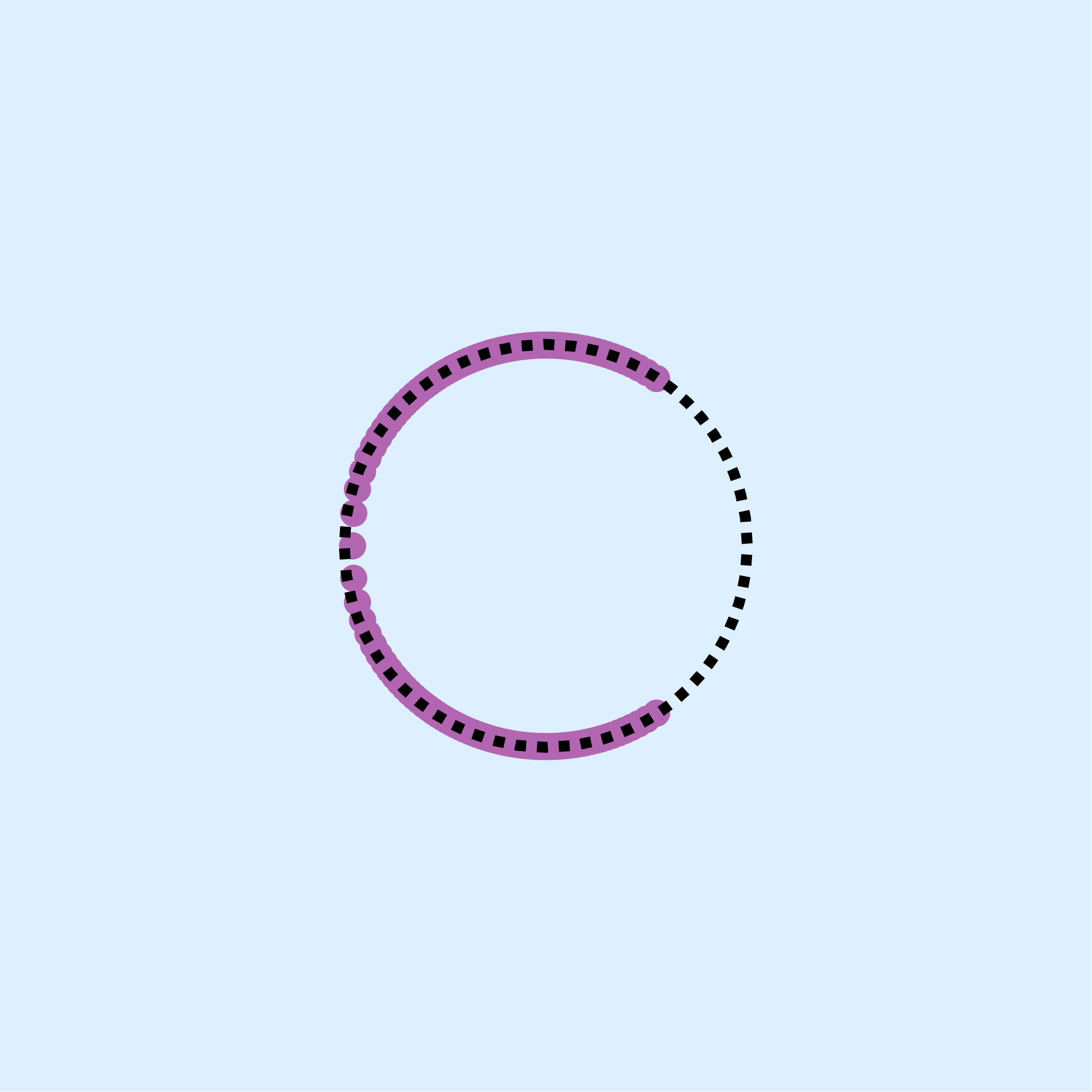}};

\node[anchor=center] at (6.1,-2) {\includegraphics[scale = 0.22]{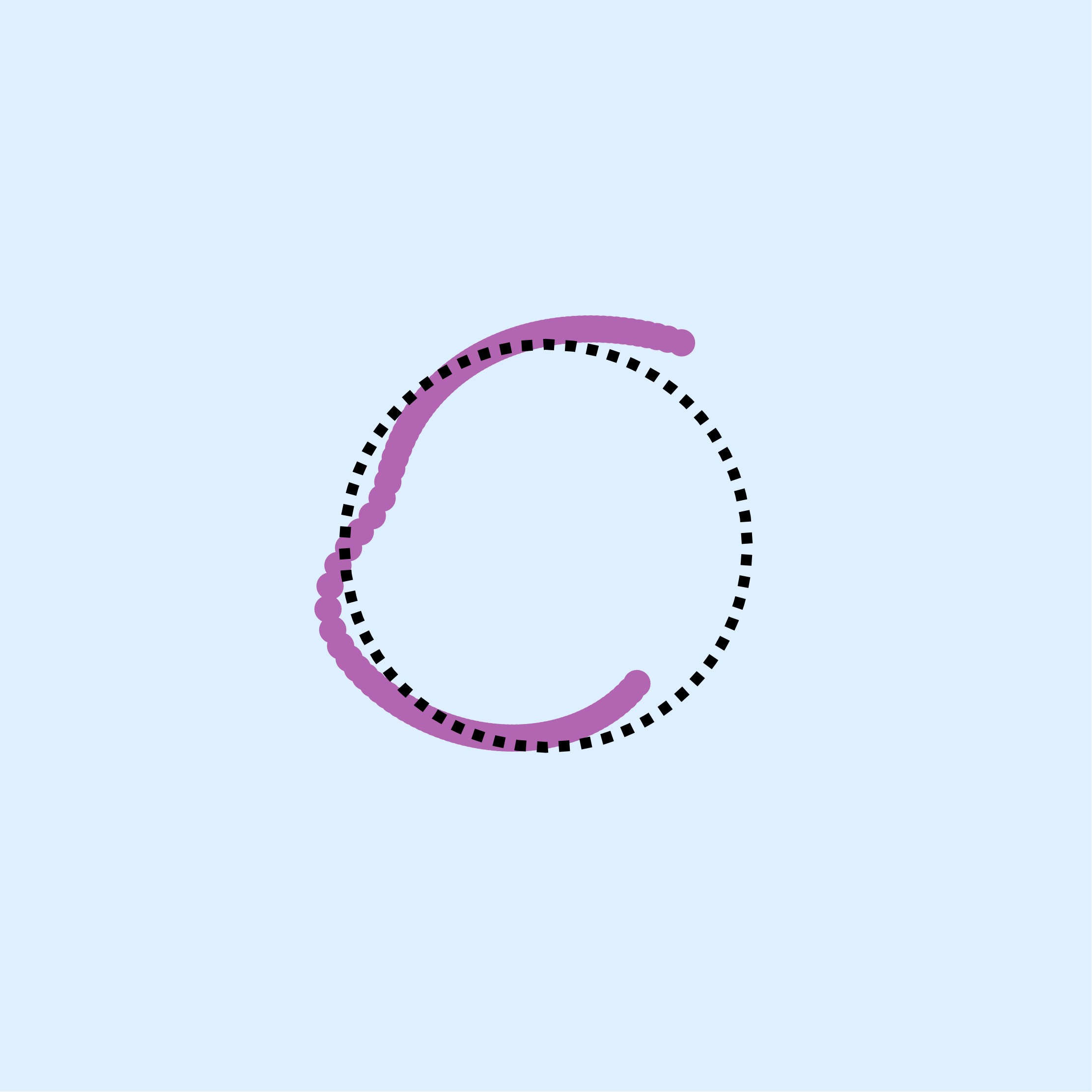}};

\node[anchor=center] at (-6.1,2) {\includegraphics[scale = 0.22]{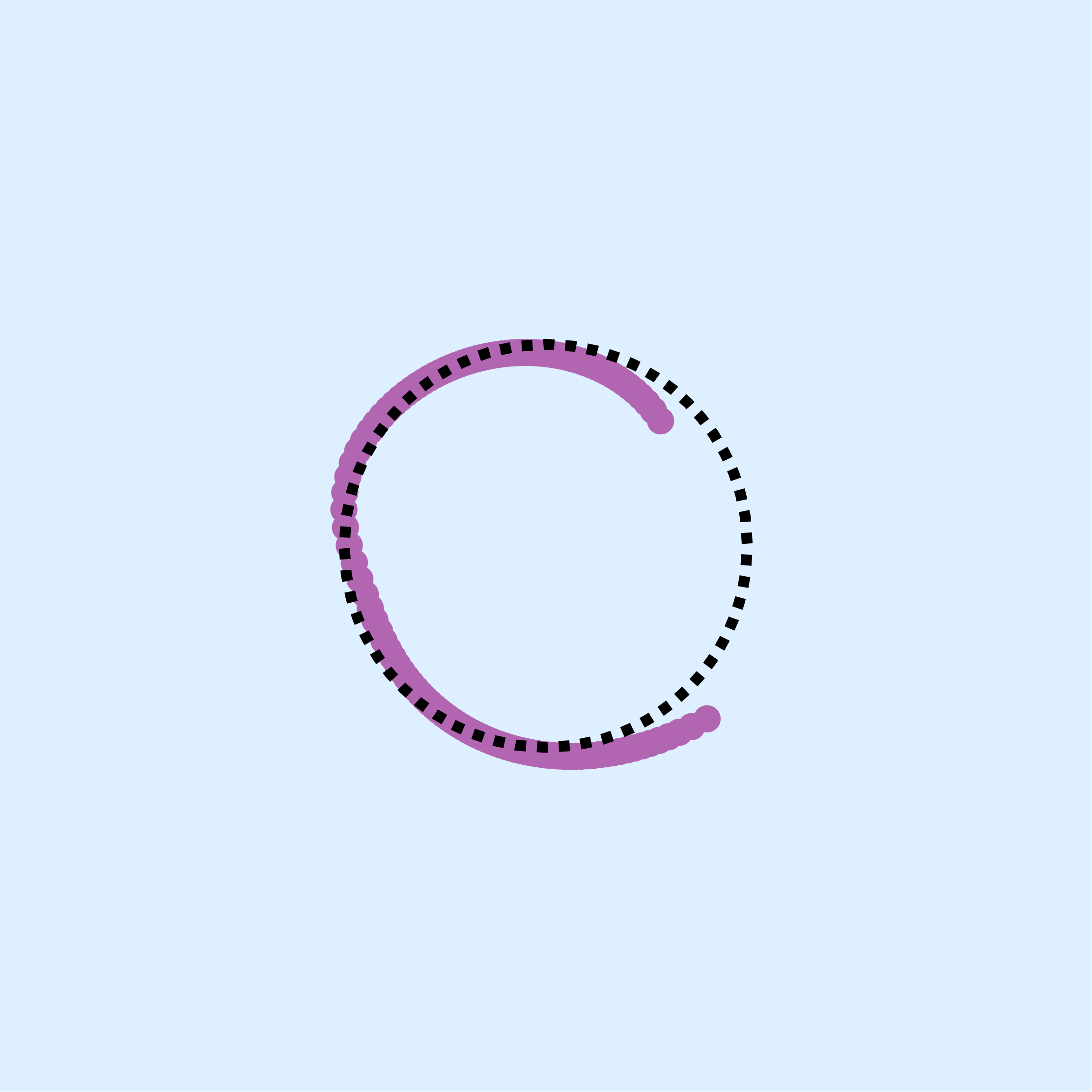}};

\node[anchor=center] at (-6.1,-2) {\includegraphics[scale = 0.22]{plots/CMM-Spectraldensityplot1-One-cut.png}};

\draw[line width = 2pt] (4.545,0.45) rectangle (4.545+3.1,3.55);

\draw[line width = 2pt] (4.545,-0.45) rectangle (4.545+3.1,-3.55);

\draw[line width = 2pt] (-4.545,0.45) rectangle (-4.545-3.1,3.55);

\draw[line width = 2pt] (-4.545,-0.45) rectangle (-4.545-3.1,-3.55);

\coordinate (Sample1) at(-0.89,-0.325);
 \coordinate (Sample2) at  (-0.89,0.07);
 \coordinate (Sample3) at (-1.21,0.405);
 \coordinate (Sample4) at (-1.57,0.07);

 \draw[line width = 2pt] (Sample1) to[out = -80, in = 180] (4.545,-0.45);
  \draw[line width = 2pt] (Sample1) to[out = -80, in = 180] (4.545,-3.55);

  \draw[line width = 2pt] (Sample2) to[out = 90, in = 180] (4.545,0.45);
  \draw[line width = 2pt] (Sample2) to[out = 90, in = 180] (4.545,3.55);

  \draw[line width = 2pt] (Sample3) to[out = 180-45, in = 0] (-4.545,0.45);
  \draw[line width = 2pt] (Sample3) to[out = 180-45, in = 0] (-4.545,3.55);

  \draw[line width = 2pt] (Sample4) to[out = -90, in = 0] (-4.545,-0.45);
  \draw[line width = 2pt] (Sample4) to[out = -90, in = 0] (-4.545,-3.55);
        
    \end{tikzpicture}
    \caption{Root plots of the orthogonal polynomial $p_n$ for $n = N = 125$, evaluated at four values of $g$ sampled from the light blue region of figure \ref{fig:CMMphasediagram}. Each root plot is displayed alongside the phase diagram, with the corresponding sample indicated by a black dot and connected to its root plot. Within each panel, the roots are shown as purple dots, and the unit circle is indicated by a black dashed line. The sample values are $g = -50,-30,-40+10i,-30-12i$.}
    \label{fig:CMMonecutsamples}
\end{figure}

\begin{figure}
    \centering
    \begin{tikzpicture}
        \node at (0,0) {\includegraphics[scale = 0.30]{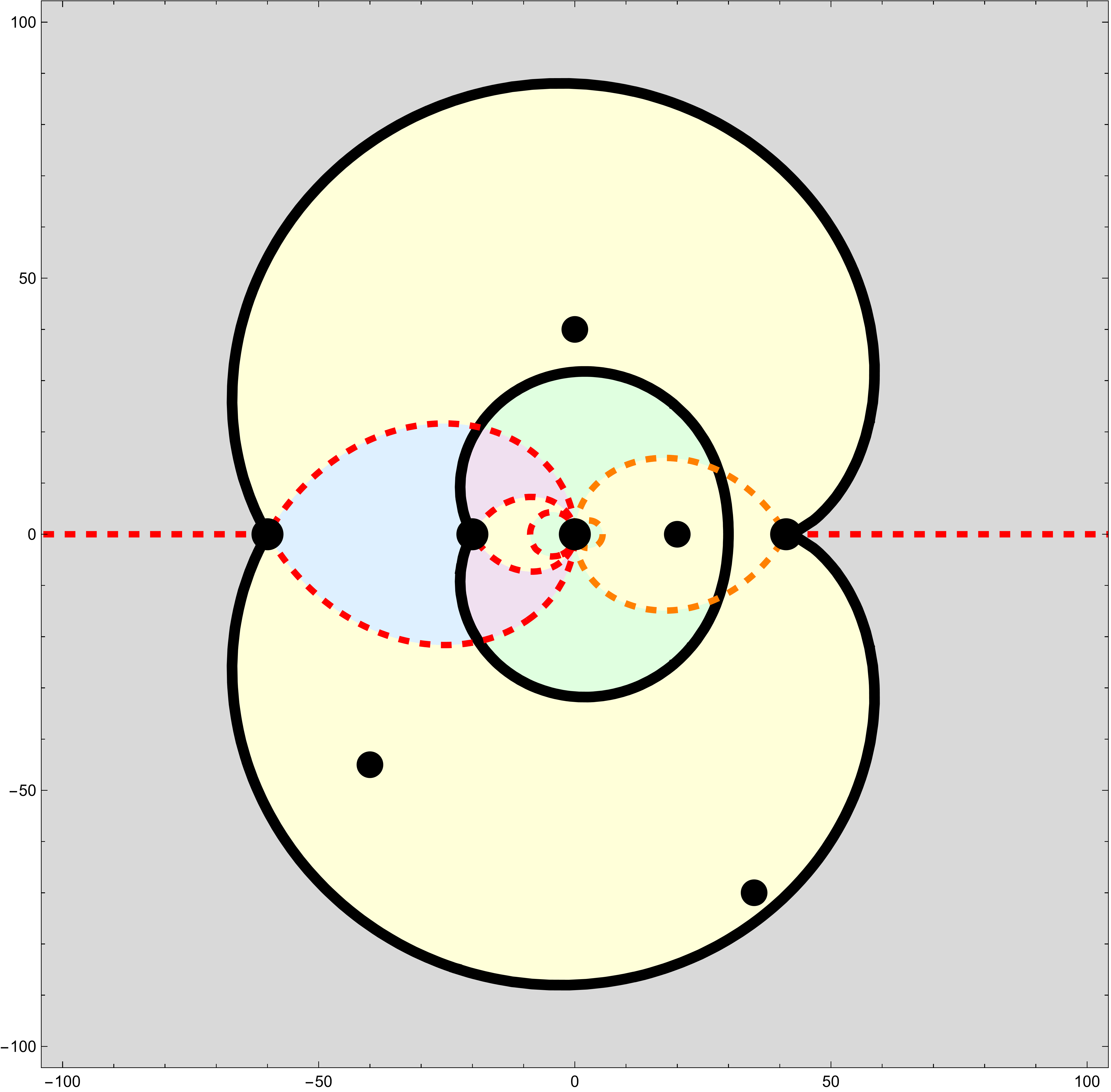}};

\node[anchor=center] at (6.1,2) {\includegraphics[scale = 0.22]{plots/CMM-Spectraldensityplot3-Two-cut.png}};

\node[anchor=center] at (6.1,-2) {\includegraphics[scale = 0.22]{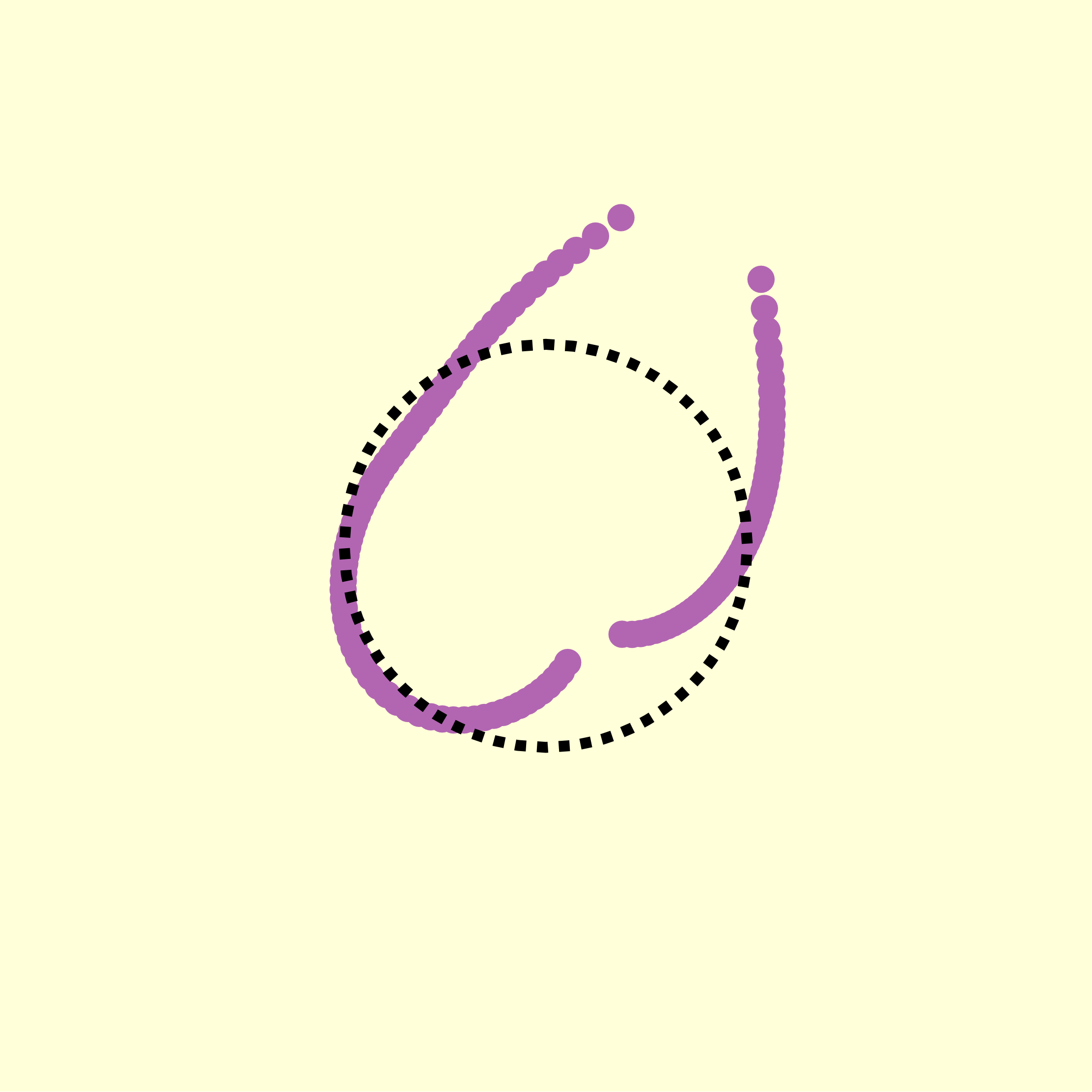}};

\node[anchor=center] at (-6.1,2) {\includegraphics[scale = 0.22]{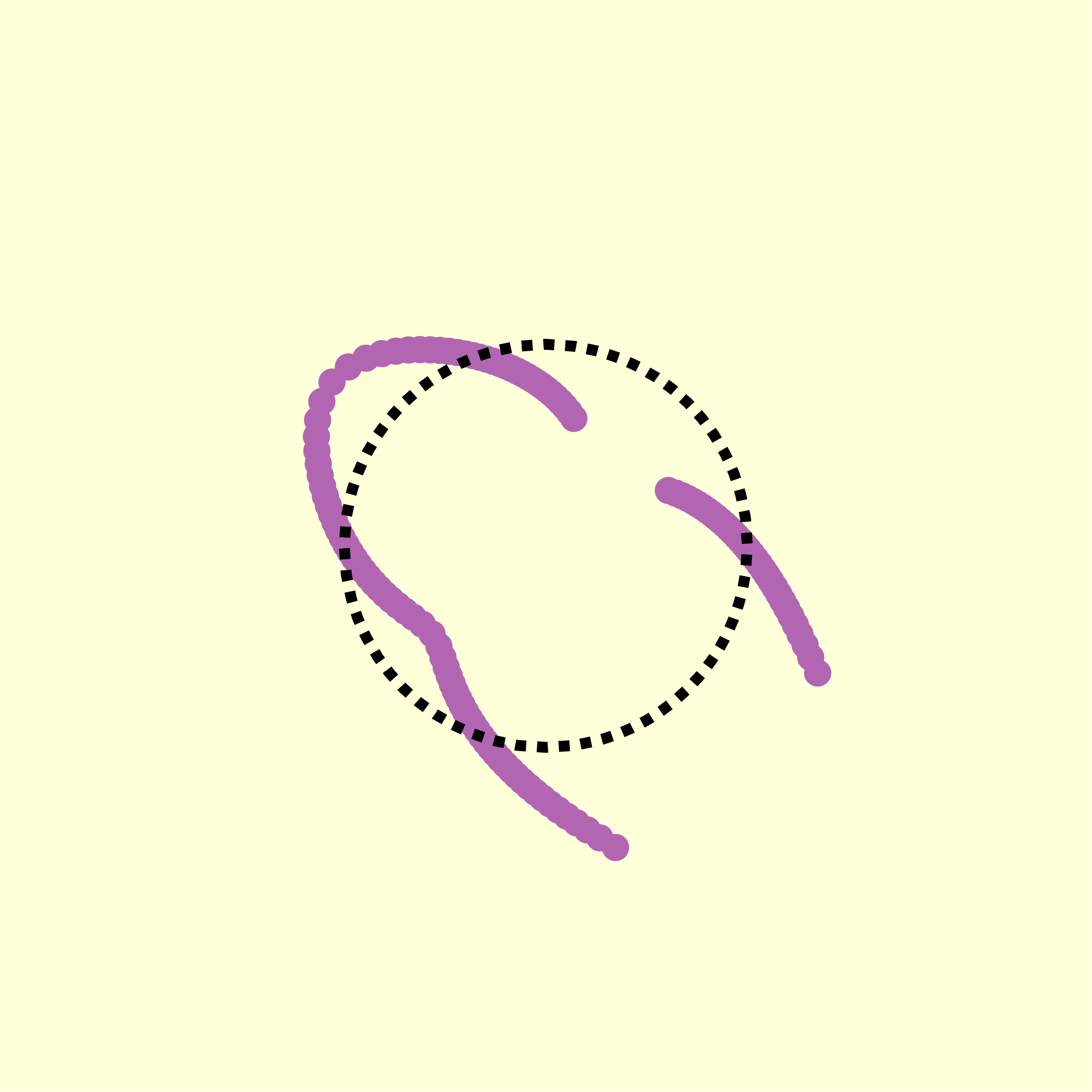}};

\node[anchor=center] at (-6.1,-2) {\includegraphics[scale = 0.22]{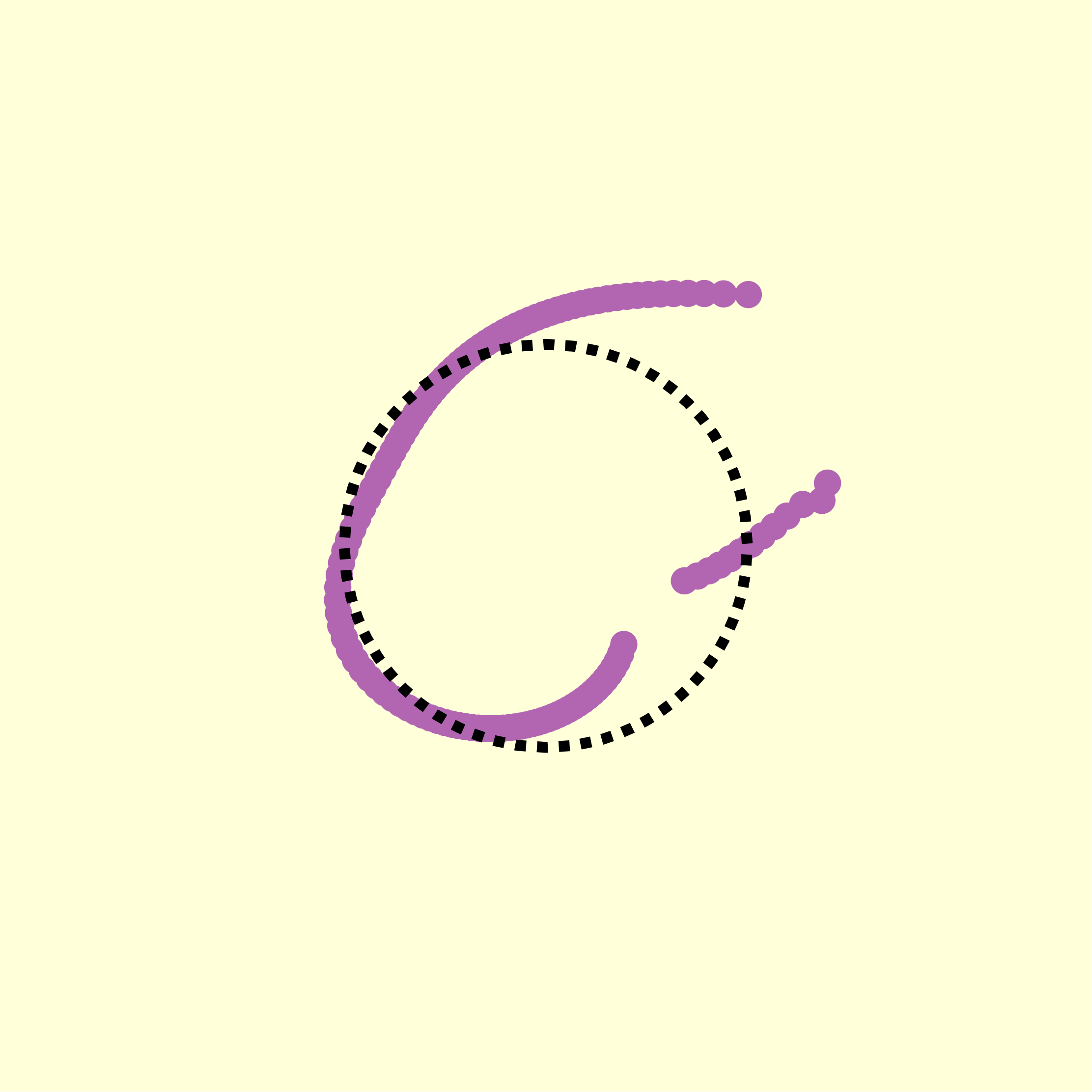}};

\draw[line width = 2pt] (4.545,0.45) rectangle (4.545+3.1,3.55);

\draw[line width = 2pt] (4.545,-0.45) rectangle (4.545+3.1,-3.55);

\draw[line width = 2pt] (-4.545,0.45) rectangle (-4.545-3.1,3.55);

\draw[line width = 2pt] (-4.545,-0.45) rectangle (-4.545-3.1,-3.55);

\coordinate (Sample1) at(1.35,-2.31);
 \coordinate (Sample2) at  (0.8,0.07);
 \coordinate (Sample3) at (0.1,1.45);
 \coordinate (Sample4) at (-1.30,-1.47);

 \draw[line width = 2pt] (Sample1) to[out = 0, in = 180] (4.545,-0.45);
  \draw[line width = 2pt] (Sample1) to[out = 0, in = 180] (4.545,-3.55);

  \draw[line width = 2pt] (Sample2) to[out = 45, in = 180] (4.545,0.45);
  \draw[line width = 2pt] (Sample2) to[out = 45, in = 180] (4.545,3.55);

  \draw[line width = 2pt] (Sample3) to[out = 180, in = 0] (-4.545,0.45);
  \draw[line width = 2pt] (Sample3) to[out = 180, in = 0] (-4.545,3.55);

  \draw[line width = 2pt] (Sample4) to[out = 180, in = 0] (-4.545,-0.45);
  \draw[line width = 2pt] (Sample4) to[out = 180, in = 0] (-4.545,-3.55);
        
    \end{tikzpicture}
    \caption{Root plots of the orthogonal polynomial $p_n$ for $n = N = 125$, evaluated at four values of $g$ sampled from the light yellow region of figure \ref{fig:CMMphasediagram}. Each root plot is displayed alongside the phase diagram, with the corresponding sample indicated by a black dot and connected to its root plot. Within each panel, the roots are shown as purple dots, and the unit circle is indicated by a black dashed line. The sample values are $g = 40i,-40-45i,20,35-70i$.}
    \label{fig:CMMtwocutsamples}
\end{figure}

\begin{figure}
    \centering
    \begin{tikzpicture}
        \node at (0,0) {\includegraphics[scale = 0.30]{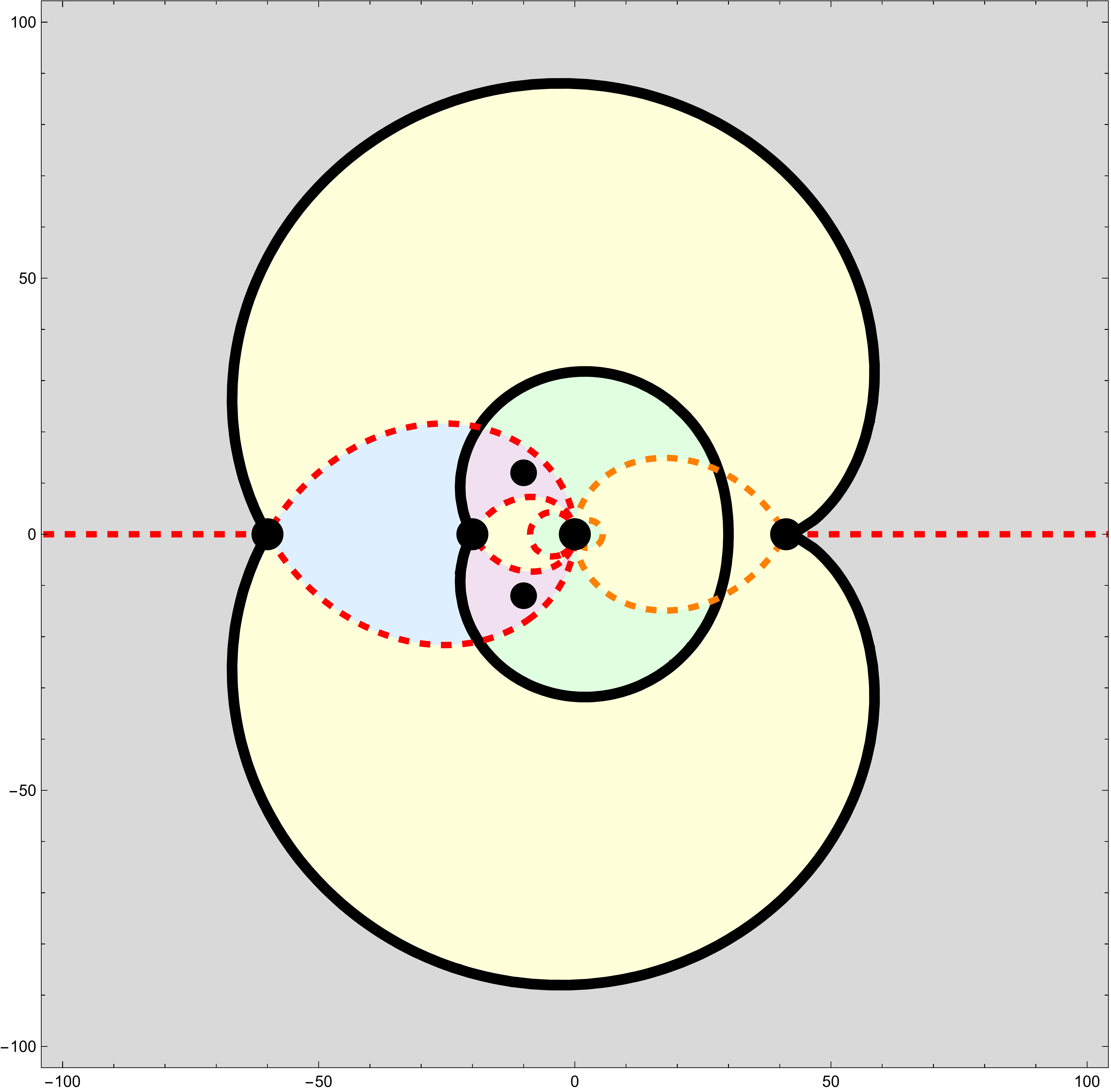}};

\node[anchor=center] at (6.1,2) {\includegraphics[scale = 0.22]{plots/CMM-Spectraldensityplot1-Three-cut.png}};



\node[anchor=center] at (-6.1,-2) {\includegraphics[scale = 0.22]{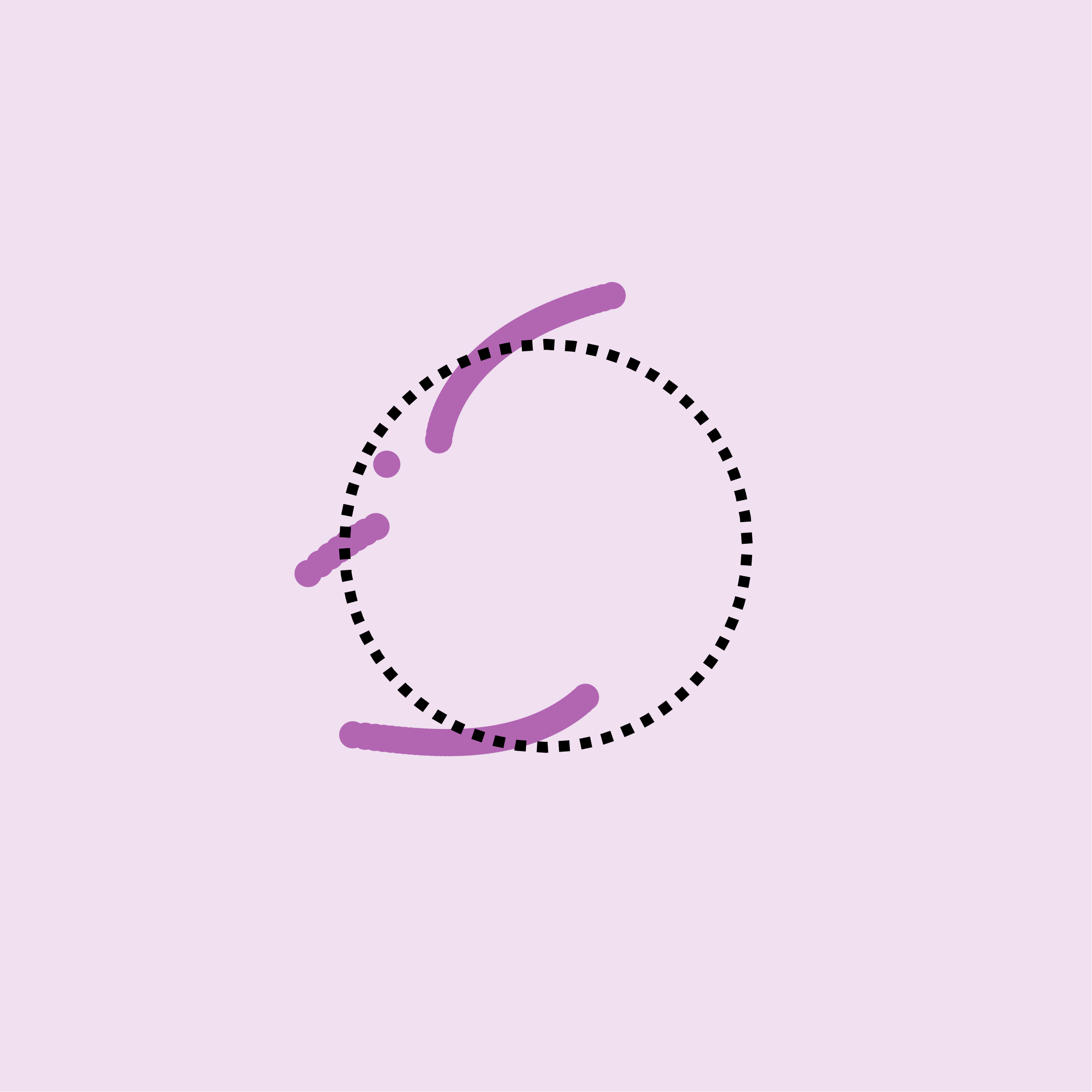}};

\draw[line width = 2pt] (4.545,0.45) rectangle (4.545+3.1,3.55);



\draw[line width = 2pt] (-4.545,-0.45) rectangle (-4.545-3.1,-3.55);

 \coordinate (Sample2) at  (-0.245,0.475);
 \coordinate (Sample4) at (-0.245,-0.475+0.14);


  \draw[line width = 2pt] (Sample2) to[out = 45, in = 180] (4.545,0.45);
  \draw[line width = 2pt] (Sample2) to[out = 45, in = 180] (4.545,3.55);


  \draw[line width = 2pt] (Sample4) to[out = 180+45, in = 0] (-4.545,-0.45);
  \draw[line width = 2pt] (Sample4) to[out = 180+45, in = 0] (-4.545,-3.55);
        
    \end{tikzpicture}
    \caption{Root plots of the orthogonal polynomial $p_n$ for $n = N = 125$, evaluated at two values of $g$ sampled from the light purple region of figure \ref{fig:CMMphasediagram}. Each root plot is displayed alongside the phase diagram, with the corresponding sample indicated by a black dot and connected to its root plot. Within each panel, the roots are shown as purple dots, and the unit circle is indicated by a black dashed line. The sample values are $g = -10\pm 12i$.}
    \label{fig:CMMthreecutsamples}
\end{figure}

\begin{figure}
    \centering
    \begin{tikzpicture}
        \node at (0,0) {\includegraphics[scale = 0.30]{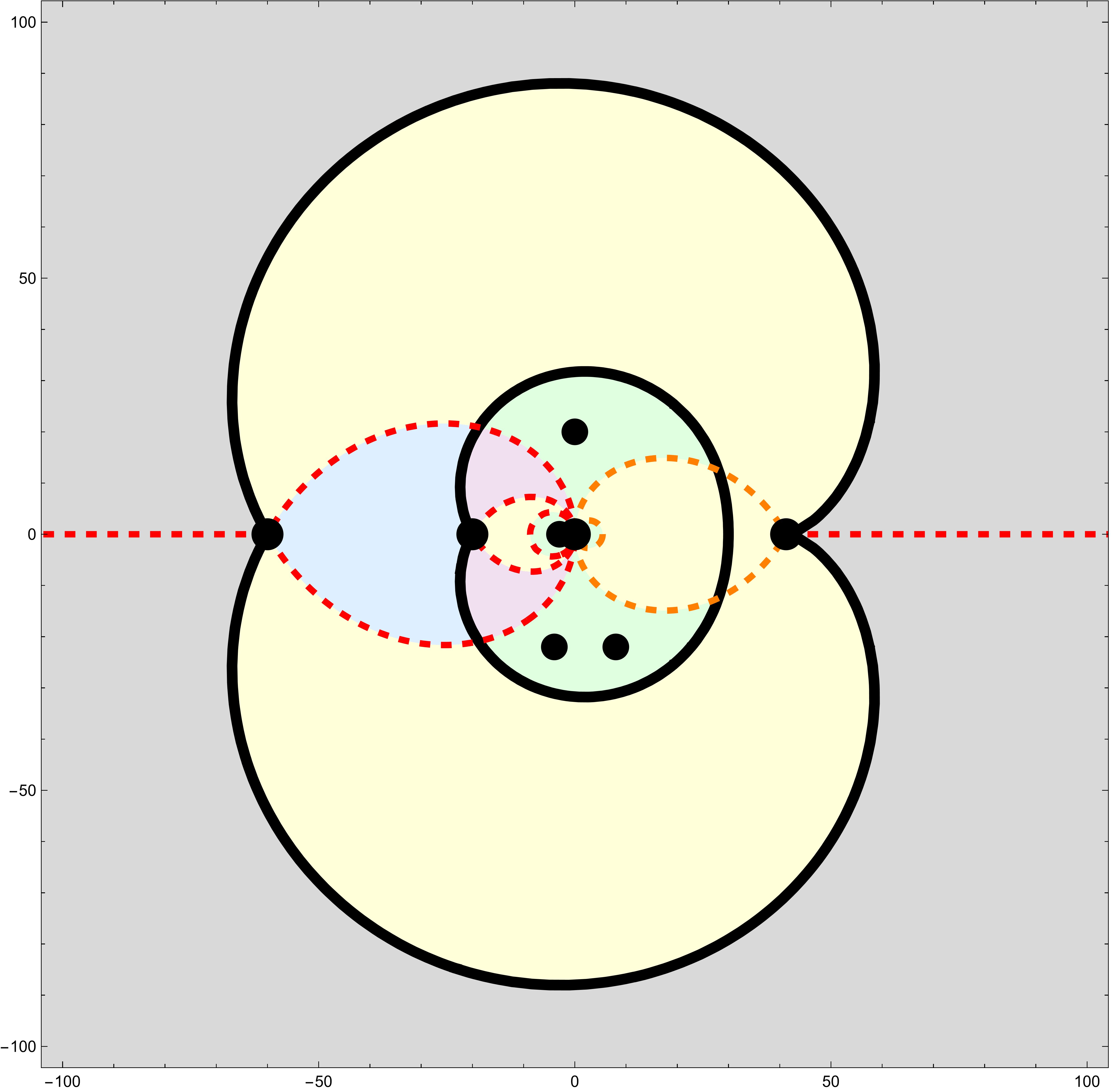}};

\node[anchor=center] at (6.1,2) {\includegraphics[scale = 0.22]{plots/CMM-Spectraldensityplot1-Four-cut.png}};

\node[anchor=center] at (6.1,-2) {\includegraphics[scale = 0.22]{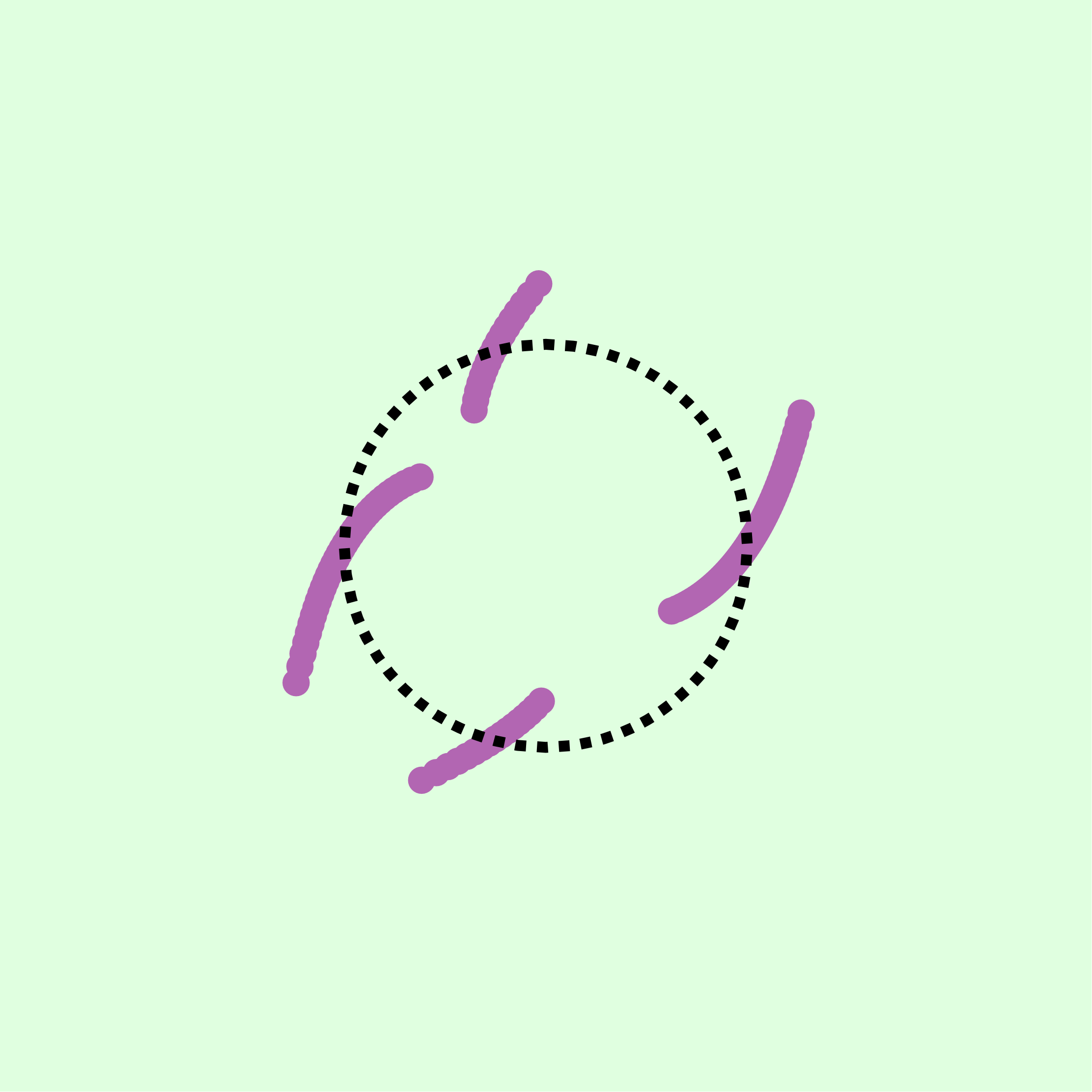}};

\node[anchor=center] at (-6.1,2) {\includegraphics[scale = 0.22]{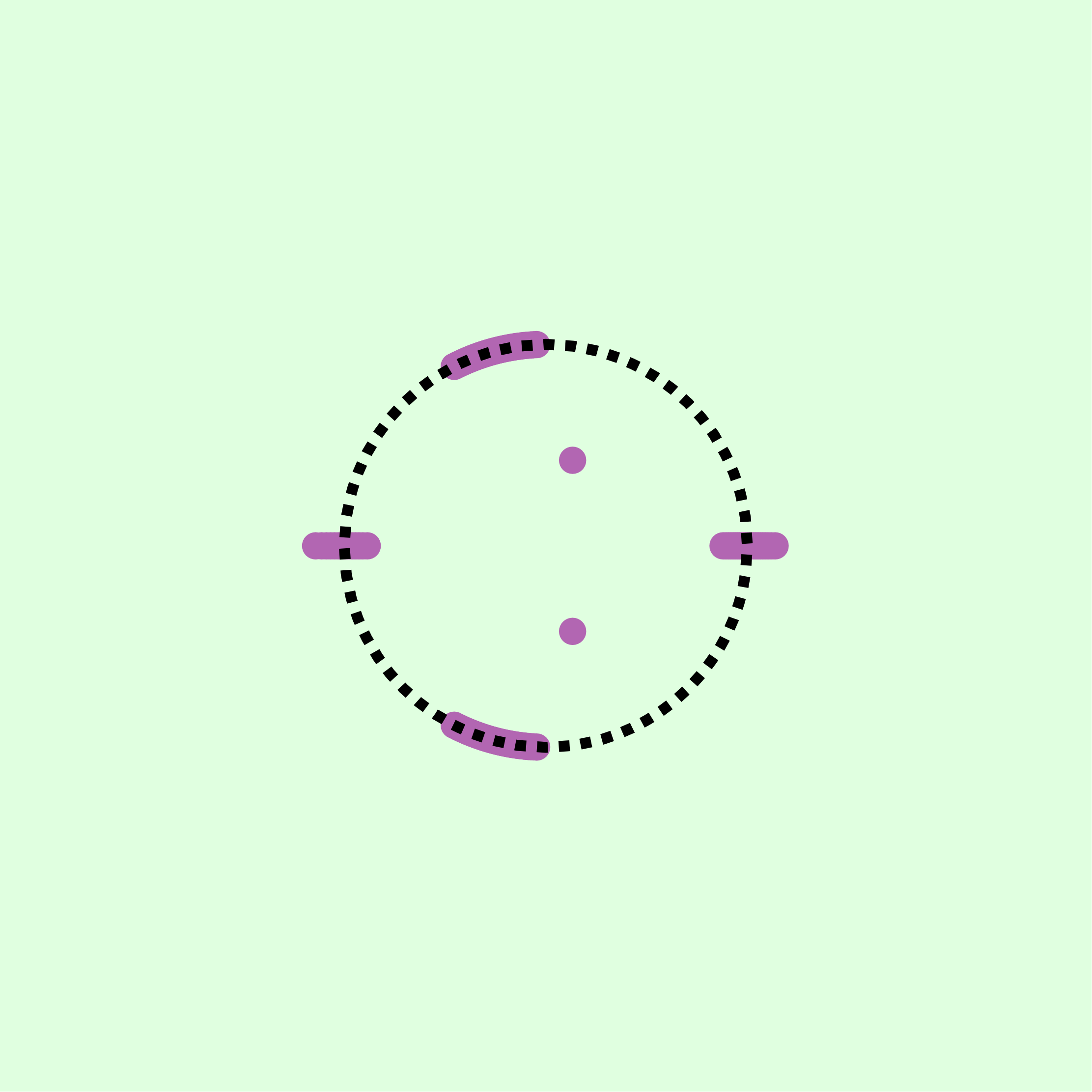}};

\node[anchor=center] at (-6.1,-2) {\includegraphics[scale = 0.22]{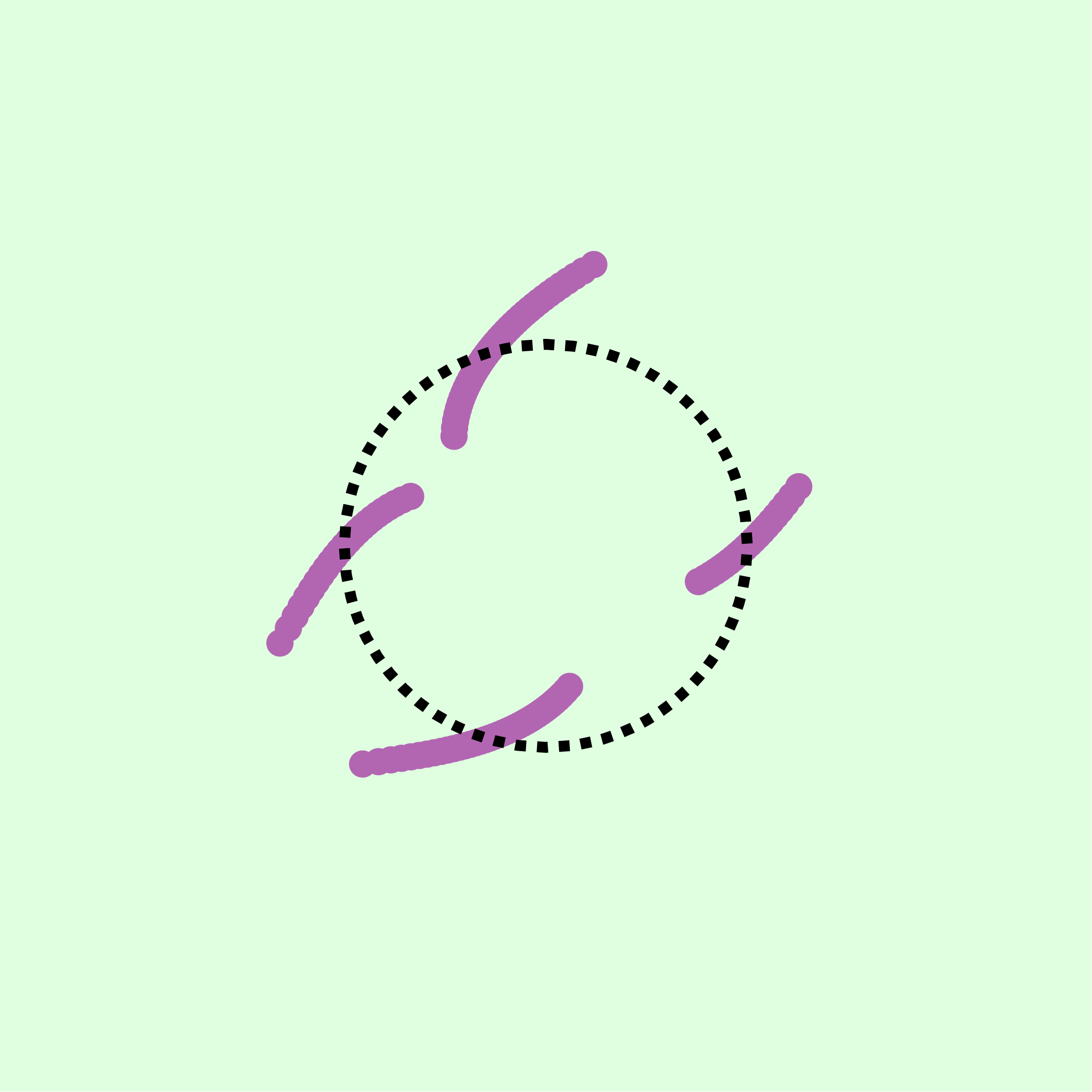}};

\draw[line width = 2pt] (4.545,0.45) rectangle (4.545+3.1,3.55);

\draw[line width = 2pt] (4.545,-0.45) rectangle (4.545+3.1,-3.55);

\draw[line width = 2pt] (-4.545,0.45) rectangle (-4.545-3.1,3.55);

\draw[line width = 2pt] (-4.545,-0.45) rectangle (-4.545-3.1,-3.55);

\coordinate (Sample1) at(0.41,-0.73);
 \coordinate (Sample2) at  (0.12,0.73);
 \coordinate (Sample3) at (0+0.02,0+0.07);
 \coordinate (Sample4) at (0,-0.73);

 \draw[line width = 2pt] (Sample1) to[out = -80, in = 180] (4.545,-0.45);
  \draw[line width = 2pt] (Sample1) to[out = -80, in = 180] (4.545,-3.55);

  \draw[line width = 2pt] (Sample2) to[out = 90, in = 180] (4.545,0.45);
  \draw[line width = 2pt] (Sample2) to[out = 90, in = 180] (4.545,3.55);

  \draw[line width = 2pt] (Sample3) to[out = 180-45, in = 0] (-4.545,0.45);
  \draw[line width = 2pt] (Sample3) to[out = 180-45, in = 0] (-4.545,3.55);

  \draw[line width = 2pt] (Sample4) to[out = -90, in = 0] (-4.545,-0.45);
  \draw[line width = 2pt] (Sample4) to[out = -90, in = 0] (-4.545,-3.55);
        
    \end{tikzpicture}
    \caption{Root plots of the orthogonal polynomial $p_n$ for $n = N = 125$, evaluated at four values of $g$ sampled from the light green region of figure \ref{fig:CMMphasediagram}. Each root plot is displayed alongside the phase diagram, with the corresponding sample indicated by a black dot and connected to its root plot. Within each panel, the roots are shown as purple dots, and the unit circle is indicated by a black dashed line. The sample values are $g = 20i,-4-22i,8-22i,-3$.}
    \label{fig:CMMfourcutsamples}
\end{figure}

\begin{figure}
    \centering
    \begin{tikzpicture}
        \node at (0,0) {\includegraphics[scale = 0.30]{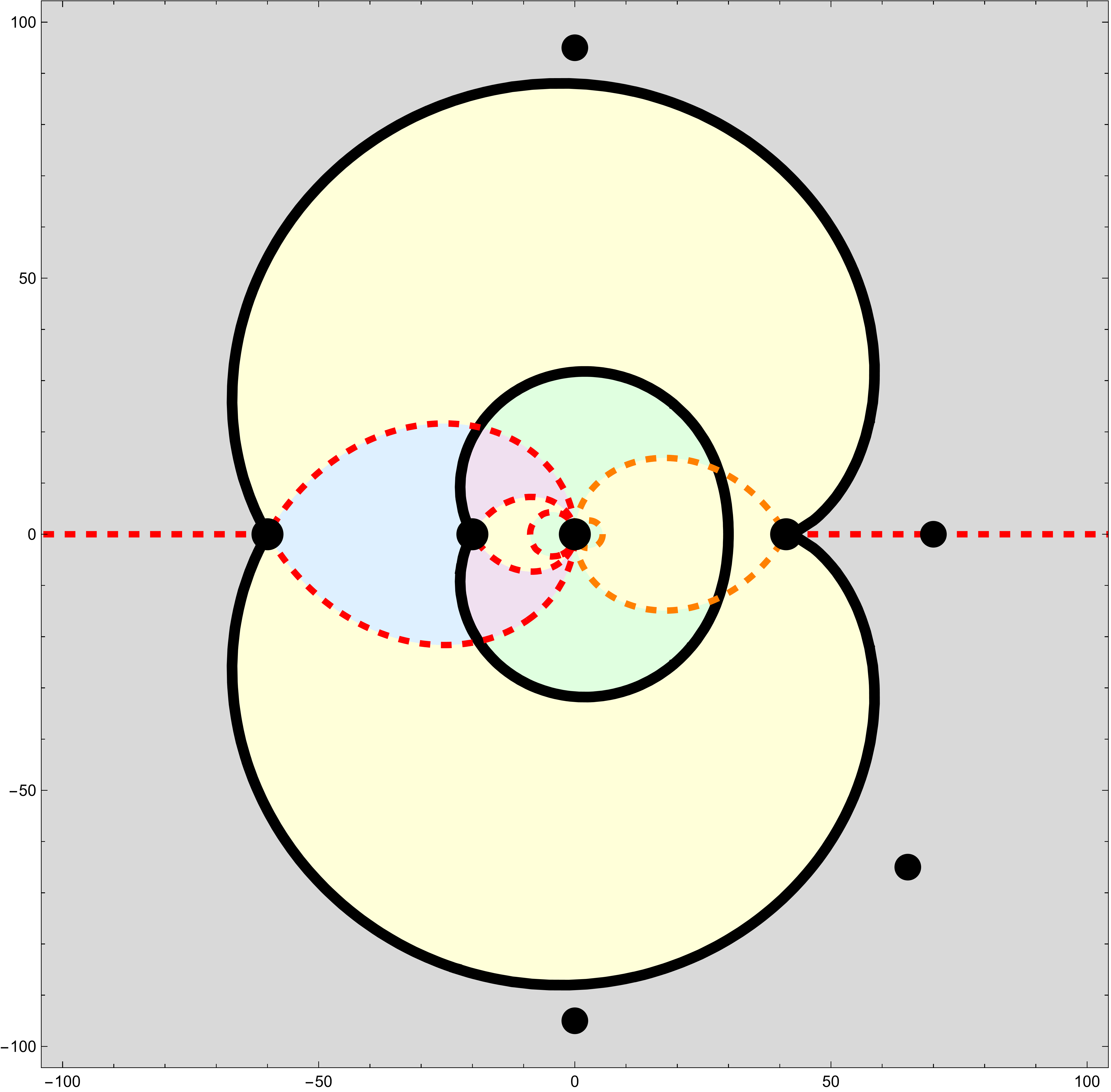}};

\node[anchor=center] at (6.1,2) {\includegraphics[scale = 0.22]{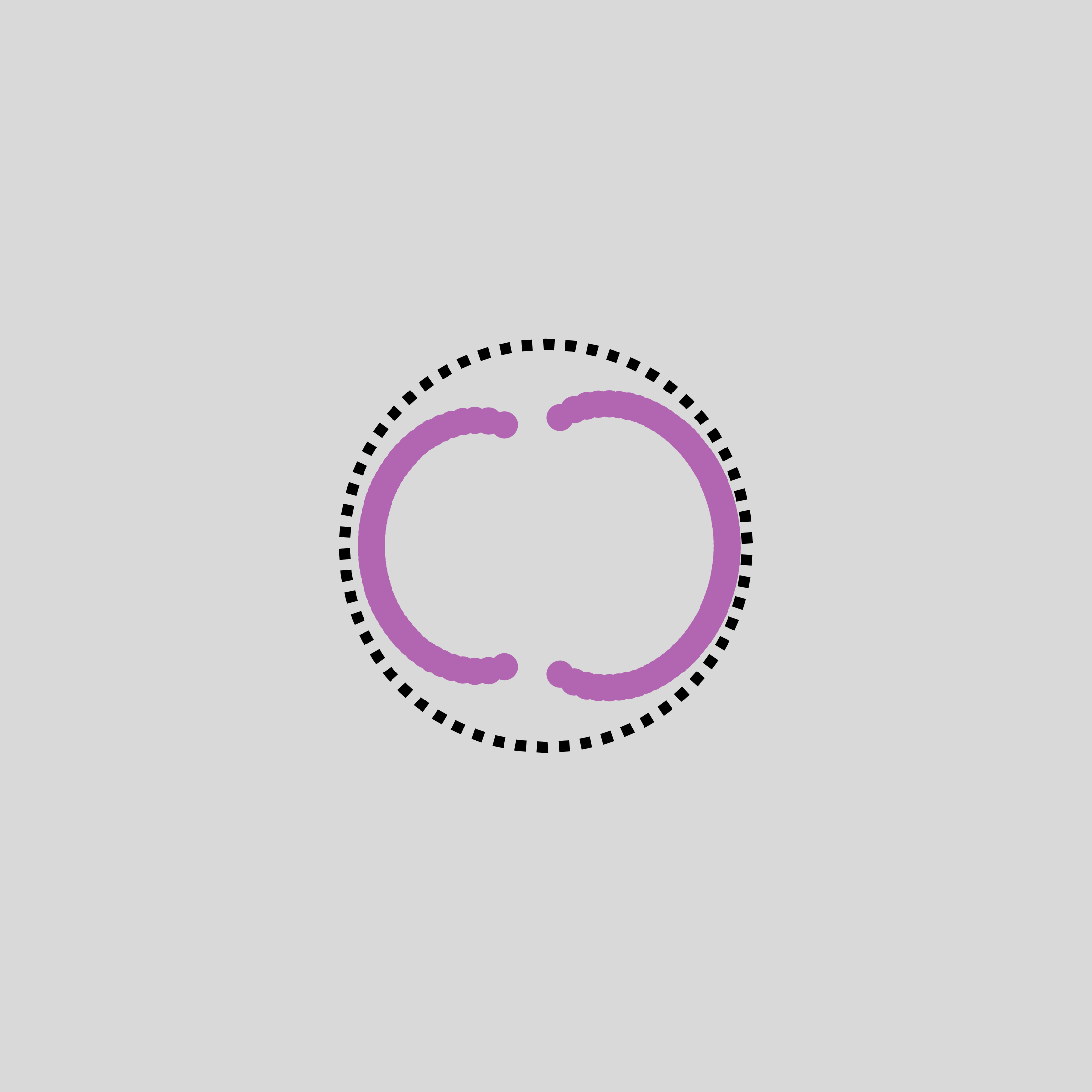}};

\node[anchor=center] at (6.1,-2) {\includegraphics[scale = 0.22]{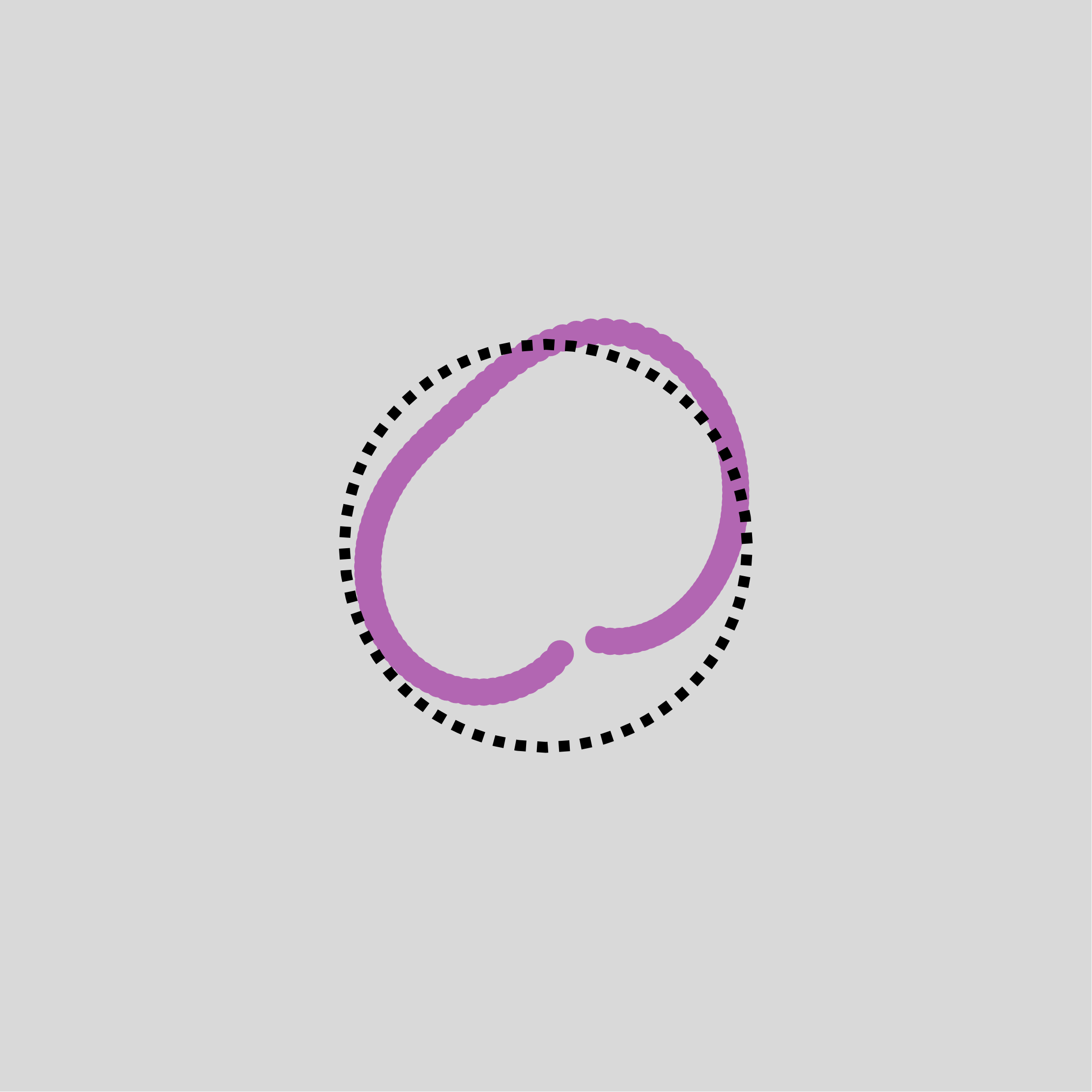}};

\node[anchor=center] at (-6.1,2) {\includegraphics[scale = 0.22]{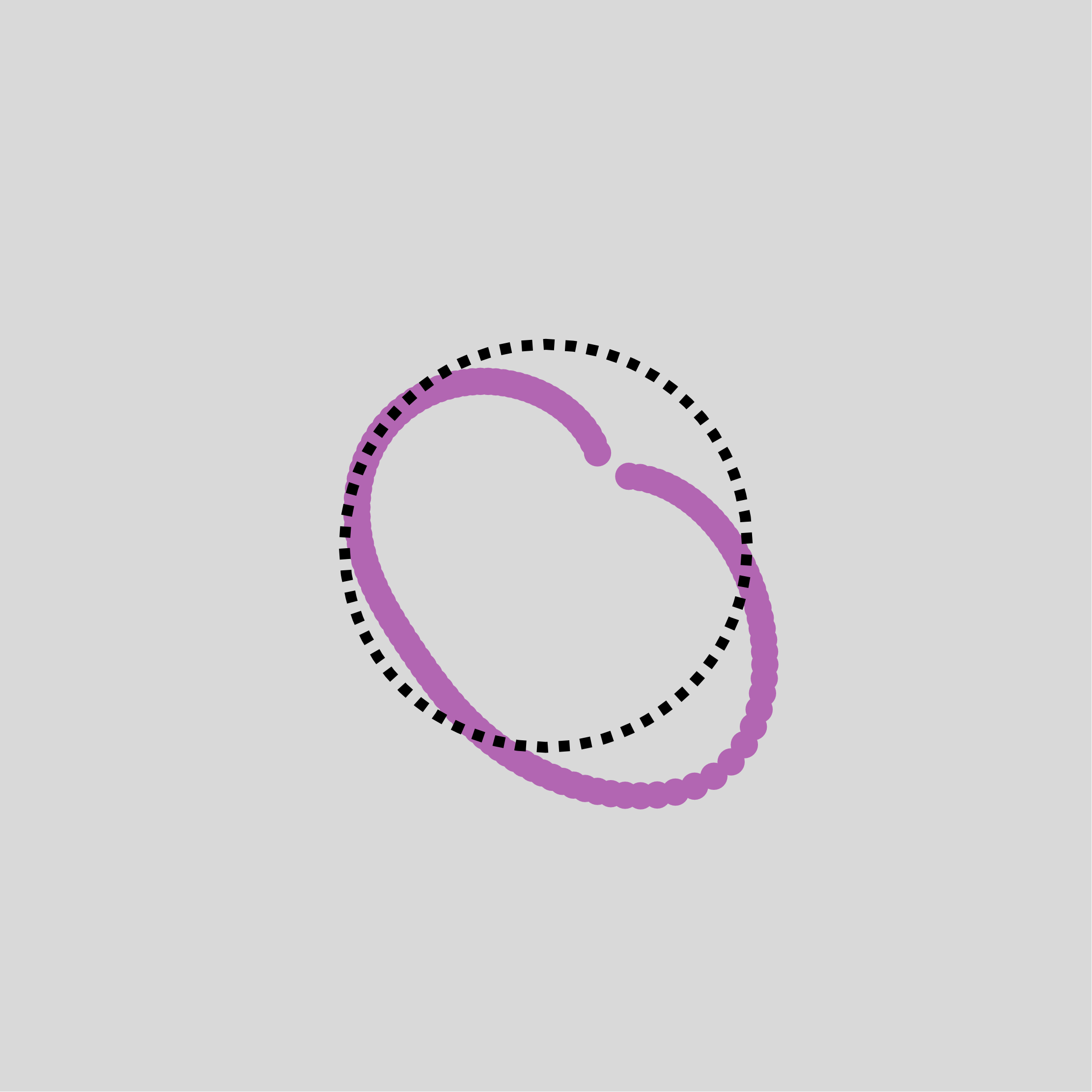}};

\node[anchor=center] at (-6.1,-2) {\includegraphics[scale = 0.22]{plots/CMM-Spectraldensityplot2-Ungapped.png}};

\draw[line width = 2pt] (4.545,0.45) rectangle (4.545+3.1,3.55);

\draw[line width = 2pt] (4.545,-0.45) rectangle (4.545+3.1,-3.55);

\draw[line width = 2pt] (-4.545,0.45) rectangle (-4.545-3.1,3.55);

\draw[line width = 2pt] (-4.545,-0.45) rectangle (-4.545-3.1,-3.55);

\coordinate (Sample1) at(2.35,-2.14);
 \coordinate (Sample2) at  (2.52,0.07);
 \coordinate (Sample3) at (0.13,3.33);
 \coordinate (Sample4) at (0.13,-3.33+0.16);

 \draw[line width = 2pt] (Sample1) to[out = 0, in = 180] (4.545,-0.45);
  \draw[line width = 2pt] (Sample1) to[out = 0, in = 180] (4.545,-3.55);

  \draw[line width = 2pt] (Sample2) to[out = 90, in = 180] (4.545,0.45);
  \draw[line width = 2pt] (Sample2) to[out = 90, in = 180] (4.545,3.55);

  \draw[line width = 2pt] (Sample3) to[out = 180, in = 0] (-4.545,0.45);
  \draw[line width = 2pt] (Sample3) to[out = 180, in = 0] (-4.545,3.55);

  \draw[line width = 2pt] (Sample4) to[out = 180, in = 0] (-4.545,-0.45);
  \draw[line width = 2pt] (Sample4) to[out = 180, in = 0] (-4.545,-3.55);
        
    \end{tikzpicture}
    \caption{Root plots of the orthogonal polynomial $p_n$ for $n = N = 125$, evaluated at four values of $g$ sampled from the light gray region of figure \ref{fig:CMMphasediagram}. Each root plot is displayed alongside the phase diagram, with the corresponding sample indicated by a black dot and connected to its root plot. Within each panel, the roots are shown as purple dots, and the unit circle is indicated by a black dashed line. The sample values are $g = \pm 95i$ and $g = 70,65-65i$.}
    \label{fig:CMMungappedsamples}
\end{figure}

In figures \ref{fig:CMMonecutsamples}, \ref{fig:CMMtwocutsamples}, \ref{fig:CMMthreecutsamples}, and \ref{fig:CMMfourcutsamples}, we plot the root distributions of $p_n$ for representative values of $g$ sampled from the regions of figure \ref{fig:CMMphasediagram} shaded in light blue, yellow, purple, and green. 
These regions are consistent with one-cut, two-cut, three-cut, and four-cut phases, respectively. 
Finally, in figure \ref{fig:CMMungappedsamples}, we display the root distributions of $p_n$ for representative values of $g$ sampled from the light gray region of figure \ref{fig:CMMphasediagram}, which corresponds to the ungapped phase.

The root distributions of the orthogonal polynomials appear to successfully capture the topology changes of the eigenvalue spectral density across the different phase approximations shown in figure \ref{fig:CMMphasediagram}. As in our numerical exploration of the GWW model phase diagram in subsection \ref{subsec:GWW}, this provides supporting evidence that the asymptotic bootstrap estimate method used to compute the orthogonal polynomials performs reliably for complex measures.

Unlike the GWW case, however, we do not have access to exact analytical predictions for the locations of the phase boundaries, making this a comparatively weaker cross-check of the method. Nevertheless, assuming the validity of the asymptotic bootstrap estimates, this analysis illustrates a powerful practical application: it enables the exploration of the phase diagram of unitary matrix models, whose phase boundaries and phase structure are otherwise inaccessible by analytical means, with relatively modest computational effort.

\section{Conclusion}
\label{sec:conclusion}

In this paper, we explored two complementary applications of the asymptotic bootstrap estimate method in the context of unitary matrix models. 

First, we applied the method to efficiently compute Wilson loop expectation values in the Gross-Witten-Wadia model \eqref{eq:GWW} and in the more complicated model \eqref{eq:potential}. To validate the numerical results, we compared them with analytical instanton predictions available in the literature \cite{emt24,cmt24}, which we independently motivated using standard Lefschetz-thimble arguments. We found excellent agreement between the analytical instanton expansions and the numerical bootstrap estimates, confirming that the asymptotic bootstrap estimate provides a highly efficient and accurate tool for computing non-perturbative observables. 

Second, we employed the asymptotic bootstrap estimate to efficiently compute orthogonal polynomials and their root distributions, which are known to accumulate along the planar eigenvalue spectral density of the matrix model and thus provide a natural probe of the phase structure. 
Guided by instanton actions, we identified several distinct phases, exactly for the GWW model \eqref{eq:GWW} and approximately for the model \eqref{eq:promotedpotential}, which were confirmed by the numerical results showing topologically different root accumulation sets. In particular, for the GWW model, we observed two one-cut phases, the ungapped phase and a two-cut phase, which is accessible only for complex ’t Hooft coupling values. These results agree well with previous ones obtained in the literature \cite{m08,cgkt22}. For the model \eqref{eq:promotedpotential}, we observed phases featuring between one and four cuts as well as the ungapped phase.

Another surprising feature is that several phase boundaries seem to be well approximated by Stokes lines. While anti-Stokes lines are typically associated with phase transitions in matrix models, our results suggest that Stokes lines can also play an important role in organizing the phase structure of unitary matrix models with complex couplings. 
We believe that this phenomenon merits further investigation.

Our calculations were done for a class of unitary matrix models in which the potential is symmetric under reflection of the angle, namely $V(\theta)=V(-\theta)$, even when $V$ is complex. This symmetry implies that the Fourier moments of the measure satisfy $a_n=a_{-n}$, and the shoestring analysis relies on the asymptotic decay $a_n \to 0 $ for $n\gg 0$.
We have not studied the case in which the potential is not symmetric. In that situation, we will also need to impose $a_{-n}\to 0$ for $n\gg 0$ (since it will not be related to $a_n$ by symmetry), and we have not yet implemented that additional condition. 
This is not an issue for real-valued potentials, as then $a_{-n}= a_n^*$ are related by complex conjugation. 
Understanding how to consistently impose both asymptotic conditions in the most general case of a complex potential remains an interesting problem and would be necessary to treat the most general unitary matrix model.

Lastly, it would be interesting to explore whether the asymptotic bootstrap estimate method can be extended to double-scaled unitary and Hermitian matrix integrals arising as dual descriptions of low-dimensional string theories.

\section*{Acknowledgements}

We would like to thank Yiming Chen, Paula Garc\'\i a Mart\'\i nez,  David Gross, Clifford V. Johnson, Ricardo Schiappa, and Maximilian Schwick for useful discussions.
DB is supported in part by the Department of Energy under grant DE-SC 0011702.
JR is supported by the FCT-Portugal scholarship UI/BD/151499/2021 and by the CAMGSD scholarship BL197/2025-IST-ID. 
VAR is supported by the University of California President’s Postdoctoral Fellowship. This
work has been supported by Fundação para a Ciência e Tecnologia through the project 2024.04456.CERN. This paper is partly a result of the ERC-SyG project, Recursive and Exact New Quantum Theory (ReNewQuantum) funded by the European Research Council (ERC) under the European Union’s Horizon 2020 research and innovation programme, grant agreement 810573.

\appendix

\section{Programming the recursion and better estimates of the error}
\label{sec:numerics_rec}

Consider the general problem given by
\begin{equation}
\int \d\mu = {\cal N} \int \d\theta \exp( 2 c_1  \cos(\theta)+c_2 \cos(2\theta))
\end{equation}
where ${\cal N}= \exp(c_0)$ is a normalization constant.
The initial conditions for the recursion are given by $a_0=1= \int \d\mu$, $a_{-1}=a_1=\int \exp(i\theta)\d\mu $ and $a_2=\int \exp(2i\theta)\d\mu$ while all other Fourier coefficients are given by $a_{n}=\int \exp(i n \theta)\d\mu $.  The following integration
by parts identity is  a linear recursion relation between the $a_n$
\begin{equation}
0=\int \d\theta \, \partial_\theta(\exp(i n\theta) \exp(2 c_1 \cos(\theta)+c_2\cos(2\theta))).
\end{equation}

The recursion in this case is
\begin{equation}
    -c_2 a_{n-2}-c_1 a_{n-1}+ n a_n +c_1 a_{n+1}+c_2 a_{n+2}=0
\end{equation}
which shows that given the data $a_0=1,a_{-1}=a_1, a_2$, they are sufficient to generate all the $a_n$. 

It is convenient to write the recursion in matrix form as follows
\begin{equation}
\begin{pmatrix}
    a_{n-1}\\
    a_n\\
    a_{n+1}\\
    a_{n+2}    
\end{pmatrix}=\begin{pmatrix} 0&1&0&0\\
0&0&1&0\\
0&0&0&1\\
1 & c_1/c_2 & -n/c_2 &-c_1/c_2    
\end{pmatrix}\begin{pmatrix}
    a_{n-2}\\
    a_{n-1}\\
    a_n\\
    a_{n+1}  
\end{pmatrix}= M_{n} . \vec a_{n-1}
\end{equation}
where $\vec a_n$ is the last vector of entries produced by acting with $M_n$.
The vector $\vec a_0$ is given by
\begin{equation}
    \vec a_0= \begin{pmatrix}a_1\\1\\a_1\\a_2\end{pmatrix}
\end{equation}

\begin{figure}[ht]
\begin{center}
\includegraphics[width=8cm]{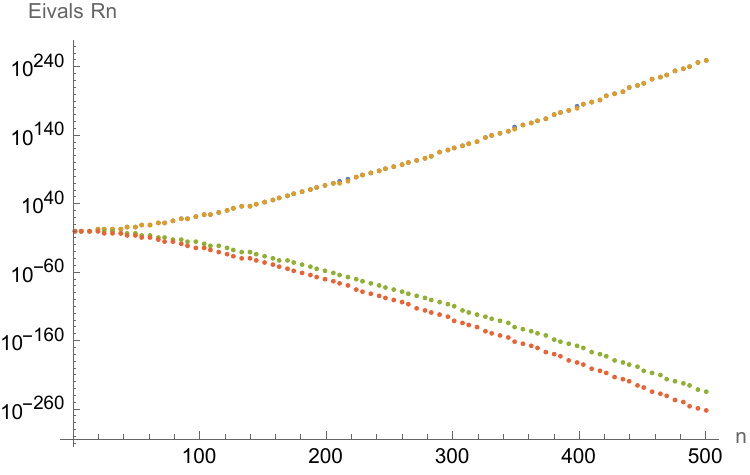}
\caption{Absolute value of the eigenvalues of the matrices $R_n$ for various values of $n$, using $c_1=10, c_2=20$. The two large eigenvalues are very close to each other and are almost indistinguishable in the graph. }
\label{fig:eivalsRn}
\end{center}
\end{figure}

Multiplying the $M$, we find that
\begin{equation}
\vec a_n =\Big(\prod_{k=n}^1 M_k\Big) \vec a_0= R_n \vec a_0
\end{equation}
which is entirely determined by the product of the matrices $M$, which do not have any free parameters (they do not depend on the $a$). We choose $c_1,c_2$ rational so that $R_n$ can be computed to arbitrary precision. We solve for $a_1,a_2$ given 
$a_{n+1}=a_{n+2}=0$. Given $R_n$, a better estimate of the growth of a solution to the recursion is given by the eigenvalues and eigenvectors of the $4\times 4$ matrix $R_n$, rather than just by the asymptotic estimate of section \ref{sec:method}.
For the case at hand, this is shown in figure \ref{fig:eivalsRn}.

To do a better estimate of the error of the various schemes than what we find in the main text, we think of the cheap bootstrap error as being related to the inverse of the large eigenvalues $\lambda_{large}(n)$ (there is a growing solution that saturates near $|a_{n+1}|,|a_{n+2}|\simeq 1$).
By contrast, the strictly decreasing true solution asymptotes to the larger of the two small eigenvalues $\lambda_{sm,+}(n)$. Setting the $a_{n+1}=a_{n+2}=0$ as in the asymptotic method makes an error of order $\lambda_{sm,+}(n)$, which must be carried by a growing solution. The correct estimate of the error is then of order $\delta {a_1,a_2} \sim \lambda_{sm,+}(n)/\lambda_{large}(n)$.

\section{The strong coupling phase: absence of perturbative $1/N$ corrections}
\label{app:strong_couplng_results}

The goal of this section is to give an intuitive picture of why the strong coupling phase of unitary matrix models of one single matrix is such that the solutions do not have any $1/N$ perturbative corrections. The first description of this property can be found in \cite{Goldschmidt:1979hq}, which was noticed with the evaluation of the strong coupling phase Gross-Witten model \cite{Gross:1980he} in terms of determinants of Bessel functions and their asymptotic analysis.

A simple starting point is pure $U(N)$ Yang Mills in $1+1$ dimensions on a circle in the Hamiltonian formalism. This is the starting point for the model considered by Wadia in \cite{Wadia:1980cp}. We will follow a path similar to \cite{Douglas:1993wy} at the beginning. The relevance to the GWW unitary matrix integral will become precise in \eqref{eq:connection with MM}.

The Hamiltonian can be written as
\begin{equation}
    \hat H = \int \d\theta \, \frac 12 \Tr{E^2}\label{eq:ungauged}
\end{equation}
where $E$ is the electric field squared. Since the field equations of motion of pure Yang Mills make $E$ covariantly constant, it is easy to prove that $E$ can be taken to be a diagonal matrix, reducing the problem to $U(1)^N$. The same can be done with the holonomy around the circle $U \simeq \hbox{diag}(\exp(i\theta_1), \dots \exp(i\theta_N))$ and one easily proves that classically the eigenvalues of $E_a$ are the conjugate coordinates to the $\theta_a$. When considering the gauge fixing problem of going to eigenvalues, the Hamiltonian becomes 
\begin{equation}
    \hat H= -\sum_{a=1}^N \frac 1{2\mu^2}\partial_{\theta_a} \mu^2 \partial_{\theta_a}
\end{equation}
where the measure term is the volume of the gauge orbit, given by
$\mu^2= \prod_{a<b}(\exp(i \theta_a)- \exp(i \theta_b))(\exp(-i \theta_a)- \exp(-i \theta_b)) = \Delta(z) \Delta(\bar z)$ which is a Vandermonde squared term of the variables $z_a= \exp(i \theta_a)$. The term $ \Delta(z) \Delta(\bar z)$ shows in the measure for orthogonality of wave functions of the $\theta_a$
\begin{equation}\label{eq:norm}
\braket{\tilde \psi}{\psi}= \int \prod \d\theta_a \tilde \psi^*\Delta(\bar z) \Delta(z) \psi
\end{equation}
The main idea is to absorb $\Delta(z)$ into the wavefunction of the ket states and the $\Delta(\bar z)$ into the wavefunctions of the bra states, $\psi_F(z)\equiv\Delta(z) \psi(z)$.
In particular, since $\Delta(z)$ is completely antisymmetric, the redefined wavefunctions $\psi_F(z)$ are fermionic, and the norm \eqref{eq:norm} becomes flat.
In addition, one can show that the Hamiltonian that acts on these fermionic wavefunctions is given by $\hat H' = -\frac12 \sum_{a}\partial_{\theta_a}^2$, yielding a system of $N$ non-interacting free fermions on a circle.
The associated free particle basis states of fixed energy (for each eigenvalue) and their wavefunctions are given by
\begin{equation}
\ket{n} : \quad \frac{1}{\sqrt{2\pi}} \exp(i n \theta) \equiv \frac{1}{\sqrt{2\pi}} z^n,
\end{equation}
where $n\in {\mathbb Z}$, and the wavefunctions are unit normalized.
We also notice that the matrix elements of $z$ are given by the obvious relation $z \ket{n}= \ket{n+1}$, so all matrix elements are $1$ or zero.
The Hilbert space of $N$ fermions has a basis that is made from Slater determinants, $\det_{ab} z_a^{n_b}$, of the single particle wavefunction basis. The individual particle wavefunctions  can be classified
in descending order $n_1\geq n_2 \dots$. It is straightforward to show that all these Slater determinants have the same norm, since the basis wave functions $z^n$ all have the same norm.  We can take the Slater determinant norms to be equal to one.

When $N=2M+1$ is odd there is a unique ground state with $n_1= M$, $n_2= M-1$, $\ldots, 0, \dots$, $n_M=-M$ (when $N$ is even, the electric fluxes are $N/2-1/2 \dots -N/2+1/2$ in steps of one and fermions are antiperiodic \cite{Douglas:1993wy}). Let us denote such state by $\ket 0$.

Consider now acting on $\ket{0} $ with $\widehat W_k =\Tr{U^k}\simeq \sum z_a^k$ with $k> 0$.
Since $z_a^k = e^{ik\theta_a}$, this operator shifts the momentum of a single fermion by $k$, sending $n_a \to n_a + k$.
Because the ground state consists of a filled Fermi sea, most such shifts lead to Pauli-forbidden states: if a fermion deep inside the Fermi sea is shifted by $k$, the new momentum $n_a+k$ coincides with an already occupied level, producing a Slater determinant with repeated entries, and hence vanishing.
Only the top $k$ fermions near the Fermi surface can be shifted without violating the Pauli exclusion principle.

Thus $\widehat W_k$ creates particle-hole excitations localized near the Fermi surface, and only $k$ independent contributions survive.
When $N$ is very large and we restrict to operators built from finitely many traces with $k$ of order one, only a finite number of fermions near the Fermi surface are modified, while the bulk of the Fermi sea remains unchanged. 
This gives rise to a small Hilbert space, which we will call the holomorphic Hilbert space. This space is spanned by multi-trace states of the form $\prod \widehat W_{s>0} \ket 0$, which we will refer to as the trace basis.

Now we want to think of $\widehat W_{s>0}$ as holomorphic coordinates in the same sense that $z$ is both a holomorphic function and an operator acting on the Hilbert space of a single particle. The adjoint operator to $z$ is $\bar z= 1/z$ and the adjoint of $\widehat W_{s>0}$ is $\widehat W_{-s}$.
Consider a correlator given by 
\begin{align}\label{eq:connection with MM}
    \bra 0 \prod_{\{s'>0\}}\widehat W_{-s'}   \prod_{\{s>0\}} \widehat W_{s}\ket 0 &= C \int \prod \d \theta_i \Delta(z) \Delta(\bar z) \prod_{\{s'>0\}}\Tr(U^{-s'}) \prod_{\{s>0\}} \Tr(U^s) \\
    &= \int \d U \prod_{\{s'>0\}}\Tr(U^{-s'}) \prod_{\{s>0\}} \Tr(U^s)
\end{align}
where $C$ is a normalization constant of the Slater determinants (usually taken to be $1/N!$), and $\d U$ is the Haar measure for $U(N)$. In the second equality, we recognize that the right hand side is exactly the matrix model computation. That is, we can interpret the Matrix model computation we want to do as a vacuum expectation value in a specific quantum mechanical system of matrices. Our goal in what follows is to get some more intuition about the small Hilbert space interpretation of the correlators and also about how to compute in the trace basis directly. For that, we need additional ingredients.

Just like we can have a trace basis of states, we can also have a basis made with characters $\chi_R(U)$ which are also gauge invariant. These are traces in higher representations of the group and therefore this character basis 
is very natural from group theory. It is also easy to show that the character basis and the Slater determinant basis are actually the same. So, even though we have three possible bases for the Hilbert space: one made of traces, one made of Slater determinants of fermions, and another built from characters, the second and third bases are the same. We need the interpretation in terms of characters to be able to compute the inner product in the trace basis more readily. 

This non trivial fact of the equality of basis follows from the Weyl character formula, which is a ratio of determinants. Each of these determinants is actually a Slater determinant. Given the fermion formulation, the Slater determinant basis is always orthogonal. What about the trace basis?

In general, matrix models that are not of this type (Unitary matrix models) have trace bases that are only approximately orthogonal, with $1/N$ corrections.
This follows from standard large $N$ counting.
For example,  in the case of half BPS states for  ${\cal N}=4 $ SYM it is a similar character basis that is orthogonal \cite{Corley:2001zk}, but not the trace basis. The fermion interpretation of the characters in that case is as fermion droplets on the plane \cite{Berenstein:2004kk}. The idea is that the trace basis 
in that case is approximately orthogonal and generates an approximate Fock space of traces where each $\widehat W_s= \Tr(Z^{s>0})$ is a raising operator
for a boson of momentum $s$. 
For half BPS states in ${\cal N}=4$ SYM, the three point correlation functions of traces do not vanish and start at order $1/N$. 

What is important for us is that since we already proved that the Slater determinant basis has all norms equal to one (because of how the Fourier modes behave on the circle), we have not only exact orthogonality of characters, but all such character states in the small Hilbert space have the same norm. One can then prove that in that case the trace basis is also exactly orthogonal \cite{Berenstein:2017abm} and is an exact Fock space in the infinite $N$ limit with no $1/N$ corrections. The mathematical reason for this exact orthogonality of the Fock can be traced in that case to orthogonality relations of character tables of the symmetric group. A mathematical formulation of this orthogonality of traces fact can also be found in \cite{diaconis1994eigenvalues}.

Furthermore, the small Hilbert space becomes exactly the Hilbert space of a chiral boson. 
Indeed, there are two chiral bosons: one from holomorphic excitations of positive electric field, and one from excitations of the negative values of the electric field. These arise from classifying Young diagrams with boxes and anti-boxes \cite{Gross:1993hu} (see also \cite{Douglas:1993wy}). We are only interested in one such chiral edge where the Hilbert space adds boxes and no anti-boxes are present.

In particular, at finite $N$ the chiral boson built this way is a free theory with a Hilbert space that has the same number of states and overlaps of the 
basic raising/lowering operators up to a cutoff. The cutoff arises because $N$ is finite and the depth of the Young diagrams (the way we think of representations) is cutoff at columns of length $N$. This is non-perturbative in the following sense: we don't see anything at all until the occupation number gets large enough and then suddenly effects occur. This is discontinuous behavior in $1/N$.

Now we want to describe the Fock space as polynomials in traces. Each of the trace variables will be assigned one such variable. 
For each $\widehat W_{s>0}$ we have a holomorphic coordinate $\xi_s$ which we think of as a raising operator $b^\dagger_s$. The definition is $\widehat W_{s>0}=\xi_s\equiv b^\dagger_s$.
We think of states as being polynomials in the $\xi_s$, which we call $p(\xi_s)$. This is just thinking of individual traces as operators in the small Hilbert space rather than functions on the group $U(N)$ and that is why we make the distinction.

The Hilbert space norm for each such oscillator is then given by
\begin{equation}
\braket q p= \int \d\bar \xi_s \d\xi_s \exp(- \xi_s\bar \xi_s/s) q^*(\bar \xi_s) p(\xi_s)\label{eq:Bargmann}
\end{equation}
which is the standard holomorphic (Bargmann) quantization of a harmonic oscillator. The factor of $1/s$ in the exponential arises from the requirement that $\braket{\bar \xi_s}{\xi_s}=s$ (this also follows from the results in \cite{Berenstein:2017abm}).
The complex conjugate variable can then be thought of the lowering operator $s\partial_{\xi_s}$ when computing matrix elements and is realized as $\widehat W_{-s}$ inside integrals. The variables $\widehat W$ are evaluated in the matrix integral and are thought of as actual functions on $U$.
The variables $\xi_s$ are to be thought as phase space collective coordinates that are useful to describe the (quantum) excitations of the system in terms of an exact oscillator basis (so long a that is a sufficiently good approximation).

Importantly, this exact approximation of the inner products of the chiral boson Hilbert space breaks down when the total occupation number is $N$ or higher (this is where Young diagrams get truncated for $U(N)$). At that point, certain states of Young diagrams that are available for $U(\infty)$ disappear and therefore the Fock space counting overcounts the states. 
All matrix elements of the Fock space with energy less than or equal to $N$ are captured exactly. It is only some states with energy larger than $N$ that are missing. The big question is if these states are important or not for the matrix computations we actually want to do.

Consider now a general matrix model correlator given by
\begin{equation}
 MM(\vec t)=   \int \d U \exp( \Tr(V(U))) 
\end{equation}
When we separate $V$ into $\widehat W_{k>0}$ and the opposite, we see that
the matrix model can be thought of exactly as an overlap between coherent states
\begin{equation}
MM(\vec t)=\bra 0  \exp(\sum_s t_{-s} \bar \xi_s/s+\sum_{s>0} t_s \xi_s/s)\ket 0
\end{equation}
A straightforward computation with coherent states gives us that
\begin{equation}
MM(\vec t)= \exp\left[\sum t_{-s} t_s/s \right]
\end{equation}
This is equivalent to doing an exact Gaussian integral with the measure given by 
\eqref{eq:Bargmann} (there the measure is normalized so that $\braket 00=1$). 
From that Gaussian integral we get exactly the saddle point contribution and nothing else. 

Now, the overlaps are exact so long as the total occupation number is less than $N$. If the $t$ are finite (like we stated in the definition of the strong coupling phase), then the occupation number is related to the amplitude $t$ and grows like $|t|^2$.
If the only few non-vanishing $t$ are finite and not scaling with $N$, the average occupation number per mode in the Fock space is finite and the result is well within the bounds of applicability of the exact chiral boson approximation. 
However, at sufficiently high occupation number this approximation breaks down, as for instance, the results of the exponential as a generating series differ from the exact Fock space.
Again, this is because the Fock space overcounts states. These effects on the tail start when we go to order $m\simeq N$ and tails that look as $|t_s|^m/m!$ which are exponentially suppressed in $N$ (because we have an $N!$ denominator). This is strictly non-perturbative (it is exponentially suppressed in $N$ rather than polynomially suppressed in $N$).
Indeed, in this analogy, it has been shown that for a single truncated harmonic oscillator, that the result is robust even when the occupation number scales like $N$ so long as the average occupation number is close to the edge of exact applicability roughly to order $\sqrt{N}$  \cite{Berenstein:2013md}, which is the variance of the occupation number from the Poisson distribution of occupation numbers.

In the case we are dealing with, since the system is second quantized, we have more states that we need to compare to, as they are generally given by Young diagrams with columns of length at most $N$. What we consider the edge of applicability is when random Young diagrams that are dominant start approaching the edge, rather than when the first possible Young diagram of all diagrams approaches the edge. In systems with more than two matrices one has robust arguments that the typical Young diagrams have a non-trivial shape distribution  (the VKLS shape) with area of order $N^2$ (see \cite{Berenstein:2018lrm,Berenstein:2023srv}). For the one matrix case, na\"\i ve thermal arguments about random Young diagrams do not have the same scaling \cite{Balasubramanian:2005mg} and more work needs to be done to have an easy argument for $N^2$ scaling of the answer. Luckily, this is provided by other ideas which we will now present. A better guess to find what Young diagrams contribute is to use the phase space description of the states as a droplet on a cylinder (the exact phase space of one eigenvalue). That is, we look at the pictures of the states as droplet on the cylinder to estimate the shapes of Young diagrams that dominate.
This is done in \cite{Chattopadhyay:2017ckc}, particularly section 5.4. We see in that case that the Young diagram reaches the allowed bottom when the projected density of eigenvalues into the position space is close to vanishing (this is what becomes the density of eigenvalues in the usual formulation). That is, the coherent states we have produced are accurate descriptions of the states up to amplitudes of order $N$ for each mode.
The overlap is therefore of order $\exp(N^2)$ before the non-perturbative truncation of the Hilbert space starts dominating the discussion.
Now we want to connect this result to other simpler methods to get the estimates of the strong coupling expectation values.

\subsection{Perturbative treatment of Wilson loop expectation values}
\label{app:pert Wilson}

As discussed in the body of the paper, there are two notions of the large $N$ (strong coupling) limit that we have used. The naive large $N$ limit where we keep the coupling constants $t_k$ in the potential fixed and finite with $N\to \infty$, and a second 't Hooft limit where we keep $\tilde t_k= t_k/N$ fixed. By convention, the strong coupling 
regime is where the $\tilde t_k$ are fixed and small (the convention originates from the Gross-Witten model and the relation between the coupling constant and the Wilson action for plaquettes). The naive large $N$ limit therefore has $\tilde k=0$ in the t'Hooft sense.

The simplest way to compute the leading order in $N$ result for the Wilson loop expectation values, and which covers both the finite  coupling large $N$ and the strong t' Hooft  coupling limits, is to use the arguments of Anderson-Kruczenski \cite{Anderson:2016rcw}.
One can write an effective action for the $W_n$, so that the matrix integral, restricted to eigenvalues, 
can be written in the t'Hooft limit as
\begin{equation}
    \int \d U \exp(N \hbox{Tr}(\widetilde V(U))) \simeq \exp(S_{eff}(W_i))
\end{equation}
where $\widetilde V(U)=V(U)/N$. One can pass to the density of eigenvalues
$\rho$ and one finds an effective action given by
\begin{equation}
    S_{eff} = \sum_{k\neq 0} N^2  \frac{\tilde t_k}{|k|} W_k-N^2 \sum_{k> 0} W_k W_{-k}/k
\end{equation}
where $\tilde t_k= t_k/N$ are the t'Hooft coupling constants that are fixed. 
For the strong coupling limit, the $\tilde t_k$ are small.
Notice the similarity to our measure \eqref{eq:Bargmann}. In our case it was derived using properties of the exact Hilbert space, whereas here the quadratic term arises from eigenvalue repulsion and an integral identity. This method is more similar to the one used by Gross-Witten \cite{Gross:1980he} in their paper on the phase transition.

Solving for $W_k$, we find that in this setup the answer for the saddle is
\begin{equation}
W_n = \tilde t_{-n} 
\end{equation}
which is the same as the one that arises from the Gaussian saddle we discussed before.
Since we are choosing potentials where $t_n=t_{-n}$, 
this is equivalent to
\begin{equation}
W_{-n}= W_n = \tilde t_{n} 
\end{equation}
If we restore factors of $N$, we find that
\begin{equation}
    W_n=  t_n/ N
\end{equation}
or equivalently
\begin{equation}
\widehat W_n = t_n
\end{equation}
Now, in the t'Hooft limit, where $\tilde t_n$ is fixed, we see that $W_n$ has a nice limit. In the strong coupling limit where we keep $t_n$ fixed, it is $\widehat W_n$ that has a nice limit. The edge of applicability of this equation as given requires that we are sufficiently deep in the strong coupling phase so that the density of eigenvalues is nowhere vanishing on the circle.

We can  argue that the free Fock space theory should be free of perturbative corrections all the way to the edge of applicability of the effective theory described by $\rho>0$. The saddles thus have action of order $N^2$ and $\tilde t_k$ finite and small.

\section{Saddle-point expressions}
\label{appendix:coefficients}

In this appendix, we display the first couple saddle-point expansion coefficients featuring in the right-hand side of equation \eqref{eq:Instantons11}. The first three read
\begin{align}
     C_1(x,y) =&\frac{V_{\text{eff}}^{(2)}(x)^{-\frac{1}{2}} \left(-V_{\text{eff}}^{(2)}(y)\right)^{-\frac{1}{2}}}{2 \pi  (y-x)^2} \\ 
    C_2(x,y)=& -\frac{V_{\text{eff}}^{(2)}(x)^{-\frac{7}{2}} \left(-V_{\text{eff}}^{(2)}(y)\right)^{-\frac{7}{2}}}{48 \pi  (x-y)^4}\big(72 V_{\text{eff}}^{(2)}(x)^2 V_{\text{eff}}^{(2)}(y)^3+5 V_{\text{eff}}^{(3)}(x)^2 \times \nonumber\\&(x-y)^2 V_{\text{eff}}^{(2)}(y)^3 + V_{\text{eff}}^{(2)}(x)^3\Big(-5 V_{\text{eff}}^{(3)}(y)^2 (x-y)^2-72 V_{\text{eff}}^{(2)}(y)^2+3 (x-y) \times \nonumber \\ &\left(V_{\text{eff}}^{(4)}(y) (x-y)+8 V_{\text{eff}}^{(3)}(y)\right) V_{\text{eff}}^{(2)}(y)\Big)-3 (x-y) \left(V_{\text{eff}}^{(4)}(x) (x-y)-8 V_{\text{eff}}^{(3)}(x)\right) \times \nonumber\\& V_{\text{eff}}^{(2)}(x) V_{\text{eff}}^{(2)}(y)^3\big) \\ 
  C_3(x,y)= &\frac{V_{\text{eff}}^{(2)}(x)^{-\frac{13}{2}} \left(-V_{\text{eff}}^{(2)}(y)\right)^{-\frac{13}{2}}}{2304 \pi  (x-y)^6}\Big(385 V_{\text{eff}}^{(3)}(x)^4 (x-y)^4 V_{\text{eff}}^{(2)}(y)^6 +144 V_{\text{eff}}^{(2)}(x)^5 \times\nonumber\\& V_{\text{eff}}^{(2)}(y)^3   \Big(-5 V_{\text{eff}}^{(3)}(y)^2 (x-y)^2 -240 V_{\text{eff}}^{(2)}(y)^2+3 (x-y) \Big(V_{\text{eff}}^{(4)}(y) (x-y)+16 V_{\text{eff}}^{(3)}(y)\Big)\times \nonumber\\& V_{\text{eff}}^{(2)}(y)\Big) +6 V_{\text{eff}}^{(2)}(x)^4 V_{\text{eff}}^{(2)}(y)^3 \Big(2880 V_{\text{eff}}^{(2)}(y)^3+5 V_{\text{eff}}^{(3)}(y)^2 (x-y)^3 \Big((V_{\text{eff}}^{(4)}(x) (x-y)-\nonumber\\&8 V_{\text{eff}}^{(3)}(x)\Big)  +3 (x-y)^2 \Big(8 V_{\text{eff}}^{(3)}(x) \left(V_{\text{eff}}^{(4)}(y) (x-y)+12 V_{\text{eff}}^{(3)}(y)\right)-V_{\text{eff}}^{(4)}(x) (x-y) \times \nonumber\\&\Big(V_{\text{eff}}^{(4)}(y) (x-y) +8 V_{\text{eff}}^{(3)}(y)\Big)\Big)V_{\text{eff}}^{(2)}(y)+72 (x-y) \left(V_{\text{eff}}^{(4)}(x) (x-y)-16 V_{\text{eff}}^{(3)}(x)\right) V_{\text{eff}}^{(2)}(y)^2\Big) \nonumber\\& -210 V_{\text{eff}}^{(3)}(x)^2 (x-y)^3 \left(3 V_{\text{eff}}^{(4)}(x) (x-y)-8 V_{\text{eff}}^{(3)}(x)\right) V_{\text{eff}}^{(2)}(x) V_{\text{eff}}^{(2)}(y)^6 +21 (x-y)^2 \times \nonumber\\&\Big(8 V_{\text{eff}}^{(3)}(x) V_{\text{eff}}^{(5)}(x) (x-y)^2+5 \left(V_{\text{eff}}^{(4)}(x) (y-x)+4 V_{\text{eff}}^{(3)}(x)\right) \Big(V_{\text{eff}}^{(4)}(x) (y-x) +12 V_{\text{eff}}^{(3)}(x)\Big)\Big) \times \nonumber\\&V_{\text{eff}}^{(2)}(x)^2 V_{\text{eff}}^{(2)}(y)^6+V_{\text{eff}}^{(2)}(x)^6 \Big(385 V_{\text{eff}}^{(3)}(y)^4 (x-y)^4+17280 V_{\text{eff}}^{(2)}(y)^4   -210 V_{\text{eff}}^{(3)}(y)^2 (x-y)^3 \times \nonumber\\&\left(3 V_{\text{eff}}^{(4)}(y) (x-y)+8 V_{\text{eff}}^{(3)}(y)\right) V_{\text{eff}}^{(2)}(y)+21 (x-y)^2 \Big(5 V_{\text{eff}}^{(4)}(y)^2 (x-y)^2 +240 V_{\text{eff}}^{(3)}(y)^2+\nonumber\\ &8 V_{\text{eff}}^{(3)}(y) (x-y) \left(V_{\text{eff}}^{(5)}(y) (x-y)+10 V_{\text{eff}}^{(4)}(y)\right)\Big) V_{\text{eff}}^{(2)}(y)^2  +24 (x-y) \Big(90 V_{\text{eff}}^{(4)}(y) (y-x)-\nonumber\\& 480 V_{\text{eff}}^{(3)}(y)+(x-y)^2 \left(V_{\text{eff}}^{(6)}(y) (y-x)-12 V_{\text{eff}}^{(5)}(y)\right)\Big) V_{\text{eff}}^{(2)}(y)^3\Big) -2 (x-y) V_{\text{eff}}^{(2)}(x)^3 \times \nonumber\\&V_{\text{eff}}^{(2)}(y)^3 \Big(25 V_{\text{eff}}^{(3)}(x)^2 V_{\text{eff}}^{(3)}(y)^2 (x-y)^3+360 V_{\text{eff}}^{(3)}(x)^2 (x-y) V_{\text{eff}}^{(2)}(y)^2 -15 V_{\text{eff}}^{(3)}(x)^2 (x-y)^2 \times \nonumber\\ &\left(V_{\text{eff}}^{(4)}(y) (x-y)+8 V_{\text{eff}}^{(3)}(y)\right) V_{\text{eff}}^{(2)}(y) +12 \Big(V_{\text{eff}}^{(6)}(x) (x-y)^3-480 V_{\text{eff}}^{(3)}(x)+6 (x-y)\times  \nonumber\\&\left(2 V_{\text{eff}}^{(5)}(x) (y-x)+15 V_{\text{eff}}^{(4)}(x)\right)\Big) V_{\text{eff}}^{(2)}(y)^3\Big).
\end{align}
The expression for $C_4(x,y)$ is rather lengthy and is therefore omitted here, but it remains available upon request from the authors.

\section{Lefschetz-thimble analysis}
\label{app:LefschetzThimbleAnalysis}

In this appendix, we perform a Lefschetz-thimble analysis of the unitary matrix integral \eqref{eq:Instantons2} in the ungapped phase. The purpose of this analysis is to support the discussion in subsection \ref{subsec:instantons} with further details and motivate the instanton expansion \eqref{eq:expansionNcontribution} from a resurgent point of view. 

For simplicity, we focus on the GWW model \eqref{eq:GWW}. This model admits two saddle-points: one located outside the unit circle, given by \eqref{eq:GWWs1}, and another located inside the unit circle, given by \eqref{eq:GWWs2}.

Following \cite{msw08,mss22}, we analyse the behaviour of the Lefschetz thimbles associated with \eqref{eq:Instantons2} as the phase of the coupling $t$ is varied. Topology changes in these thimbles are expected to encode the Stokes phenomena (see \cite{abs18} for a pedagogical approach to this correspondence or \cite{msw08,mss22} for a more modern take), thereby providing guidance on which non-perturbative instanton contributions must be included in order to construct a suitable transseries completion. 

\begin{figure}
    \centering
    \hspace{-50pt}
    \begin{tikzpicture}[scale = 0.75]


        \filldraw[decorate, decoration={snake, segment length=5pt, amplitude=3pt},color=ForestGreen, fill=Green!5, line width=2pt](0,0) circle (3);

        \draw[color=Black, line width=2pt] (-3,0) to[out = 180, in = -90] (-5,2);

        \draw[color=Black, line width=2pt] (-3,0) to[out = 0, in = 90] (-1,-0.2) to[out = -90, in = 180] (-1+0.8,-0.2-0.6);

\filldraw[color=Black, line width=2pt](-3,0) circle (0.1);

        \draw[color=Orange, line width=2pt] (-5.5,0) arc(180:170:15);

        \draw[color=Orange, line width=2pt] (-5.5,0) arc(180:190:15);

        \draw[color=blue, line width=2pt] (-0.6,0) arc(180:90:0.5);
        \draw[color=blue, line width=2pt] (-0.6,0) arc(180:270:0.5);

        \filldraw[color= cornellred,  line width=2pt](-0.6,0) circle (0.1);

        \filldraw[color= cornellred,  line width=2pt](-5.5,0) circle (0.1);


\def\hspace{13-2}
\def\vspace{7.5}

\filldraw[decorate, decoration={snake, segment length=5pt, amplitude=3pt},color= ForestGreen, fill=Green!5, line width=2pt](0+\hspace,0-7.5+\vspace) circle (3);

        \draw[color=Black, line width=2pt] (-3+\hspace,0-7.5+\vspace) to[out = 180, in = 90] (-5+\hspace,-2-7.5+\vspace);

        \draw[color=Black, line width=2pt] (-3+\hspace,0-7.5+\vspace) to[out = 0, in = -90] (-1+\hspace,+0.2-7.5+\vspace) to[out = 90, in = 180] (-1+0.8+\hspace,0.2+0.6-7.5+\vspace);

\filldraw[color=Black,  line width=2pt](-3+\hspace,0-7.5+\vspace) circle (0.1);

        \draw[color=Orange, line width=2pt] (-5.5+\hspace,0-7.5+\vspace) arc(180:170:15);

        \draw[color=Orange, line width=2pt] (-5.5+\hspace,0-7.5+\vspace) arc(180:190:15);

        \draw[color=blue, line width=2pt] (-0.6+\hspace,0-7.5+\vspace) arc(180:90:0.5);
        \draw[color=blue, line width=2pt] (-0.6+\hspace,0-7.5+\vspace) arc(180:270:0.5);

             \filldraw[color= cornellred, line width=2pt](-0.6+\hspace,0-7.5+\vspace) circle (0.1);

          \filldraw[color= cornellred, line width=2pt](-5.5+\hspace,0-7.5+\vspace) circle (0.1);

         \node at (5-5,3+1.3){\scalebox{1.3}{$\text{Im}(t) >0$}};
         \node at (5+\hspace-5,3-7.5+\vspace+1.3){\scalebox{1.3}{$\text{Im}(t) <0$}};

         \node[color=cornellred] at (-6,0){\scalebox{1.3}{$z_1^\star$}};

         \node[color=cornellred] at (-0.07,0){\scalebox{1.3}{$z_2^\star$}};

         \node[color=cornellred] at (-6+\hspace,-7.5+\vspace){\scalebox{1.3}{$z_1^\star$}};

         \node[color=cornellred] at (-0.07+\hspace,-7.5+\vspace){\scalebox{1.3}{$z_2^\star$}};



\node[color=Orange] at (-5+\hspace+0.1,1-7.5+\vspace){\scalebox{1.3}{$\bar{\mathcal{C}}_1^\star$}};

\node[color=blue] at (-0.5+\hspace,-0.8-7.5+\vspace-0.2){\scalebox{1.3}{$\mathcal{C}_2^\star$}};


\node[color=blue] at (-0.5,0.8+0.2){\scalebox{1.3}{$\mathcal{C}_2^\star$}};

\node[color=Orange] at (-5+0.1,-1){\scalebox{1.3}{$\bar{\mathcal{C}}_1^\star$}};

    \end{tikzpicture}
    \caption{Pictorial representation of the Lefschetz-thimble topology change associated with an eigenvalue (black dot) taking place on the point of the unit circle distribution (denoted as the green wavy line) closest to both saddles (marked by red dots). The steepest-ascent contour associated with the saddle $z_1^\star$ is denoted by $\bar{\mathcal{C}}_1^\star$ and shown as an orange line while the steepest-descent contour associated with the saddle $z_2^\star$ is denoted by $\mathcal{C}_2^\star$ and shown as a blue line.}
    \label{fig:Instantons1}
\end{figure}
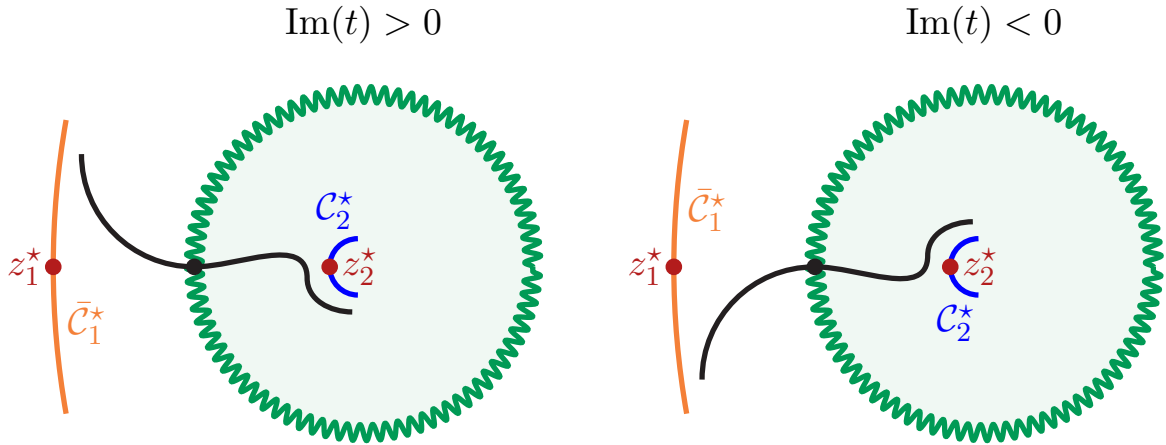

The two main players (Lefschetz-thimbles) in this analysis are the steepest-descent and steepest-ascent contours associated with eigenvalues $z^\star$ taking place on the unit circle distribution or the saddles. Both are defined by the set of points such that
\begin{equation}
    \text{Im}\left(V_{\text{eff}}(z)-V_{\text{eff}}(z^\star)\right) = 0 \,,
    \label{eq:Instantons4}
\end{equation}
while the former is further defined by a growing real part of $V_{\text{eff}}(z)$ away from $z^\star$ and the latter is defined by decreasing real part. Expecting a topology change of the Lefschetz-thimble associated with the eigenvalue in the cut closest to the saddles (in analogy with what was considered for the Hermitian case in \cite{msw08,mss22}), we depict in figure \ref{fig:Instantons1} the set defined by the equation above, for $z^\star = -1$. 
As can be seen, crossing the line $\mathrm{Im}(t)=0$ induces a sharp topology change in the Lefschetz thimble. This change translates into an asymptotic discontinuity $\text{Disc}_0$ of the partition function \eqref{eq:Instantons2}, obtained by performing the same integral for a background in which several of the eigenvalue integrations domains are shifted from $S^1$ to the contours capturing the topology change.\footnote{The subscript in the discontinuity indicates we are considering the Stokes phenomena taking place along the direction $\theta = \arg(t)= 0$.} Due to holomorphicity of the integrand, these contours can be deformed into two Lefschetz thimbles: a steepest-descent contour $\mathcal{C}_2^\star$ associated with the inner saddle $z_2^\star$, and a steepest-ascent contour $\bar{\mathcal{C}}_1^\star$ associated with the outer saddle $z_1^\star$ (see figure \ref{fig:Instantons1}). 

Assuming we can treat the unitary matrix model (away from the unit circle) as an Hermitian matrix model  with effective potential given by \eqref{eq:Instantons13}, we can borrow the result of equation (3.10) of \cite{mss22} to write the leading exponential contribution to the discontinuity $\text{Disc}_0$ as
\begin{align}
    2\times Z_N^{(0)}(t) \int_{\mathcal{C}_2^\star}\frac{\rmd z}{2\pi}\int_{\bar{\mathcal{C}}^\star_1}\frac{\rmd \bar{z}}{2\pi} \exp\left(-N\left(V(z)-V(\bar{z})\right)\right)\left\langle\frac{\det^2(z-U)}{\det^2(\bar{z}-U)}\right\rangle \frac{1}{(z-\bar{z})^2} \,.
    \label{eq:leading-instanton}
\end{align}
Following \cite{mss22}, the term above is to be understood as the non-perturbative instanton contribution associated with a background in which one eigenvalue and one anti-eigenvalue (eigenvalue tunneling in the second branch of the spectral curve \ref{fig:Instantons4}) have departed the unit circle distribution and tunneled to the saddles $z_2^\star$ and $z_1^\star$, respectively. The factor of $2$ multiplying the expression above accounts for the fact that, prior to the tunneling process, the eigenvalue and anti-eigenvalue originate from two indistinguishable eigenvalues in the unit-circle distribution, whose roles may be interchanged.

Now, we can  write
\begin{equation}
    \left\langle\frac{\det^2(z-U)}{\det^2(\bar{z}-U)}\right\rangle = \exp\left(2 \left\langle\text{Tr}\left[\log\left(z-U\right)\right]\right\rangle-2\left\langle\text{Tr}\left[\log\left(\bar{z}-U\right)\right]\right\rangle+\cdots\hspace{2pt}\right) \,,
\end{equation}
where the dots represent terms proportional to correlation functions involving multiple powers of the traces appearing on the right-hand side of the equation above. It can be shown, using, for instance, Theorem 2 of \cite{diaconis1994eigenvalues}, that in the ungapped phase all such correlation functions vanish perturbatively (see Appendix C of \cite{Eniceicu:2023cxn} for a proof), turning the equality above exact. Using equation \eqref{eq:effpotential}, we can then rewrite the term \eqref{eq:leading-instanton} as
\begin{align}
    2\times Z_N^{(0)}(t) \int_{\mathcal{C}_2^\star}\frac{\rmd z}{2\pi}\int_{\bar{\mathcal{C}}^\star_1}\frac{\rmd \bar{z}}{2\pi} \exp\left(-N\left(V_{\text{eff}}(z)-V_{\text{eff}}(\bar{z})\right)\right) \frac{1}{(z-\bar{z})^2} \,.
  \label{eq:finalleading-instanton}
\end{align}
Interestingly, up to the combinatorial factor of $2$, this term exactly matches the leading instanton correction to the GWW model obtained by setting $n = 1$ in equation \eqref{eq:GWWInstanton}. 

The term above is merely the leading contribution to the discontinuity $\text{Disc}_0$. Indeed, higher instanton corrections can be obtained by considering backgrounds in which more eigenvalues and anti-eigenvalues tunnel to the saddles. Interestingly enough, due to the nature of the topology change depicted in figure \ref{fig:Instantons1}, it seems eigenvalues and anti-eigenvalues are bound to tunnel in pairs. This greatly reduces the number of allowed eigenvalue configurations, in which case the discontinuity is fully accounted for by considering backgrounds in which a generic number $n \in \mathbb{N}$ of eigenvalues and anti-eigenvalues tunnel in pairs to the saddles $z_2^\star$ and $z_1^\star$, respectively. Resorting again to equation (3.10) of \cite{mss22}, we can write the associated contribution as 
\begin{align}
     &2^n\times\frac{Z_N^{(0)}(t)}{\left(n!\right)^2}\left[\prod_{m=1}^n \int_{\mathcal{C}_2^\star}\frac{\rmd z_m}{2\pi}\int_{\bar{\mathcal{C}}^\star_1}\frac{\rmd \bar{z}_m}{2\pi} \exp\left(-N\left(V(z_m)-V(\bar{z}_m)\right)\right)\right] \times \nonumber \\ &\Delta^2(\boldsymbol{z})\Delta^2(\boldsymbol{\bar{z}}) \left\langle \prod_{m=1}^n\frac{\det^2(z_m-U)}{\det^2(\bar{z}_m-U)}\right\rangle\prod_{m=1}^{n}\prod_{\bar{m}=1}^{n}\frac{1}{(z_m-\bar{z}_{\bar{m}})^2} \,,
\end{align}
which we can be recasted as
\begin{align}
     &2^n\times\frac{Z_N^{(0)}(t)}{\left(n!\right)^2}\left[\prod_{m=1}^n \int_{\mathcal{C}_2^\star}\frac{\rmd z_m}{2\pi}\int_{\bar{\mathcal{C}}^\star_1}\frac{\rmd \bar{z}_m}{2\pi} \exp\left(-N\left(V_{\text{eff}}(z_m)-V_{\text{eff}}(\bar{z}_m)\right)\right)\right] \times \nonumber \\ &\Delta^2(\boldsymbol{z})\Delta^2(\boldsymbol{\bar{z}}) \prod_{m=1}^{n}\prod_{\bar{m}=1}^{n}\frac{1}{(z_m-\bar{z}_{\bar{m}})^2} \,,
\end{align}
by using an argument entirely analogous to the one leading up to \eqref{eq:finalleading-instanton}. The combinatorial factor of $2^n$ multiplying the expression above generalizes the factor of $2$ in \eqref{eq:leading-instanton}, accounting for the indistinguishability of the eigenvalues that give rise to each of the $n$ pairs. 

Using equation \eqref{eq:GWWInstanton}, we can then write the complete discontinuity as
\begin{equation}\label{eq:discZN}
    \text{Disc}_0 = \sum_{n=1}^{+\infty} 2^n Z^{(n)}_N(t) \,.
\end{equation}
In \cite{msw08,mss22}, it was understood that matrix model discontinuities induced by topology changes of Lefschetz-thimbles accurately reproduce the action of the Stokes automorphism $\underline{\mathfrak{S}}_{0}$ (see, for instance, \cite{abs18,s14} for definitions and further details). In our case, however, this correspondence appears to encounter a subtle tension, as the perturbative sector \eqref{eq:Instantons3} is not asymptotic and should therefore admit a trivial Stokes automorphism action. 
In what follows, we proceed by following \cite{mss22} in a strict sense: we redefine the action of the Stokes automorphism so that it reproduces the Lefschetz-thimble topology change.\footnote{A natural way to justify this prescription is to deform the potential by introducing an infinitesimal source term coupled to the trace of a higher power of $U$, and then tune it so as to (infinitesimally) split the unit-circle distribution into a gapped configuration, without qualitatively altering the Lefschetz-thimble analysis carried out above. In this regime, the perturbative sector becomes asymptotic, thereby resolving the tension and hopefully restoring the correspondence between the Lefschetz-thimble topology change and the action of the Stokes automorphism. 

This constitutes a particular instance of Cheshire cat resurgence \cite{kstu16}. See, for example, \cite{dg18} for a concrete application of this framework in the context of the $\mathbb{C}\mathbb{P}^{N-1}$ model with $\mathcal{N}=(2,2)$ supersymmetry. In this case, the partition function on $S^2$ decomposes into topological sectors, each of which exhibits a non-asymptotic (convergent) perturbative expansion together with non-perturbative instanton-anti-instanton contributions that appear to lie beyond the reach of standard resurgence techniques. The authors introduce a non-supersymmetric deformation (akin to our infinitesimal deformation) to lift the supersymmetric cancellations and restore the asymptotic character of the perturbative expansions. Once asymptoticity is recovered, standard (Cheshire cat) resurgence techniques allow them to reconstruct the non-perturbative corrections, before removing the deformation altogether.}


Concretely, if the partial transseries completion, accounting for the topology change depicted in figure \ref{fig:Instantons1}, of the GWW model is written as
\begin{align}
     Z_N(t;\sigma) &= Z_N^{(0)}(t) + \sum_{n=1}^{+\infty}\sigma^n Z_N^{(n)}(t) \,,
    \label{eq:transseries} 
\end{align}
for some transseries parameter $\sigma \in \mathbb{C}$ capturing the non-perturbative ambiguity, then
\begin{equation}
\underline{\mathfrak{S}}_{0}\left[Z_N(t;0)\right] =   Z_N(t;\sigma=S)  = Z_N^{(0)}(t) + \text{Disc}_0 \,,
\end{equation}
where $S \in \mathbb{C}$ is the Stokes constant associated with the Stokes phenomena underlying the topology change depicted in figure \ref{fig:Instantons1} (see \cite{abs18} for a pedagogical introduction to resurgence and Stokes constants). 
In particular, in view of \eqref{eq:discZN} this implies that
\begin{equation}
    S = 2 \,.
\end{equation}

Obtaining a transseries representation such as \eqref{eq:transseries} is not sufficient to determine the full non-perturbative completion, since a residual ambiguity remains, encoded in the unfixed transseries parameter $\sigma$. A standard way to constrain this ambiguity is to impose reality of the Borel resummation of the transseries for $t>0$ (see \cite{abs18,s14} for pedagogical introductions to summability and resurgence). 
In our case, however, a Stokes line lies precisely along this direction. As a result, the corresponding prescription --- known as the median resummation \cite{dp99,m08,ar15} --- is more subtle and does not fully remove the ambiguity. Instead, it yields the condition
\begin{equation}
    \sigma = C+\frac{S}{2} \,,
\end{equation}
for some $C \in \mathbb{R}$.\footnote{This condition is uniquely fixed by requiring that the Borel resummation of the transseries and that of its image under the Stokes automorphism $\underline{\mathfrak{S}}_0$ be complex conjugates of one another. Since the resurgent gluing condition along the positive real axis $t > 0$ demands that these two resummations coincide \cite{abs18}, it follows immediately that the resulting expression is real. For further discussion, see \cite{dp99,m08,ar15}.} 
It is interesting to realize that, upon choosing the ``simplest'' transseries parameter by setting $C=0$, one obtains the non-perturbative completion
\begin{equation}
    Z_N\left(t;\frac{S}{2}\right) = Z_N^{(0)}(t) + \sum_{n=1}^{+\infty}  Z^{(n)}_N(t) \,.
\end{equation}
This observation motivates the ungapped-phase instanton expansion \eqref{eq:Instantons7} for the GWW model. In particular, the resulting instanton corrections can be recast as \eqref{eq:GWWInstanton}, exactly matching those above. 

\begin{figure}
    \centering
    \begin{tikzpicture}[scale = 0.75]


\filldraw[decorate, decoration={snake, segment length=5pt, amplitude=3pt},color=ForestGreen, fill=Green!5, line width=2pt](0,0) circle (3);


\draw[color=Black, line width=2pt] (-0.95,-2.85) to[out = -90-20, in = 0] (-0.95-2,-2.85-0.8);
\draw[color=Black, line width=2pt] (-0.95,-2.85) to[out = -90-20+180, in = 180] (-0.95+1.2,-2.85+0.8)to[out = 0, in = -90-60](-0.95+1.2+0.5,-2.85+0.8+0.1);
\draw[color=Black, line width=2pt] (-0.95,2.85) to[out = 90+20, in = 180] (-0.95+2,2.85+0.8);
\draw[color=Black, line width=2pt] (-0.95,2.85) to[out = 90+20+180, in = 20] (-0.95-0.5,2.85-1.2) to[out = 20+180, in = 30] (-0.95-0.5-0.3,2.85-1.2-0.1);


\draw[color=Orange, line width=2pt](-1.5*0.9,-4.5*0.9) arc(-90:-90-10:15);
\draw[color=Orange, line width=2pt](-1.5*0.9,-4.5*0.9) arc(-90:-90+10:15);
\draw[color=Orange, line width=2pt](-1.5*0.9,4.5*0.9) arc(90:90+10:15);
\draw[color=Orange, line width=2pt](-1.5*0.9,4.5*0.9) arc(90:90-10:15);
\draw[color=blue, line width=2pt](-1.5*0.38,4.5*0.38) arc(90:90+60:1);
\draw[color=blue, line width=2pt](-1.5*0.38,4.5*0.38) arc(90:90-60:1);
\draw[color=blue, line width=2pt](-1.5*0.38,-4.5*0.38) arc(-90:-90+60:1);
\draw[color=blue, line width=2pt](-1.5*0.38,-4.5*0.38) arc(-90:-90-60:1);


\filldraw[color=black, line width=2pt](-0.95,2.85) circle (0.1);
\filldraw[color= cornellred, line width=2pt](-1.5*0.9,4.5*0.9) circle (0.1);
\filldraw[color= cornellred, line width=2pt](-1.5*0.38,4.5*0.38) circle (0.1);

\filldraw[color=black, line width=2pt](-0.95,-2.85) circle (0.1);
\filldraw[color= cornellred,  line width=2pt](-1.5*0.9,-4.5*0.9) circle (0.1);
\filldraw[color= cornellred,  line width=2pt](-1.5*0.38,-4.5*0.38) circle (0.1);


\def\hspace{13-2-1}
\def\picspacing{0}

\filldraw[decorate, decoration={snake, segment length=5pt, amplitude=3pt},color=ForestGreen, fill=green!5, line width=2pt](0+\hspace,\picspacing) circle (3);


\draw[color=black, line width=2pt] (-0.95+\hspace,-2.85+\picspacing) to[out = -90-20, in = 180] (-0.95+2+\hspace,-2.85-0.8+\picspacing);
\draw[color=black, line width=2pt] (-0.95+\hspace,-2.85+\picspacing) to[out = -90-20+180, in = -60] (-0.95-0.9+\hspace,-2.85+1.4+\picspacing);
\draw[color=black, line width=2pt] (-0.95+\hspace,2.85+\picspacing) to[out = 90+20, in = 0] (-0.95-2+\hspace,2.85+0.8+\picspacing);
\draw[color=black, line width=2pt] (-0.95+\hspace,2.85+\picspacing) to[out = 90+20+180, in = 90+30] (-0.95+1.5+\hspace,2.85-1.5+\picspacing);


\draw[color=Orange, line width=2pt](-1.5*0.9+\hspace,-4.5*0.9+\picspacing) arc(-90:-90-10:15);
\draw[color=Orange, line width=2pt](-1.5*0.9+\hspace,-4.5*0.9+\picspacing) arc(-90:-90+10:15);
\draw[color=Orange, line width=2pt](-1.5*0.9+\hspace,4.5*0.9+\picspacing) arc(90:90+10:15);
\draw[color=Orange, line width=2pt](-1.5*0.9+\hspace,4.5*0.9+\picspacing) arc(90:90-10:15);
\draw[color=blue, line width=2pt](-1.5*0.38+\hspace,4.5*0.38+\picspacing) arc(90:90+60:1);
\draw[color=blue, line width=2pt](-1.5*0.38+\hspace,4.5*0.38+\picspacing) arc(90:90-60:1);
\draw[color=blue, line width=2pt](-1.5*0.38+\hspace,-4.5*0.38+\picspacing) arc(-90:-90+60:1);
\draw[color=blue, line width=2pt](-1.5*0.38+\hspace,-4.5*0.38+\picspacing) arc(-90:-90-60:1);


\filldraw[color=black,  line width=2pt](-0.95+\hspace,2.85+\picspacing) circle (0.1);
\filldraw[color= cornellred,  line width=2pt](-1.5*0.9+\hspace,4.5*0.9+\picspacing) circle (0.1);
\filldraw[color= cornellred,  line width=2pt](-1.5*0.38+\hspace,4.5*0.38+\picspacing) circle (0.1);

\filldraw[color=black, line width=2pt](-0.95+\hspace,-2.85+\picspacing) circle (0.1);
\filldraw[color= cornellred,  line width=2pt](-1.5*0.9+\hspace,-4.5*0.9+\picspacing) circle (0.1);
\filldraw[color= cornellred,  line width=2pt](-1.5*0.38+\hspace,-4.5*0.38+\picspacing) circle (0.1);


\node at (5-5,3+3){\scalebox{1.3}{$\text{Im}\left(N\right) >0$}};
         \node at (5+\hspace-5,3+\picspacing+3){\scalebox{1.3}{$\text{Im}\left(N\right) <0$}};

\node[color=cornellred] at(-1.5*0.38,4.5*0.38-0.4-0.15){\scalebox{1.3}{$z_2^\star$}};
\node[color=blue] at(-1.5*0.38+0.6,4.5*0.38+0.2+0.15){\scalebox{1.3}{$\mathcal{C}_2^\star$}};
\node[color=cornellred] at(-1.5*0.38,-4.5*0.38+0.4+0.15){\scalebox{1.3}{$z_4^\star$}};
\node[color=blue] at(-1.5*0.38+1+0.35,-4.5*0.38+0.05+0.2){\scalebox{1.3}{$\mathcal{C}_4^\star$}};
\node[color=cornellred] at(-1.5*0.9,4.5*0.9+0.4+0.15){\scalebox{1.3}{$z_1^\star$}};
\node[color=Orange] at(-1.5*0.9+1.2,4.5*0.9+0.3+0.2){\scalebox{1.3}{$\bar{\mathcal{C}}_1^\star$}};
\node[color=cornellred] at(-1.5*0.9,-4.5*0.9-0.4-0.15){\scalebox{1.3}{$z_3^\star$}};
\node[color=Orange] at(-1.5*0.9+1.2,-4.5*0.9-0.3-0.25){\scalebox{1.3}{$\bar{\mathcal{C}}_3^\star$}};


\node[color=cornellred] at(-1.5*0.38+\hspace,4.5*0.38-0.4+\picspacing-0.15){\scalebox{1.3}{$z_2^\star$}};
\node[color=blue] at(-1.5*0.38-0.6+\hspace-0.7,4.5*0.38+0.2+\picspacing-0.4){\scalebox{1.3}{$\mathcal{C}_2^\star$}};
\node[color=cornellred] at(-1.5*0.38+\hspace,-4.5*0.38+0.4+\picspacing+0.15){\scalebox{1.3}{$z_4^\star$}};
\node[color=blue] at(-1.5*0.38+1+\hspace+0.15,-4.5*0.38+0.05+\picspacing-0.2){\scalebox{1.3}{$\mathcal{C}_4^\star$}};
\node[color=cornellred] at(-1.5*0.9+\hspace,4.5*0.9+0.4+\picspacing+0.15){\scalebox{1.3}{$z_1^\star$}};
\node[color=Orange] at(-1.5*0.9+1.2+\hspace,4.5*0.9+0.3+\picspacing+0.2){\scalebox{1.3}{$\bar{\mathcal{C}}_1^\star$}};
\node[color=cornellred] at(-1.5*0.9+\hspace,-4.5*0.9-0.4+\picspacing-0.15){\scalebox{1.3}{$z_3^\star$}};
\node[color=Orange] at(-1.5*0.9+1.2+\hspace,-4.5*0.9-0.3+\picspacing-0.25){\scalebox{1.3}{$\bar{\mathcal{C}}_3^\star$}};

    \end{tikzpicture}
    \caption{Pictorial representation of the Lefschetz-thimble topology change associated with two eigenvalues (black dots) taking place on the points of the unit circle distribution (denoted as the green wavy line) closest to the saddles (marked by red dots). The steepest-ascent contours associated with the saddles $z_1^\star,z_3^\star$ are denoted by $\bar{\mathcal{C}}_1^\star,\bar{\mathcal{C}}_3^\star$ and shown as orange lines while the steepest-descent contours associated with the saddles $z_2^\star,z_4^\star$ are denoted by $\mathcal{C}_2^\star,\mathcal{C}_4^\star$ and shown as blue lines.}
    \label{fig:Instantons2}
\end{figure}
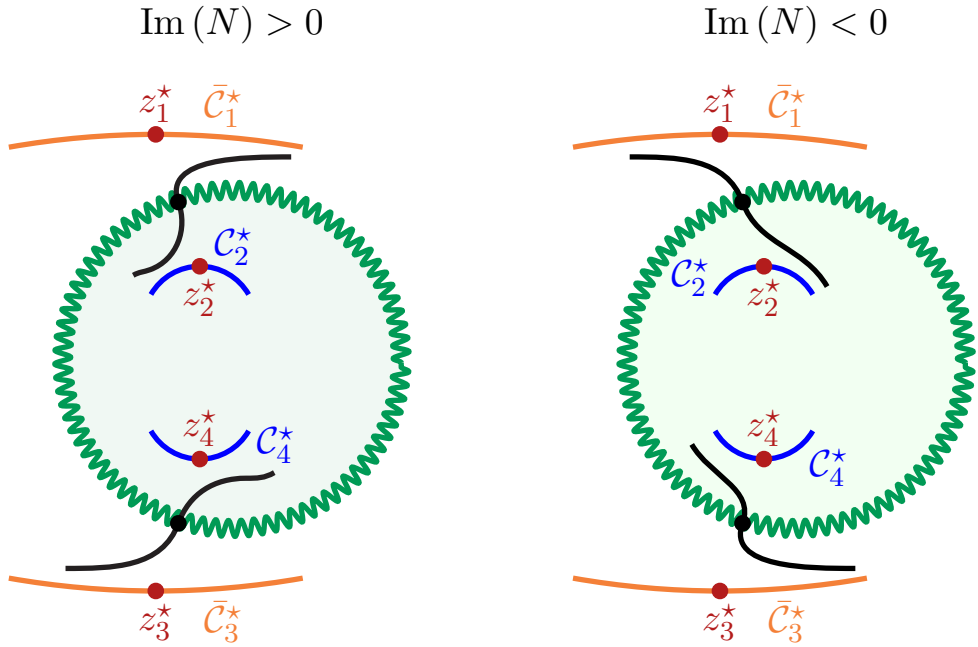

One expects the arguments of this appendix to generalize for more complicated single-trace unitary matrix models, such as \eqref{eq:potential}. The main novelty and added complication, is the existence of multiple saddle-points, both inside and outside the unit circle. As argued for in subsection \ref{subsec:instantons}, one expects to be able to deform the circular contours $\mathcal{C},\bar{\mathcal{C}}$ into disjoin unions of steepest-descent and steepest-ascent contours associated with saddles taking place inside and outside the unit circle, respectively (see figure \ref{fig:deformation}). Generalizing the argument of this appendix would entail studying the Lefschetz-thimble topology changes in these more general models and, in particular, verifying whether the tunneling to multiple saddles occurs simultaneously at $\text{Im}(t) = 0$. If so, a similar argument, based on the median resummation procedure, should be applicable to these cases as well. We depict in figure \ref{fig:Instantons2} the  Lefschetz-thimble topology change associated with the model \eqref{eq:potential}, where we can see the tunneling of eigenvalues and anti-eigenvalues happening simultaneously to all saddles, as expected.\footnote{In this figure, we are varying the argument of $N$, as there is no coupling $t$ in the model \eqref{eq:potential}.} 

Finally, we remark that the Lefschetz–thimble approach considered in the appendix directly motivates an instanton expansion that is already amenable to saddle-point analysis, since no contour deformation is required. This is in contrast with the instanton expansion \eqref{eq:expansionNcontribution}, whose saddle-point analysis necessitates a detailed understanding of how the circular contours $\mathcal{C}, \bar{\mathcal{C}}$ are deformed into the various steepest-descent and ascent contours.


\bibliographystyle{JHEP}
\bibliography{Bibliography.bib}

@article{Berenstein:2025itw,
    author = "Berenstein, David and Rodriguez, Victor A.",
    title = "{Goldilocks and the bootstrap}",
    eprint = "2503.00104",
    archivePrefix = "arXiv",
    primaryClass = "hep-th",
    doi = "10.1007/JHEP09(2025)109",
    journal = "JHEP",
    volume = "09",
    pages = "109",
    year = "2025"
}

@article{Belavin:1984vu,
    author = "Belavin, A. A. and Polyakov, Alexander M. and Zamolodchikov, A. B.",
    editor = "Khalatnikov, I. M. and Mineev, V. P.",
    title = "{Infinite Conformal Symmetry in Two-Dimensional Quantum Field Theory}",
    reportNumber = "CERN-TH-3827",
    doi = "10.1016/0550-3213(84)90052-X",
    journal = "Nucl. Phys. B",
    volume = "241",
    pages = "333--380",
    year = "1984"
}

@article{Lin:2020mme,
    author = "Lin, Henry W.",
    title = "{Bootstraps to strings: solving random matrix models with positivity}",
    eprint = "2002.08387",
    archivePrefix = "arXiv",
    primaryClass = "hep-th",
    doi = "10.1007/JHEP06(2020)090",
    journal = "JHEP",
    volume = "06",
    pages = "090",
    year = "2020"
}

@article{Kazakov:2021lel,
    author = "Kazakov, Vladimir and Zheng, Zechuan",
    title = "{Analytic and numerical bootstrap for one-matrix model and \textquotedblleft{}unsolvable\textquotedblright{} two-matrix model}",
    eprint = "2108.04830",
    archivePrefix = "arXiv",
    primaryClass = "hep-th",
    doi = "10.1007/JHEP06(2022)030",
    journal = "JHEP",
    volume = "06",
    pages = "030",
    year = "2022"
}

@article{Kazakov:2022xuh,
    author = "Kazakov, Vladimir and Zheng, Zechuan",
    title = "{Bootstrap for lattice Yang-Mills theory}",
    eprint = "2203.11360",
    archivePrefix = "arXiv",
    primaryClass = "hep-th",
    doi = "10.1103/PhysRevD.107.L051501",
    journal = "Phys. Rev. D",
    volume = "107",
    number = "5",
    pages = "L051501",
    year = "2023"
}

@article{Cho:2022lcj,
    author = "Cho, Minjae and Gabai, Barak and Lin, Ying-Hsuan and Rodriguez, Victor A. and Sandor, Joshua and Yin, Xi",
    title = "{Bootstrapping the Ising Model on the Lattice}",
    eprint = "2206.12538",
    archivePrefix = "arXiv",
    primaryClass = "hep-th",
    month = "6",
    year = "2022"
}

@article{Kazakov:2024ool,
    author = "Kazakov, Vladimir and Zheng, Zechuan",
    title = "{Bootstrap for finite N lattice Yang-Mills theory}",
    eprint = "2404.16925",
    archivePrefix = "arXiv",
    primaryClass = "hep-th",
    doi = "10.1007/JHEP03(2025)099",
    journal = "JHEP",
    volume = "03",
    pages = "099",
    year = "2025"
}

@article{Poland:2018epd,
    author = "Poland, David and Rychkov, Slava and Vichi, Alessandro",
    title = "{The Conformal Bootstrap: Theory, Numerical Techniques, and Applications}",
    eprint = "1805.04405",
    archivePrefix = "arXiv",
    primaryClass = "hep-th",
    doi = "10.1103/RevModPhys.91.015002",
    journal = "Rev. Mod. Phys.",
    volume = "91",
    pages = "015002",
    year = "2019"
}

@article{Rychkov:2023wsd,
    author = "Rychkov, Slava and Su, Ning",
    title = "{New developments in the numerical conformal bootstrap}",
    eprint = "2311.15844",
    archivePrefix = "arXiv",
    primaryClass = "hep-th",
    doi = "10.1103/RevModPhys.96.045004",
    journal = "Rev. Mod. Phys.",
    volume = "96",
    number = "4",
    pages = "045004",
    year = "2024"
}

@article{Rattazzi:2008pe,
    author = "Rattazzi, Riccardo and Rychkov, Vyacheslav S. and Tonni, Erik and Vichi, Alessandro",
    title = "{Bounding scalar operator dimensions in 4D CFT}",
    eprint = "0807.0004",
    archivePrefix = "arXiv",
    primaryClass = "hep-th",
    doi = "10.1088/1126-6708/2008/12/031",
    journal = "JHEP",
    volume = "12",
    pages = "031",
    year = "2008"
}

@article{El-Showk:2012cjh,
    author = "El-Showk, Sheer and Paulos, Miguel F. and Poland, David and Rychkov, Slava and Simmons-Duffin, David and Vichi, Alessandro",
    title = "{Solving the 3D Ising Model with the Conformal Bootstrap}",
    eprint = "1203.6064",
    archivePrefix = "arXiv",
    primaryClass = "hep-th",
    reportNumber = "LPTENS-12-07",
    doi = "10.1103/PhysRevD.86.025022",
    journal = "Phys. Rev. D",
    volume = "86",
    pages = "025022",
    year = "2012"
}

@article{Han:2020bkb,
    author = "Han, Xizhi and Hartnoll, Sean A. and Kruthoff, Jorrit",
    title = "{Bootstrapping Matrix Quantum Mechanics}",
    eprint = "2004.10212",
    archivePrefix = "arXiv",
    primaryClass = "hep-th",
    doi = "10.1103/PhysRevLett.125.041601",
    journal = "Phys. Rev. Lett.",
    volume = "125",
    number = "4",
    pages = "041601",
    year = "2020"
}

@article{Lin:2023owt,
    author = "Lin, Henry W.",
    title = "{Bootstrap bounds on D0-brane quantum mechanics}",
    eprint = "2302.04416",
    archivePrefix = "arXiv",
    primaryClass = "hep-th",
    doi = "10.1007/JHEP06(2023)038",
    journal = "JHEP",
    volume = "06",
    pages = "038",
    year = "2023"
}

@article{Lin:2025srf,
    author = "Lin, Henry W. and Zheng, Zechuan",
    title = "{High-precision bootstrap of multi-matrix quantum mechanics}",
    eprint = "2507.21007",
    archivePrefix = "arXiv",
    primaryClass = "hep-th",
    month = "7",
    year = "2025"
}

@article{Cho:2024owx,
    author = "Cho, Minjae and Nancarrow, Colin Oscar and Tadi{\'c}, Petar and Xin, Yuan and Zheng, Zechuan",
    title = "{Coarse-grained Bootstrap of Quantum Many-body Systems}",
    eprint = "2412.07837",
    archivePrefix = "arXiv",
    primaryClass = "hep-th",
    month = "12",
    year = "2024"
}

@article{Aikawa:2021eai,
    author = "Aikawa, Yu and Morita, Takeshi and Yoshimura, Kota",
    title = "{Application of bootstrap to a {\ensuremath{\theta}} term}",
    eprint = "2109.02701",
    archivePrefix = "arXiv",
    primaryClass = "hep-th",
    doi = "10.1103/PhysRevD.105.085017",
    journal = "Phys. Rev. D",
    volume = "105",
    number = "8",
    pages = "085017",
    year = "2022"
}

@article{Corley:2001zk,
    author = "Corley, Steve and Jevicki, Antal and Ramgoolam, Sanjaye",
    title = "{Exact correlators of giant gravitons from dual N=4 SYM theory}",
    eprint = "hep-th/0111222",
    archivePrefix = "arXiv",
    reportNumber = "BROWN-HET-1292",
    doi = "10.4310/ATMP.2001.v5.n4.a6",
    journal = "Adv. Theor. Math. Phys.",
    volume = "5",
    pages = "809--839",
    year = "2002"
}

@article{Berenstein:2013md,
    author = "Berenstein, David",
    title = "{Giant gravitons: a collective coordinate approach}",
    eprint = "1301.3519",
    archivePrefix = "arXiv",
    primaryClass = "hep-th",
    doi = "10.1103/PhysRevD.87.126009",
    journal = "Phys. Rev. D",
    volume = "87",
    number = "12",
    pages = "126009",
    year = "2013"
}

@article{Berenstein:2017abm,
    author = "Berenstein, David and Miller, Alexandra",
    title = "{Superposition induced topology changes in quantum gravity}",
    eprint = "1702.03011",
    archivePrefix = "arXiv",
    primaryClass = "hep-th",
    doi = "10.1007/JHEP11(2017)121",
    journal = "JHEP",
    volume = "11",
    pages = "121",
    year = "2017"
}

@article{Berenstein:2018lrm,
    author = "Berenstein, David",
    title = "{Submatrix deconfinement and small black holes in AdS}",
    eprint = "1806.05729",
    archivePrefix = "arXiv",
    primaryClass = "hep-th",
    doi = "10.1007/JHEP09(2018)054",
    journal = "JHEP",
    volume = "09",
    pages = "054",
    year = "2018"
}

@article{Berenstein:2004kk,
    author = "Berenstein, David",
    title = "{A Toy model for the AdS / CFT correspondence}",
    eprint = "hep-th/0403110",
    archivePrefix = "arXiv",
    doi = "10.1088/1126-6708/2004/07/018",
    journal = "JHEP",
    volume = "07",
    pages = "018",
    year = "2004"
}

@article{Berenstein:2023srv,
    author = "Berenstein, David and Yan, Kai",
    title = "{The endpoint of partial deconfinement}",
    eprint = "2307.06122",
    archivePrefix = "arXiv",
    primaryClass = "hep-th",
    doi = "10.1007/JHEP12(2023)030",
    journal = "JHEP",
    volume = "12",
    pages = "030",
    year = "2023"
}

@article{Goldschmidt:1979hq,
    author = "Goldschmidt, Yadin Y.",
    title = "{1/$N$ Expansion in Two-dimensional Lattice Gauge Theory}",
    reportNumber = "SACLAY-DPh-T 79/153",
    doi = "10.1063/1.524600",
    journal = "J. Math. Phys.",
    volume = "21",
    pages = "1842",
    year = "1980"
}

@inproceedings{Douglas:1993wy,
    author = "Douglas, Michael R.",
    title = "{Conformal field theory techniques in large N Yang-Mills theory}",
    booktitle = "{NATO Advanced Research Workshop on New Developments in String Theory, Conformal Models and Topological Field Theory}",
    eprint = "hep-th/9311130",
    archivePrefix = "arXiv",
    reportNumber = "RU-93-57",
    month = "5",
    year = "1993"
}

@article{diaconis1994eigenvalues,
  title={On the eigenvalues of random matrices},
  author={Diaconis, Persi and Shahshahani, Mehrdad},
  journal={Journal of Applied Probability},
  volume={31},
  number={A},
  pages={49--62},
  year={1994},
  publisher={Cambridge University Press}
}

@article{Gross:1993hu,
    author = "Gross, David J. and Taylor, Washington",
    title = "{Two-dimensional QCD is a string theory}",
    eprint = "hep-th/9301068",
    archivePrefix = "arXiv",
    reportNumber = "LBL-33458, PUPT-1376, UCB-PTH-93-02",
    doi = "10.1016/0550-3213(93)90403-C",
    journal = "Nucl. Phys. B",
    volume = "400",
    pages = "181--208",
    year = "1993"
}

@article{Chattopadhyay:2017ckc,
    author = "Chattopadhyay, Arghya and Dutta, Parikshit and Dutta, Suvankar",
    title = "{Emergent Phase Space Description of Unitary Matrix Model}",
    eprint = "1708.03298",
    archivePrefix = "arXiv",
    primaryClass = "hep-th",
    doi = "10.1007/JHEP11(2017)186",
    journal = "JHEP",
    volume = "11",
    pages = "186",
    year = "2017"
}

@article{Gross:1980he,
    author = "Gross, D. J. and Witten, Edward",
    title = "{Possible Third Order Phase Transition in the Large N Lattice Gauge Theory}",
    doi = "10.1103/PhysRevD.21.446",
    journal = "Phys. Rev. D",
    volume = "21",
    pages = "446--453",
    year = "1980"
}

@article{Balasubramanian:2005mg,
    author = "Balasubramanian, Vijay and de Boer, Jan and Jejjala, Vishnu and Simon, Joan",
    title = "{The Library of Babel: On the origin of gravitational thermodynamics}",
    eprint = "hep-th/0508023",
    archivePrefix = "arXiv",
    reportNumber = "UPR-1127-7, ITFA-2005-37, DCTP-05-33",
    doi = "10.1088/1126-6708/2005/12/006",
    journal = "JHEP",
    volume = "12",
    pages = "006",
    year = "2005"
}

@article{Wadia:1980cp,
    author = "Wadia, Spenta R.",
    title = "{$N$ = Infinity Phase Transition in a Class of Exactly Soluble Model Lattice Gauge Theories}",
    reportNumber = "EFI-80/15-CHICAGO",
    doi = "10.1016/0370-2693(80)90353-6",
    journal = "Phys. Lett. B",
    volume = "93",
    pages = "403--410",
    year = "1980"
}

@article{Huang:2025sua,
    author = "Huang, Zhijian and Li, Wenliang",
    title = "{Bootstrapping periodic quantum systems}",
    eprint = "2507.02386",
    archivePrefix = "arXiv",
    primaryClass = "hep-th",
    month = "7",
    year = "2025"
}

@article{Berenstein:2021dyf,
    author = "Berenstein, David and Hulsey, George",
    title = "{Bootstrapping Simple QM Systems}",
    eprint = "2108.08757",
    archivePrefix = "arXiv",
    primaryClass = "hep-th",
    month = "8",
    year = "2021"
}

@article{Tchoumakov:2021mnh,
    author = "Tchoumakov, Serguei and Florens, Serge",
    title = "{Bootstrapping Bloch bands}",
    eprint = "2109.06600",
    archivePrefix = "arXiv",
    primaryClass = "cond-mat.mes-hall",
    doi = "10.1088/1751-8121/ac3c82",
    journal = "J. Phys. A",
    volume = "55",
    number = "1",
    pages = "015203",
    year = "2022"
}

@article{Berenstein:2021loy,
    author = "Berenstein, David and Hulsey, George",
    title = "{Bootstrapping more QM systems}",
    eprint = "2109.06251",
    archivePrefix = "arXiv",
    primaryClass = "hep-th",
    doi = "10.1088/1751-8121/ac7118",
    journal = "J. Phys. A",
    volume = "55",
    number = "27",
    pages = "275304",
    year = "2022"
}

@article{Fawzi2024,
  author={Fawzi, Hamza and Fawzi, Omar and Scalet, Samuel O.},
  title        = {Certified algorithms for equilibrium states of local quantum Hamiltonians},
  journal      = {Nature Communications},
  year         = {2024},
  volume       = {15},
  number       = {1},
  pages        = {7394},
  doi          = {10.1038/s41467-024-51592-3},
  url          = {https://doi.org/10.1038/s41467-024-51592-3},
  issn         = {2041-1723},
  date         = {2024-08-27},
  eprint = {2311.18706}
}

@article{Cho:2024kxn,
    author = "Cho, Minjae and Gabai, Barak and Sandor, Joshua and Yin, Xi",
    title = "{Thermal bootstrap of matrix quantum mechanics}",
    eprint = "2410.04262",
    archivePrefix = "arXiv",
    primaryClass = "hep-th",
    doi = "10.1007/JHEP04(2025)186",
    journal = "JHEP",
    volume = "04",
    pages = "186",
    year = "2025"
}

@article{Lawrence:2024mnj,
    author = "Lawrence, Scott and McPeak, Brian and Neill, Duff",
    title = "{Bootstrapping time-evolution in quantum mechanics}",
    eprint = "2412.08721",
    archivePrefix = "arXiv",
    primaryClass = "hep-th",
    reportNumber = "LA-UR-24-33001",
    month = "12",
    year = "2024"
}

@article{Cho:2025vws,
    author = "Cho, Minjae and Gabai, Barak and Lin, Henry W. and Yeh, Jessica and Zheng, Zechuan",
    title = "{Bootstrapping Euclidean Two-point Correlators}",
    eprint = "2511.08560",
    archivePrefix = "arXiv",
    primaryClass = "hep-th",
    month = "11",
    year = "2025"
}

@article{Gliozzi:2013ysa,
    author = "Gliozzi, Ferdinando",
    title = "{More constraining conformal bootstrap}",
    eprint = "1307.3111",
    archivePrefix = "arXiv",
    primaryClass = "hep-th",
    doi = "10.1103/PhysRevLett.111.161602",
    journal = "Phys. Rev. Lett.",
    volume = "111",
    pages = "161602",
    year = "2013"
}

@article{Anderson:2016rcw,
    author = "Anderson, Peter D. and Kruczenski, Martin",
    title = "{Loop Equations and bootstrap methods in the lattice}",
    eprint = "1612.08140",
    archivePrefix = "arXiv",
    primaryClass = "hep-th",
    doi = "10.1016/j.nuclphysb.2017.06.009",
    journal = "Nucl. Phys. B",
    volume = "921",
    pages = "702--726",
    year = "2017"
}

@article{Bender:2023ttu,
    author = "Bender, Carl M. and Karapoulitidis, Christos and Klevansky, S. P.",
    title = "{Dyson-Schwinger equations in zero dimensions and polynomial approximations}",
    eprint = "2307.01008",
    archivePrefix = "arXiv",
    primaryClass = "math-ph",
    doi = "10.1103/PhysRevD.108.056002",
    journal = "Phys. Rev. D",
    volume = "108",
    number = "5",
    pages = "056002",
    year = "2023"
}

@article{Li:2017ukc,
    author = "Li, Wenliang",
    title = "{New method for the conformal bootstrap with OPE truncations}",
    eprint = "1711.09075",
    archivePrefix = "arXiv",
    primaryClass = "hep-th",
    month = "11",
    year = "2017"
}

@article{Hu:2025yrs,
    author = "Hu, Runzhe and Li, Wenliang",
    title = "{Accurate boundary bootstrap for the three-dimensional O($N$) normal universality class}",
    eprint = "2508.20854",
    archivePrefix = "arXiv",
    primaryClass = "hep-th",
    month = "8",
    year = "2025"
}

@article{MARCHESINI1985225,
title = {Convergence of the iterative solution of loop equations in planar QCD2},
journal = {Nuclear Physics B},
volume = {249},
number = {2},
pages = {225-243},
year = {1985},
issn = {0550-3213},
doi = {https://doi.org/10.1016/0550-3213(85)90016-1},
url = {https://www.sciencedirect.com/science/article/pii/0550321385900161},
author = {G. Marchesini and E. Onofri},
}

@article{Eniceicu:2023cxn,
    author = "Eniceicu, Dan Stefan and Mahajan, Raghu and Murdia, Chitraang",
    title = "{Complex eigenvalue instantons and the Fredholm determinant expansion in the Gross-Witten-Wadia model}",
    eprint = "2308.06320",
    archivePrefix = "arXiv",
    primaryClass = "hep-th",
    doi = "10.1007/JHEP01(2024)129",
    journal = "JHEP",
    volume = "01",
    pages = "129",
    year = "2024"
}

@article{mss22,
      title={New Instantons for Matrix Models}, 
      author={Marcos Marino and Ricardo Schiappa and Maximilian Schwick},
      year={2022},
      eprint={2210.13479},
      archivePrefix={arXiv},
      primaryClass={hep-th},
      url={https://arxiv.org/abs/2210.13479}, 
}

@article{emt24,
      title={The complete non-perturbative partition function of minimal superstring theory and JT supergravity}, 
      author={Dan Stefan Eniceicu and Chitraang Murdia and Andrii Torchylo},
      year={2025},
      eprint={2412.08698},
      archivePrefix={arXiv},
      primaryClass={hep-th},
      url={https://arxiv.org/abs/2412.08698}, 
}

@article{cmt24,
      title={Giant graviton expansion from eigenvalue instantons}, 
      author={Yiming Chen and Raghu Mahajan and Haifeng Tang},
      year={2024},
      eprint={2407.08155},
      archivePrefix={arXiv},
      primaryClass={hep-th},
      url={https://arxiv.org/abs/2407.08155}, 
}

@article{e23,
      title={Comments on the Giant-Graviton Expansion of the Superconformal Index}, 
      author={Dan Stefan Eniceicu},
      year={2023},
      eprint={2302.04887},
      archivePrefix={arXiv},
      primaryClass={hep-th},
      url={https://arxiv.org/abs/2302.04887}, 
}

@article{bt16,
   title={On Asymptotic Regimes of Orthogonal Polynomials with Complex Varying Quartic Exponential Weight},
   ISSN={1815-0659},
   url={http://dx.doi.org/10.3842/SIGMA.2016.118},
   DOI={10.3842/sigma.2016.118},
   journal={Symmetry, Integrability and Geometry: Methods and Applications},
   publisher={SIGMA (Symmetry, Integrability and Geometry: Methods and Application)},
   author={Bertola, Marco and Tovbis, Alexander},
   year={2016},
   month=dec,
   eprint={1612.08732}
   }

@article{abs18,
   title={A primer on resurgent transseries and their asymptotics},
   volume={809},
   ISSN={0370-1573},
   url={http://dx.doi.org/10.1016/j.physrep.2019.02.003},
   DOI={10.1016/j.physrep.2019.02.003},
   journal={Physics Reports},
   publisher={Elsevier BV},
   author={Aniceto, Inês and Başar, Gökçe and Schiappa, Ricardo},
   year={2019},
   month=may, pages={1–135},
   eprint = {1802.10441}}

@article{d93,
   title={Non-perturbative effects in matrix models and vacua of two dimensional gravity},
   volume={302},
   ISSN={0370-2693},
   url={http://dx.doi.org/10.1016/0370-2693(93)90417-G},
   DOI={10.1016/0370-2693(93)90417-g},
   number={4},
   journal={Physics Letters B},
   publisher={Elsevier BV},
   author={David, François},
   year={1993},
   month=apr, pages={403–410},
   eprint = {hep-th/9212106}}

@article{msw08,
    author = "Marino, Marcos and Schiappa, Ricardo and Weiss, Marlene",
    title = "{Nonperturbative Effects and the Large-Order Behavior of Matrix Models and Topological Strings}",
    eprint = "0711.1954",
    archivePrefix = "arXiv",
    primaryClass = "hep-th",
    reportNumber = "CERN-PH-TH-2007-218",
    doi = "10.4310/CNTP.2008.v2.n2.a3",
    journal = "Commun. Num. Theor. Phys.",
    volume = "2",
    pages = "349--419",
    year = "2008",
     
}

@article{krsst25a,
      title={Exact Solutions to Matrix Models and String Theories: The Local Construction}, 
      author={Jasper Kager and João Rodrigues and Ricardo Schiappa and Maximilian Schwick and Noam Tamarin},
      year={2026},
      eprint={2602.15101},
      archivePrefix={arXiv},
      primaryClass={hep-th},
      url={https://arxiv.org/abs/2602.15101}, 
}

@article{cgkt22,
    author = "Copetti, Christian and Grassi, Alba and Komargodski, Zohar and Tizzano, Luigi",
    title = "{Delayed deconfinement and the Hawking-Page transition}",
    eprint = "2008.04950",
    archivePrefix = "arXiv",
    primaryClass = "hep-th",
    doi = "10.1007/JHEP04(2022)132",
    journal = "JHEP",
    volume = "04",
    pages = "132",
    year = "2022"
}

@article{d90a,
    author = "David, Francois",
    title = "{Phases of the large N matrix model and nonperturbative effects in 2-d gravity}",
    reportNumber = "SACLAY-SPH-T-90-090",
    doi = "10.1016/0550-3213(91)90202-9",
    journal = "Nucl. Phys. B",
    volume = "348",
    pages = "507--524",
    year = "1991"
}

@article{ai19,
  author  = {Arai, Ryohei and Imamura, Yuji},
  title   = {Finite {$N$} Corrections to the Superconformal Index of {$S$}-fold Theories},
  journal = {Prog. Theor. Exp. Phys.},
  year    = {2019},
  number  = {8},
  pages   = {083B04},
  eprint  = {1904.09776},
  archivePrefix = {arXiv},
  primaryClass  = {hep-th}
}

@article{afim19,
  author  = {Arai, Ryohei and Fujiwara, Shun and Imamura, Yuji and Mori, Takuya},
  title   = {Finite {$N$} corrections to the superconformal index of orbifold quiver gauge theories},
  journal = {JHEP},
  year    = {2019},
  volume  = {10},
  pages   = {243},
  eprint  = {1907.05660},
  archivePrefix = {arXiv},
  primaryClass  = {hep-th}
}

@article{afim20a,
  author  = {Arai, Ryohei and Fujiwara, Shun and Imamura, Yuji and Mori, Takuya},
  title   = {Finite {$N$} corrections to the superconformal index of toric quiver gauge theories},
  journal = {Prog. Theor. Exp. Phys.},
  year    = {2020},
  number  = {4},
  pages   = {043B09},
  eprint  = {1911.10794},
  archivePrefix = {arXiv},
  primaryClass  = {hep-th}
}

@article{afim20b,
  author  = {Arai, Ryohei and Fujiwara, Shun and Imamura, Yuji and Mori, Takuya},
  title   = {Schur index of the {$\mathcal{N}=4$} {$U(N)$} supersymmetric {Y}ang--{M}ills theory via the {AdS}/{CFT} correspondence},
  journal = {Phys. Rev. D},
  year    = {2020},
  volume  = {101},
  number  = {8},
  pages   = {086017},
  eprint  = {2001.11667},
  archivePrefix = {arXiv},
  primaryClass  = {hep-th}
}

@article{afimy20,
  author  = {Arai, Ryohei and Fujiwara, Shun and Imamura, Yuji and Mori, Takuya and Yokoyama, Daisuke},
  title   = {Finite-{$N$} corrections to the {M}-brane indices},
  journal = {JHEP},
  year    = {2020},
  volume  = {11},
  pages   = {093},
  eprint  = {2007.05213},
  archivePrefix = {arXiv},
  primaryClass  = {hep-th}
}

@article{fim21,
  author  = {Fujiwara, Shun and Imamura, Yuji and Mori, Takuya},
  title   = {Flavor symmetries of six-dimensional {$\mathcal{N}=(1,0)$} theories from the {AdS}/{CFT} correspondence},
  journal = {JHEP},
  year    = {2021},
  volume  = {05},
  pages   = {221},
  eprint  = {2103.16094},
  archivePrefix = {arXiv},
  primaryClass  = {hep-th}
}

@article{i21,
  author  = {Imamura, Yuji},
  title   = {Finite-{$N$} superconformal index via the {AdS}/{CFT} correspondence},
  journal = {Prog. Theor. Exp. Phys.},
  year    = {2021},
  number  = {12},
  pages   = {123B05},
  eprint  = {2108.12090},
  archivePrefix = {arXiv},
  primaryClass  = {hep-th}
}

@article{gl24,
    author = "Gaiotto, Davide and Lee, Ji Hoon",
    title = "{The giant graviton expansion}",
    eprint = "2109.02545",
    archivePrefix = "arXiv",
    primaryClass = "hep-th",
    doi = "10.1007/JHEP08(2024)025",
    journal = "JHEP",
    volume = "08",
    pages = "025",
    year = "2024"
}

@article{m22,
    author = "Murthy, Sameer",
    title = "{Unitary matrix models, free fermions, and the giant graviton expansion}",
    eprint = "2202.06897",
    archivePrefix = "arXiv",
    primaryClass = "hep-th",
    doi = "10.4310/PAMQ.2023.v19.n1.a12",
    journal = "Pure Appl. Math. Quart.",
    volume = "19",
    number = "1",
    pages = "299--340",
    year = "2023"
}

@article{s14,
    author = "Sauzin, David",
    title = "{Introduction to 1-summability and resurgence}",
    eprint = "1405.0356",
    archivePrefix = "arXiv",
    primaryClass = "math.DS",
    month = "5",
    year = "2014"
}

@article{ar15,
    author = "Aniceto, In{\^e}s and Schiappa, Ricardo",
    title = "{Nonperturbative Ambiguities and the Reality of Resurgent Transseries}",
    eprint = "1308.1115",
    archivePrefix = "arXiv",
    primaryClass = "hep-th",
    doi = "10.1007/s00220-014-2165-z",
    journal = "Commun. Math. Phys.",
    volume = "335",
    number = "1",
    pages = "183--245",
    year = "2015"
}

@article{m08,
   title={Nonperturbative effects and nonperturbative definitions in matrix models and topological strings},
   volume={2008},
   ISSN={1029-8479},
   url={http://dx.doi.org/10.1088/1126-6708/2008/12/114},
   DOI={10.1088/1126-6708/2008/12/114},
   number={12},
   journal={Journal of High Energy Physics},
   publisher={Springer Science and Business Media LLC},
   author={Mariño, Marcos},
   year={2008},
   month=dec, pages={114–114},
   eprint = {0805.3033}}

@article{o99,
  title={Infinite wedge and random partitions},
  author={Andrei Okounkov},
  journal={Selecta Mathematica},
  year={1999},
  volume={7},
  pages={57-81},
  url={https://api.semanticscholar.org/CorpusID:119176413},
  eprint = {math/9907127} 
}

@article{bo99,
  title={A Fredholm determinant formula for Toeplitz determinants},
  author={Alexei Borodin and Andrei Okounkov},
  journal={Integral Equations and Operator Theory},
  year={1999},
  volume={37},
  pages={386-396},
  url={https://api.semanticscholar.org/CorpusID:17727009},
  eprint = {math/9907165}
}

@article{lr23,
   title={Finite N indices and the giant graviton expansion},
   volume={2023},
   ISSN={1029-8479},
   url={http://dx.doi.org/10.1007/JHEP04(2023)078},
   DOI={10.1007/jhep04(2023)078},
   number={4},
   journal={Journal of High Energy Physics},
   publisher={Springer Science and Business Media LLC},
   author={Liu, James T. and Rajappa, Neville Joshua},
   year={2023},
   month=apr, eprint = {2212.05408}}

@article{b09,
title = {Commuting difference operators, spinor bundles and the asymptotics of orthogonal polynomials with respect to varying complex weights},
journal = {Advances in Mathematics},
volume = {220},
number = {1},
pages = {154-218},
year = {2009},
issn = {0001-8708},
doi = {https://doi.org/10.1016/j.aim.2008.09.001},
url = {https://www.sciencedirect.com/science/article/pii/S000187080800251X},
author = {M. Bertola and M.Y. Mo},
eprint = {math-ph/0605043}
}

@article{b07,
      title={Boutroux curves with external field: equilibrium measures without a minimization problem}, 
      author={Marco Bertola},
      year={2007},
      eprint={0705.3062},
      archivePrefix={arXiv},
      primaryClass={nlin.SI},
      url={https://arxiv.org/abs/0705.3062}, 
}

@article{s07,
  title={EQUILIBRIUM MEASURES AND CAPACITIES IN SPECTRAL THEORY},
  author={Barry Simon},
  journal={Inverse Problems and Imaging},
  year={2007},
  volume={1},
  pages={713-772},
  url={https://api.semanticscholar.org/CorpusID:463738},
  eprint = {0711.2700}
}

@article{bt11,
  title={Asymptotics of Orthogonal Polynomials with Complex Varying Quartic Weight: Global Structure, Critical Point Behavior and the First Painlev{\'e} Equation},
  author={Marco Bertola and Alexander Tovbis},
  journal={Constructive Approximation},
  year={2011},
  volume={41},
  pages={529-587},
  url={https://api.semanticscholar.org/CorpusID:119637028},
  eprint={1108.0321}
}

@article{aam13,
  title={Determination of S-curves with applications to the theory of non-Hermitian orthogonal polynomials},
  author={Gabriel {\'A}lvarez and Luis Mart{\'i}nez Alonso and Elena Medina},
  journal={Journal of Statistical Mechanics: Theory and Experiment},
  year={2013},
  volume={2013},
  url={https://api.semanticscholar.org/CorpusID:52970317},
  eprint={1305.3028}
}

@article{hkl14,
title = {Zero distribution of complex orthogonal polynomials with respect to exponential weights},
journal = {Journal of Approximation Theory},
volume = {184},
pages = {28-54},
year = {2014},
issn = {0021-9045},
doi = {https://doi.org/10.1016/j.jat.2014.05.002},
url = {https://www.sciencedirect.com/science/article/pii/S0021904514000860},
author = {Daan Huybrechs and Arno B.J. Kuijlaars and Nele Lejon},
eprint={1312.4376}
}

@article{m89,
    author = "Mandal, Gautam",
    title = "{Phase Structure of Unitary Matrix Models}",
    reportNumber = "IASSNS-HEP-89/68",
    doi = "10.1142/S0217732390001281",
    journal = "Mod. Phys. Lett. A",
    volume = "5",
    pages = "1147--1158",
    year = "1990"
}

@article{bdv16,
   title={Complex Path Integrals and Saddles in Two-Dimensional Gauge Theory},
   volume={116},
   ISSN={1079-7114},
   url={http://dx.doi.org/10.1103/PhysRevLett.116.132001},
   DOI={10.1103/physrevlett.116.132001},
   number={13},
   journal={Physical Review Letters},
   publisher={American Physical Society (APS)},
   author={Buividovich, P. V. and Dunne, Gerald V. and Valgushev, S. N.},
   year={2016},
   month=mar,
   eprint = {1512.09021}}

@article{sst23,
    author = "Schiappa, Ricardo and Schwick, Maximilian and Tamarin, Noam",
    title = "{All the D-Branes of Resurgence}",
    eprint = "2301.05214",
    archivePrefix = "arXiv",
    primaryClass = "hep-th",
    month = "1",
    year = "2023"
}

@article{dp99,
  TITLE = {{Resurgent methods in semi-classical asymptotics}},
  AUTHOR = {Delabaere, Eric and Pham, Fr{\'e}d{\'e}ric},
  URL = {https://hal.science/hal-01886535},
  JOURNAL = {{Annales de l'Institut Henri Poincar{\'e} (A). Physique Theorique}},
  PUBLISHER = {{Birkh{\"a}user}},
  VOLUME = {71},
  NUMBER = {1},
  PAGES = {1--94},
  YEAR = {1999},
  HAL_ID = {hal-01886535},
  HAL_VERSION = {v1},
}

@ARTICLE{ps92,
       author = {{Pisani}, C. and {Smith}, E.~R.},
        title = "{Lee-Yang zeros and stokes phenomenon in a model with a wetting transition}",
      journal = {Journal of Statistical Physics},
         year = 1993,
        month = jul,
       volume = {72},
       number = {1-2},
        pages = {51-78},
          doi = {10.1007/BF01048040},
       adsurl = {https://ui.adsabs.harvard.edu/abs/1993JSP....72...51P}
}

@article{dg18,
    author = "Dorigoni, Daniele and Glass, Philip",
    title = "{The grin of Cheshire cat resurgence from supersymmetric localization}",
    eprint = "1711.04802",
    archivePrefix = "arXiv",
    primaryClass = "hep-th",
    reportNumber = "DCPT-17-31",
    doi = "10.21468/SciPostPhys.4.2.012",
    journal = "SciPost Phys.",
    volume = "4",
    number = "2",
    pages = "012",
    year = "2018"
}

@article{kstu16,
   title={Cheshire Cat Resurgence, Self-Resurgence and Quasi-Exact Solvable Systems},
   volume={364},
   ISSN={1432-0916},
   url={http://dx.doi.org/10.1007/s00220-018-3281-y},
   DOI={10.1007/s00220-018-3281-y},
   number={3},
   journal={Communications in Mathematical Physics},
   publisher={Springer Science and Business Media LLC},
   author={Kozçaz, Can and Sulejmanpasic, Tin and Tanizaki, Yuya and Ünsal, Mithat},
   year={2018},
   month=oct, pages={835–878} }

\end{document}